\documentclass[12pt]{article}
\usepackage{graphicx}
\def\beq{\begin{equation}}
\def\eeq{\end{equation}}
\def\ba{\begin{eqnarray}}
\def\ea{\end{eqnarray}}
\def\ga{\mathrel{\raise.3ex\hbox{$>$\kern-.75em\lower1ex\hbox{$\sim$}}}}
\def\la{\mathrel{\raise.3ex\hbox{$<$\kern-.75em\lower1ex\hbox{$\sim$}}}}
\def\wt#1{{\widetilde {#1}}}
\def\t#1{{\tilde {#1}}}

\def\NP{{\it Nucl.Phys.} }
\def\PL{{\it Phys.Lett.} }
\def\PR{{\it Phys.Rev.} }
\def\PRL{{\it Phys.Rev.Lett.} }

\begin{document}
\begin{titlepage} 
\pagestyle{empty}
\rightline{UMN--TH--1824/99}
\rightline{TPI--MINN--99/49}
\rightline{hep-ph/9911307}
\rightline{November 1999}  
\vspace{1.5cm}
\begin{center}
\baselineskip=21pt
{\large {\bf Introduction to Supersymmetry:\\[1mm]
Astrophysical and Phenomenological Constraints}\footnote{Based  on lectures
delivered at the Les Houches Summer School, July 1999}}
\vspace{1cm}

{Keith A. Olive}

{\it Theoretical Physics Institute, \\ School of Physics and
Astronomy, \\ University of Minnesota, \\ Minneapolis, MN 55455, USA}\\
\vspace*{.5in}
{\bf Abstract}
\end{center}

These lectures contain an introduction to supersymmetric theories and the
minimal supersymmetric standard model.  Phenomenological and cosmological
consequences of supersymmetry are also discussed.

\end{titlepage}


\section{Introduction}

It is often called the last great symmetry of nature.  Rarely has
so much effort, both theoretical and experimental, been spent to understand
and discover a symmetry of nature, which up to the present time lacks 
concrete evidence. Hopefully,  in these lectures, where I will give a
pedagogical description of supersymmetric theories, it will become clear why
there is so much excitement concerning supersymmetry's role in nature. 

After some preliminary background on the standard electroweak model, and
some motivation for supersymmetry, I will introduce the notion of
supersymmetric charges and the supersymmetric transformation laws. The second
lecture will present the simplest supersymmetric model (the non-interacting
massless Wess-Zumino model) and develop the properties of chiral superfields,
auxiliary fields, the superpotential, gauge multiplets and interactions.
The next two lectures focus on the minimal supersymmetric standard model
(MSSM) and its constrained version which is motivated by supergravity. The
last two lectures will look primarily at the cosmological and phenomenological
consequences of supersymmetry.  

\subsection{Some Preliminaries}
Why Supersymmetry? If for no other reason, it would be nice to understand
the origin of the fundamental difference between the two classes of particles
distinguished by their spin, fermions and bosons. If such a symmetry exists,
one might expect that it is represented by an operator which relates the two
classes of particles.  For example, 
\ba
Q | {\rm Boson} \rangle & = & | {\rm Fermion} \rangle \nonumber \\
Q | {\rm Fermion} \rangle & = & | {\rm Boson} \rangle
\label{ss}
\ea
As such, one could claim a raison d'etre for fundamental
scalars in nature.  Aside from the Higgs boson (which remains  to
be discovered), there are no fundamental scalars known to exist. A symmetry
as potentially powerful as that in eq. (\ref{ss}) is reason enough for its
study. However, without a connection to experiment, supersymmetry would
remain a mathematical curiosity and a subject of a very theoretical nature as
indeed it stood from its initial description in the early 1970's 
\cite{GL,WZ} until its incorporation into a realistic theory of physics at
the electroweak scale. 

One  of the first break-throughs came with the realization that 
supersymmetry could help resolve the difficult problem of mass hierarchies
\cite{hier}, namely the stability of the electroweak scale with respect to
radiative corrections. With precision experiments at the electroweak scale,
it has also become apparent that grand unification is not possible in the
absence of supersymmetry \cite{EKN}. These issues will be discussed in more
detail below.

Because one of our main goals is to discuss the MSSM, it will be useful to
first describe some key features of the standard model if for no other reason
than to establish the notation used below. The standard model is described by
the SU(3)$_{\rm c} \times$ SU(2)$_{\rm L} \times$ U(1)$_{\rm Y}$ gauge group.
For the most part, however, I will restrict the present discussion to the
electroweak sector.  The Lagrangian for the gauge sector of the theory can be
written as
\beq
{\cal L}_G = -\frac{1}{4}~G^i_{\mu\nu} G^{i\mu\nu} - {1\over 4}
F_{\mu\nu}F^{\mu\nu}
\label{ewg}
\eeq
where $G^i_{\mu\nu} \equiv \partial_\mu W^i_\nu - \partial_\nu W^i_\mu +
g \epsilon^{ijk}W^j_\mu W^k_\nu$ is the field strength for the $SU(2)$
gauge boson $W^i_\mu$, and $F_{\mu\nu} \equiv \partial_\mu B_\nu - 
\partial_\nu B_\mu$ is the field strength for the $U(1)$ gauge boson
$B_\mu$.
The fermion kinetic terms are included in
\beq
{\cal L}_F = -\sum_f i~~\left[ \bar f_L \gamma^\mu D_\mu f_L +
\bar f_R \gamma^\mu D_\mu f_R \right]
\label{fkin}
\eeq
where the gauge covariant derivative is given by
\beq
D_\mu \equiv \partial_\mu - i~ g ~{\sigma_i\over 2}~ W^i_\mu - i~g^\prime~
{Y\over 2}~ B_\mu
\label{gcov}
\eeq
The $\sigma_i$ are the Pauli matrices (representations of SU(2)) and Y is the
hypercharge. $g$ and $g^\prime$ are the SU(2)$_{\rm L}$ and  U(1)$_{\rm Y}$
gauge couplings respectively.

Fermion mass terms are generated through the coupling of the left- and
right-handed fermions to a scalar doublet Higgs boson $\phi$.
\beq
{\cal L}_Y = -\sum_f \left[ G_f \phi \bar f_L  f_R \right] + h.c.
\label{fyuk}
\eeq
The Lagrangian for the Higgs field is
\beq
{\cal L}_\phi = -\vert D_\mu \phi\vert^2 - V(\phi)
\label{lh}
\eeq
where the (unknown) Higgs potential is commonly written as
\beq
 V(\phi ) = -\mu^2\phi^{\dagger}\phi +
{\lambda} (\phi^{\dagger}\phi)^2
\label{hpot}
\eeq
The vacuum state corresponds to a Higgs expectation value\footnote{Note
that the convention used here differs by a factor of $\sqrt{2}$ from that in
much of the standard model literature.  This is done so as to conform with
the MSSM conventions used below.}
\beq
\langle \phi \rangle = \langle \phi^* \rangle = \left(\matrix{0\cr
{v}} \right) \qquad {\rm with} \qquad  v^2 = {\mu^2 \over
2 \lambda}
\eeq
The non-zero expectation value and hence the spontaneous breakdown of the
electroweak gauge symmetry generates masses for the gauge bosons (through the
Higgs kinetic term in (\ref{lh}) and fermions (through (\ref{fyuk})). In a
mass eigenstate basis, the charged $W$-bosons ($W^{\pm} \equiv (W^1 \pm i
W^2)/\sqrt{2})$ receive masses
\beq 
M_W = {1 \over \sqrt{2}} g v
\label{wmass}
\eeq
The neutral gauge bosons are defined by
\beq
Z_\mu = {gW^3_\mu - g^\prime B_\mu\over \sqrt{g^2+g^{\prime 2}}}~ \qquad 
A_\mu = {g^\prime W^3_\mu + g B_\mu\over \sqrt{g^2+g^{\prime 2}}} 
\label{neut}
\eeq
with masses
\beq
M_Z = {1\over \sqrt{2}} \sqrt{g^2+g^{\prime 2}}~ v = M_W/\cos \theta_W \qquad
m_\gamma = 0
\label{zmass}
\eeq
where the weak mixing angle is defined by
\beq
\sin \theta_W = g^\prime/\sqrt{g^2+g^{\prime 2}}
\eeq
Fermion masses are
\beq
m_f = G_f v 
\eeq

As one can see, there is a direct relationship between particle masses 
and the Higgs expectation value, $v$.  Indeed, we know from (\ref{wmass}) and
(\ref{zmass}) that $v \sim M_W \sim O(100)$ GeV. 
We can then pose the question, why is $M_W \ll M_P = 1.2 \times 10^{19}$ GeV
or equivalently why is $G_F \gg G_N$?

\subsection{The hierarchy problem}

\begin{figure}[hbtp]
	\centering
	\includegraphics[width=6.truecm]{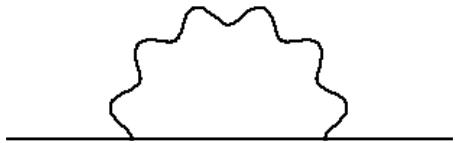}
\vskip -.5in
	\caption{1-loop correction to a fermion mass.}
	\label{floop}
\end{figure}

The mass hierarchy problem stems from the fact that masses,
in particular scalar masses, are not stable to radiative corrections
\cite{hier}.  While fermion masses also receive radiative corrections
from diagrams of the form in Figure \ref{floop}, these are only
logarithmically divergent (see for example \cite{bd}), 
\beq
\delta m_f \simeq {3\alpha \over 4\pi} m_f \ln (\Lambda^2/m_f^2)
\eeq
$\Lambda$ is an ultraviolet cutoff, where we expect new physics to play
an important role.  As one can see, even for $\Lambda \sim M_P$, these
corrections are small, $\delta m_f \la m_f$.

\begin{figure}[hbtp]
	\centering
	\includegraphics[width=10.5truecm]{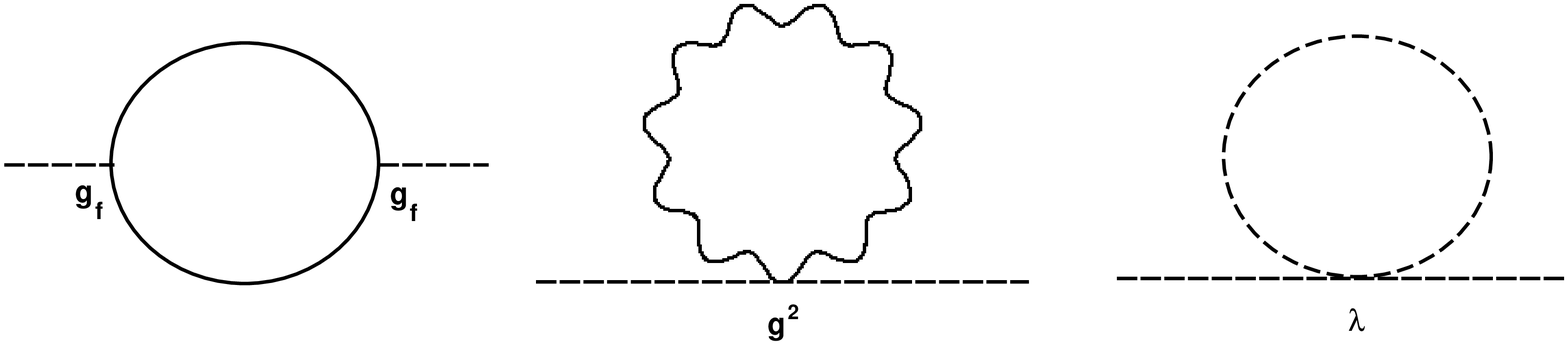}
	\caption{1-loop corrections to a scalar mass.}
	\label{bloop}
\end{figure}

In contrast, scalar masses are quadratically divergent.
1--loop contributions to scalars masses, such as those shown in Figure
\ref{bloop} are readily computed
\beq
\delta m_H^2 \simeq g_f^2,g^2,\lambda \int d^4k {1 \over k^2} \sim
O({\alpha\over 4\pi}) \Lambda^2
\eeq
due to contributions from fermion loops with coupling $g_f$, from gauge boson
loops with coupling $g^2$, and from quartic scalar-couplings $\lambda$.
From the relation (\ref{wmass}) and the fact that the Higgs mass is related
to the expectation value, $m_H^2 = 4 v^2 \lambda$, we expect $M_W \sim m_H$.
However, if new physics enters in at the GUT or Planck scale so that 
$\Lambda \gg M_W$, the 1--loop corrections destroy the stability of the weak
scale. That is, 
\beq
\Lambda \gg M_W \rightarrow \delta m_H^2 \gg m_H^2
\eeq

Of course, one can tune the bare mass $m_H$ so that it contains a large
negative term which almost exactly cancels the 1--loop correction leaving a
small electroweak scale mass$^2$. For a Planck scale correction, this
cancellation must be accurate to 32 significant digits. Even so, the 2-loop
corrections should be of order $\alpha^2 \Lambda^2$ so these too must be
accurately canceled.  Although such a series of cancellations is technically
feasible, there is hardly a sense of satisfaction that the hierarchy problem
is under control.

An alternative and by far simpler solution to this problem exists if one
postulates that there are new particles with similar masses and equal
couplings to those responsible for the radiatively induced masses but with a
difference (by a half unit) in spin. Then, because the contribution to $\delta
m_H^2$ due to a fermion loop comes with a relative minus sign, the total
contribution to the 1-loop corrected mass$^2$ is 
\beq
\delta m_{H}^2 \simeq  O({\alpha\over 4\pi}) (\Lambda^2 + m_B^2) 
- O({\alpha\over 4\pi}) (\Lambda^2 + m_F^2) = O({\alpha\over 4\pi}) (m^2_B -
m^2_F)
\label{su1loop}
\eeq
If in addition, the bosons and fermions all have the same masses, then
the radiative corrections vanish identically.
The stability of the hierarchy only requires that the weak scale is preserved
so that we need only require that
\beq
\vert m^2_B - m^2_F\vert \la 1~{\rm TeV}^2
\label{nat}
\eeq
As we will see in the lectures that follow, supersymmetry offers just the
framework for including the necessary new particles and the absence of these
dangerous radiative corrections \cite{FF}. 

Before we embark, I would like to call attention to some excellent additional
resources on supersymmetry.  These are the classic by Bagger and Wess on
supersymmetry, \cite{BW}, the book by Ross on Grand Unification \cite{ross}
and two recent reviews by Martin \cite{martin} and Ellis \cite{ellis}.

\subsection{Supersymmetric operators and transformations}

Prior to the introduction of supersymmetry, operators were generally regarded
as bosonic.  That is, they were either scalar, vector, or tensor operators.
The momentum operator, $P_\mu$, is a common example of a vector operator.
However, the types of bosonic charges are greatly limited, as was shown by
Coleman and Mandula \cite{CM}. Given a tensorial operator, $\Sigma_{\mu\nu}$,
its diagonal matrix elements can be decomposed as
\beq
<a\vert\Sigma_{\mu\nu}\vert a> = \alpha p^a_\mu p^a_\nu + \beta g_{\mu\nu}
\label{decomp}
\eeq
One can easily see that unless $\alpha = 0$, 2 to 2 scattering
process allow only forward scattering.

Operators of the form expressed in (\ref{ss}) however, are necessarily
non-diagonal as they require a change between the initial and final state by
at least a half unit of spin. Indeed, such operators, if they exist must be
fermionic in nature and carry a spinor index $Q_\alpha$. There may in fact be
several such operators, $Q_\alpha^i$ with $i = 1, ..., N$, (though for the
most part we will restrict our attention to $N=1$ here). 
As a symmetry operator, $Q$ must commute with the Hamiltonian $H$, as must its
anti-commutator. So we have
\beq
[Q^i_\alpha , H] = 0 \qquad
[\{Q^i_\alpha , Q^{j}_\beta \},H] = 0
\label{comH}
\eeq
By extending the Coleman-Mandula theorem \cite{HLS}, one can show that
\beq
\{Q^i , Q^{j\dagger} \} \propto \delta^{ij}
P_\mu + Z^{ij}
\label{HLSanti}
\eeq
where $Z^{ij}$ is antisymmetric in the supersymmetry indices $\{i,j\}$.
Thus, this so-called ``central charge" vanishes for $N = 1$.
More precisely, we have in a Weyl basis
\ba
\{Q_\alpha , {Q^\dagger}_{\dot{\beta}} \} & = & 2 \sigma^\mu_{\alpha
\dot{\beta}} P_\mu \nonumber \\
\{Q_\alpha , Q_{{\beta}} \}  =  \{Q^\dagger_{\dot{\alpha}} ,
{Q^\dagger}_{\dot{\beta}} \} & = & 0 \nonumber \\
\ [ Q_\alpha , P_\mu ] = [{Q^\dagger}_{\dot{\alpha}} , P_\mu] & = & 0
\label{anticom}
\ea

Before setting up the formalism of the supersymmetric transformations, it will
be useful to establish some notation for dealing with spinors in the Dirac
and Weyl bases. The Lagrangian for a four-component Dirac fermion with mass
$M$, can be written as
\beq
{\cal L}_D
= -i \overline\Psi_{D} \gamma^\mu \partial_\mu \Psi_{D}
-  M \overline \Psi_{D}
\Psi_{D}
\label{LD} 
\eeq
where
\beq
\gamma_\mu = \pmatrix{ 0 & \sigma_\mu \cr
                       {\overline \sigma}_\mu & 0\cr} \qquad
\qquad\gamma_5 = \pmatrix{-1 & 0\cr 0 & 1\cr}
\eeq
and $\sigma_\mu = (1,\sigma_i)$, ${\overline \sigma}_\mu = (1,-\sigma_i)$,
$\sigma_i$ are the ordinary $2\times2$ Pauli matrices. I am taking the
Minkowski signature to be $(-,+,+,+)$.  We can write the Dirac spinor
$\Psi_D$ in terms of 2 two-component Weyl spinors
\beq
\Psi_{D} =
\pmatrix{\xi_\alpha\cr {\chi^{\dagger\dot{\alpha}}}\cr}
\qquad\qquad \overline\Psi_{D}  = \pmatrix{\chi^\alpha &
                           {\xi^\dagger}_{\dot{\alpha}}\cr }
\label{Dwey}
\eeq
Note that the spinor indices ($\alpha, \dot{\alpha}$) are raised and lowered
by $\epsilon_{ij}$ where $\{ij\}$ can be either both dotted or both undotted
indices. $\epsilon$ is totally antisymmetric and $\epsilon_{ij} = -
\epsilon^{ij}$ with $\epsilon^{12} = 1$.
It is also useful to define projection operators, $P_L$ and $P_R$ with 
\beq
P_L \Psi_D = {(1-\gamma_5) \over 2} \Psi_D = \pmatrix{1 & 0\cr 0 & 0\cr}
\Psi_D = \pmatrix{\xi_\alpha\cr 0 \cr}
\eeq
with a similar expression for $P_R$.
In this way we can interpret $\xi_\alpha$ as a left-handed Weyl spinor
and $\chi^{\dagger\dot{\alpha}}$ as a right-handed Weyl spinor.
The Dirac Lagrangian (\ref{LD}) can now be written in terms of the
two-component Weyl spinors as 
\beq
{\cal L}_D
 = -i \xi^\dagger \overline \sigma^\mu \partial_\mu \xi -
i \chi^\dagger \overline \sigma^\mu \partial_\mu \chi -
M(\xi\chi + \xi^\dagger \chi^\dagger)
\eeq
having used the identity, $-\chi \sigma^\mu \xi^\dagger = \xi^\dagger
\overline\sigma^\mu \chi$.

 Instead, it is sometimes convenient to consider a
four-component Majorana spinor. This can be done rather easily from the above
conventions and taking
$\xi = \chi$, so that
\beq
\Psi_{M} =
\pmatrix{\xi_\alpha\cr {\xi^{\dagger\dot{\alpha}}}\cr}
\qquad\qquad \overline\Psi_{M}  = \pmatrix{\xi^\alpha &
                           {\xi^\dagger}_{\dot{\alpha}}\cr }
\label{Mwey}
\eeq
and the Lagrangian can be written as
\ba
{\cal L}_M &
= & -{i \over 2} \overline\Psi_{M} \gamma^\mu \partial_\mu \Psi_{M}
- {1 \over 2} M \overline \Psi_{M}
\Psi_{M} \nonumber \\
& = & -i \xi^\dagger \overline \sigma^\mu \partial_\mu \xi -
{1 \over 2} M (\xi\xi + \xi^\dagger \xi^\dagger)
\label{LM} 
\ea

The massless representations for supersymmetry are now easily constructed.
Let us consider here  $N=1$ supersymmetry, i.e.,  a single
supercharge $Q_\alpha$. For the massless case, we can choose the momentum to
be of the form $P_\mu = {1\over 4}(-1,0,0,1)$.  As can be readily found from
the anticommutation relations (\ref{anticom}), the only non-vanishing
anticommutation relation is $\{Q_1, {Q^\dagger}_{\dot 1}\} = 1$. Consider then
a state of given spin, $|\lambda \rangle$ such that ${Q^\dagger}_{\dot 1} |
\lambda \rangle = 0$. (If it is not 0, then due to the anticommutation
relations, acting on it again with ${Q^\dagger}_{\dot 1}$ will vanish.)
From the state $|\lambda \rangle$, it is possible to construct only one other
nonvanishing state, namely $Q_1 |\lambda \rangle$ - the action of any of
the other components of $Q_\alpha$ will vanish as well.  Thus, if the state
$|\lambda \rangle$ is a scalar, then the state  $Q_1 |\lambda \rangle$ will be
a fermion of spin 1/2.  This (super)multiplet will be called a chiral
multiplet. If $|\lambda \rangle$ is spin 1/2, then $Q_1 |\lambda \rangle$ is
a vector of spin 1, and we have a vector multiplet. In the absence of gravity
(supergravity), these are the only two types of multiplets of interest.

For $N>1$, one can proceed in an analogous way.  For example, with 
$N=2$, we begin with two supercharges $Q^1, Q^2$. Starting with a state
$|\lambda\rangle$, we can now obtain the following: $Q^1_1 | \lambda \rangle$,
$Q^2_1 | \lambda \rangle, Q^1_1 Q^2_1| \lambda \rangle$.  In this case,
starting with a complex scalar, one obtains two fermion states, and one
vector, hence the vector (or gauge) multiplet. One could also start with a
fermion state (say left-handed) and obtain two complex scalars, and a
right-handed fermion. This matter multiplet however, is clearly not chiral
and is not suitable for phenomenology. This problem persists for all
supersymmetric theories with $N>1$, hence the predominant interest in $N=1$
supersymmetry. 

Before we go too much further, it will be useful to make a brief connection
with the standard model. We can write all of the standard model fermions in a
two-component Weyl basis. The standard model fermions are therefore
\ba
Q_i & = & \pmatrix{u \cr d \cr}_L ,\qquad \pmatrix{c \cr s \cr}_L ,
\qquad \pmatrix{t \cr b \cr}_L \nonumber \\ \nonumber \\
u^c_i & = &~~ u^c_L , \qquad \qquad c^c_L , \qquad\qquad  t^c_L \nonumber \\
\nonumber \\
d^c_i & = &~~ d^c_L , \qquad \qquad s^c_L , \qquad\qquad  b^c_L \nonumber \\
\nonumber \\
L_i & = & \pmatrix{\nu_e \cr e \cr}_L ,\qquad \pmatrix{\nu_\mu \cr \mu \cr}_L
, \qquad \pmatrix{\nu_\tau \cr \tau \cr}_L \nonumber \\ \nonumber \\
e^c_i & = &~~ e^c_L , \qquad \qquad \mu^c_L , \qquad\qquad  \tau^c_L 
\label{chiralf}
\ea
Note that the fields above are all left-handed. Color indices have been
suppressed. From (\ref{LM}), we see that we would write the fermion kinetic
terms as 
\beq
{\cal L}_{kin} = -i Q_i^\dagger \overline \sigma^\mu \partial_\mu Q_i -
i u_i^{c\dagger} \overline \sigma^\mu \partial_\mu u^c_i - \cdots
\eeq
As indicated above and taking the electron as an example, we can form a
Dirac spinor 
\beq
\Psi_e = \pmatrix{e_L \cr e^{c\dagger}_L\cr} = \pmatrix{e_L \cr e_R\cr}
\eeq
A typical Dirac mass term now becomes
\beq
\overline \Psi_e \Psi_e = e^c_L e_L + e_L^\dagger {e^c_L}^\dagger =
e_R^\dagger e_L + e_L^\dagger e_R
\eeq
As we introduce supersymmetry, the field content of the standard model will
necessarily be extended. All of the standard model matter fields listed above
become members of chiral multiplets in $N=1$ supersymmetry. Thus, to each of
the (Weyl) spinors, we assign a complex scalar superpartner. This will be
described in more detail when we consider the MSSM.

\medskip

To introduce the notion of a supersymmetric transformation, let us consider
an infinitesimal spinor $\xi^\alpha$ with the properties that $\xi$
anticommutes with itself and the supercharge $Q$, but commutes with the
momentum operator
\beq
\{\xi^\alpha, \xi^\beta\} = \{\xi^\alpha, Q_\beta\} = [P_\mu, \xi^\alpha] = 0
\eeq
It then follows that since both $\xi$ and $Q$ commute with $P_\mu$, the
combination $\xi Q$ also commutes with $P_\mu$ or
\beq
[P_\mu,\xi Q] = [P_\mu,\xi^\dagger Q^\dagger] = 0
\eeq
where by $\xi Q$ we mean 
$\xi Q = \xi^\alpha Q_\alpha = \epsilon_{\alpha \beta} \xi^\alpha Q^\beta
= - \epsilon_{\alpha \beta}  Q^\beta \xi^\alpha = \epsilon_{\beta \alpha } 
Q^\beta \xi^\alpha = Q \xi$. Similarly, $\xi^\dagger Q^\dagger =
\xi^\dagger_{\dot \alpha} {Q^\dagger}^{\dot \alpha}$. Also note that
$\xi^\alpha Q_\alpha = -
\xi_\alpha Q^\alpha$. Finally, we can compute the commutator of
$\xi Q$ and $\xi^\dagger Q^\dagger$,
\ba
[\xi Q, \xi^\dagger Q^\dagger] & = & \xi Q  \xi^\dagger Q^\dagger -
\xi^\dagger Q^\dagger \xi Q = \xi^\alpha Q_\alpha  {\xi^\dagger}_{\dot
\beta} {Q^\dagger}^{\dot \beta} - \xi^\dagger_{\dot \beta}
{Q^\dagger}^{\dot \beta} \xi^\alpha Q_\alpha
\nonumber \\
& = &  \xi^\alpha Q_\alpha {Q^\dagger}_{\dot \beta} \xi^{\dagger {\dot
\beta}} - \xi^{\dagger {\dot \beta}} {Q^\dagger}_{\dot \beta} Q_\alpha
\xi^\alpha
\nonumber \\
& = & 2 \xi^\alpha \sigma_{\alpha {\dot\beta}}^\mu \xi^{\dagger {\dot \beta}}
P_\mu - \xi^\alpha  {Q^\dagger}_{\dot \beta} Q_\alpha \xi^{\dagger {\dot
\beta}} - \xi^{\dagger {\dot \beta}} {Q^\dagger}_{\dot \beta} Q_\alpha
\xi^\alpha
\nonumber \\
& = & 2 \xi \sigma^{\mu} \xi^{\dagger} P_\mu 
\ea

We next consider the transformation property of a scalar field, $\phi$, under
the infinitesimal $\xi$
\beq
\delta_\xi \phi = (\xi^\alpha Q_\alpha + {\xi^\dagger}_{\dot \beta}
{Q^\dagger}^{\dot \beta}) \phi
\eeq
As described above, we can pick a basis so that 
${Q^\dagger}^{\dot \beta} \phi = 0$.  Let call the spin 1/2 fermion 
$Q_\alpha \phi = \sqrt{2} \psi_\alpha$, so 
\beq
\delta_\xi \phi = \xi^\alpha Q_\alpha \phi = \sqrt{2} \xi^\alpha \psi_\alpha
\eeq
To further specify the supersymmetry transformation, we must define the
action of $Q$ and $Q^\dagger$ on $\psi$. Again, viewing $Q$ as a ``raising"
operator, we will write $Q_\alpha \psi_\gamma = -\sqrt{2} \epsilon_{\alpha
\gamma} F$ and ${Q^\dagger}^{\dot \beta} \psi_\gamma = - \sqrt{2} i
{\sigma^{\mu}_\gamma}^{\dot\beta} \partial_\mu \phi$, where $F$, as we will
see, is an auxiliary field to be determined later. Even though we had
earlier argued that $Q$ acting on the spin 1/2 component of the chiral
multiplet should vanish, we must keep it here, though as we will see, it
does not correspond to a physical degree of freedom. To understand the action
of ${Q^\dagger}$, we know that it must be related to the scalar $\phi$, and
on dimensional grounds (${Q^\dagger} \lambda$ is of mass dimension 3/2) it
must be proportional to $P_\mu \phi$. Then
\ba
\delta_\xi \psi_\gamma & = & (\xi^\alpha Q_\alpha + {\xi^\dagger}_{\dot \beta}
{Q^\dagger}^{\dot \beta}) \psi_\gamma \nonumber \\
& =  &  -\sqrt{2} \xi^\alpha \epsilon_{\alpha \gamma} F + 
\sqrt{2} i \sigma_{\gamma {\dot\beta}}^{\mu} \xi^{\dagger {\dot
\beta}} \partial_\mu
\phi \nonumber \\
& = & \sqrt{2} \xi_\gamma F + \sqrt{2} i (\sigma^\mu \xi^\dagger)_\gamma
\partial_\mu \phi
\label{cftrans}
\ea

Given these definitions, let consider the successive action of two
supersymmetry transformations $\delta_\xi$ and $\delta_\eta$.
\ba
\delta_\eta \delta_\xi \phi & = & \sqrt{2} \delta_\eta (\xi^\alpha
\psi_\alpha) \nonumber \\
& = & \sqrt{2} \bigl[ -\sqrt{2} \xi^\alpha \eta^\gamma \epsilon_{\gamma
\alpha} F + \sqrt{2} i \xi^\alpha \sigma_{\alpha {\dot\beta}}^{\mu}
\eta^{\dagger {\dot \beta}} \partial_\mu \phi \bigr]
\label{double}
\ea
If we take the difference $(\delta_\eta \delta_\xi - \delta_\xi
\delta_\eta) \phi$, we see that the first term in 
(\ref{double}) cancels if we write $\xi^\alpha \eta^\gamma \epsilon_{\gamma
\alpha} = -\xi^\alpha \eta_\alpha$ and note that $\xi^\alpha \eta_\alpha = 
\eta^\alpha \xi_\alpha$.
Therefore the difference can be written as
\beq
(\delta_\eta \delta_\xi - \delta_\xi \delta_\eta) \phi =
2 ( \eta \sigma^\mu \xi^\dagger - \xi \sigma^\mu \eta^\dagger) P_\mu \phi
\label{diff}
\eeq
In fact it is not difficult to show that (\ref{diff}) is a result of the
general operator relation 
\beq
\delta_\eta \delta_\xi - \delta_\xi \delta_\eta = 
2 ( \eta \sigma^\mu \xi^\dagger - \xi \sigma^\mu \eta^\dagger) P_\mu
\label{genop}
\eeq

Knowing the general relation (\ref{genop}) and applying it to a fermion
$\psi_\gamma$ will allow us to determine the transformation properties of the
auxiliary field $F$. 
Starting with
\beq
\delta_\eta \delta_\xi \psi_\gamma = - \sqrt{2} \xi^\alpha \epsilon_{\alpha
\gamma} \delta_\eta F + 2 i
\sigma^\mu_{\gamma {\dot \beta}} \xi^{\dagger {\dot \beta}} \eta^\alpha
\partial_\mu \psi_\alpha
\eeq
we use the Fierz identity $\chi_\gamma \eta^\alpha \zeta_\alpha + \eta_\gamma
\zeta^\alpha \chi_\alpha + \zeta_\gamma \chi^\alpha \eta_\alpha = 0$,
and making the substitutions, $\chi_\gamma = \sigma^\mu_{\gamma {\dot \beta}}
\xi^{\dagger {\dot \beta}}, \eta = \eta$, and $\zeta = \partial_\mu \psi$,
we have
\beq
\delta_\eta \delta_\xi \psi_\gamma = - \sqrt{2} \xi^\alpha \epsilon_{\alpha
\gamma} \delta_\eta F - 2 i \bigl\{ \eta_\gamma \partial_\mu \psi^\alpha 
\sigma^\mu_{\alpha {\dot \beta}} \xi^{\dagger {\dot \beta}} + \partial_\mu
\psi_\gamma {\sigma^{\mu \alpha}}_{{\dot \beta}} \  \xi^{\dagger {\dot
\beta}} \eta_\alpha \bigr\}
\eeq
Next we use the spinor identity, $\chi^\alpha \ 
\sigma^\mu_{\alpha {\dot \beta}} \  \xi^{\dagger {\dot \beta}} = -
{\xi^\dagger}_{\dot \beta} \ 
\overline\sigma^{\mu{\dot \beta} \alpha} \  \chi_\alpha$ along with
$\eta^\gamma \chi_\gamma = \chi^\gamma \eta_\gamma$ (from above)
to get
\beq
\delta_\eta \delta_\xi \psi_\gamma = - \sqrt{2} \xi^\alpha \epsilon_{\alpha
\gamma} \delta_\eta F + 2 i \bigl\{ \eta_\gamma {\xi^\dagger}_{\dot \beta} \ 
\overline\sigma^{\mu{\dot \beta} \alpha} \partial_\mu \psi_\alpha - 
\eta^\alpha \sigma^\mu_{\alpha {\dot \beta}} \xi^{\dagger {\dot \beta}} 
\partial_\mu \psi_\gamma \bigr\}
\eeq
It is not hard to see then that the difference of the double transformation
becomes
\ba
(\delta_\eta \delta_\xi - \delta_\xi \delta_\eta) \psi_\gamma  & = &
2 ( \eta \sigma^\mu \xi^\dagger - \xi \sigma^\mu \eta^\dagger) P_\mu
\psi_\gamma
\nonumber \\
& & - 2 \bigl[ \eta_\gamma (\xi^\dagger \overline\sigma^\mu P_\mu \psi) - 
\xi_\gamma (\eta^\dagger \overline\sigma^\mu P_\mu \psi) \bigr] \nonumber \\
& & + \sqrt{2} \bigl( \xi_\gamma \delta_\eta F - \eta_\gamma \delta_\xi
F\bigr)
\ea
Thus, the operator relation (\ref{genop}) will be satisfied only if 
\beq
\delta_\xi F = -\sqrt{2} (\xi^\dagger \overline\sigma^\mu P_\mu \psi)
\label{varf}
\eeq
and we have the complete set of transformation rules for the chiral
multiplet.

\section{The Simplest Models}

\subsection{The massless non-interacting Wess-Zumino model}

We begin by writing down the Lagrangian for a single chiral multiplet
containing a complex scalar field and a Majorana fermion
\ba
{\cal L} & = & -\partial_\mu \phi^* \partial^\mu \phi - i \psi^\dagger
\overline\sigma^\mu\partial_\mu \psi \nonumber \\
& = & -\partial_\mu \phi^* \partial^\mu \phi - {i \over 2} (
\psi^\dagger \overline\sigma^\mu\partial_\mu \psi - \partial_\mu \psi^\dagger
\overline\sigma^\mu \psi)
\label{lwz}
\ea
where the second line in (\ref{lwz}) is obtained by a partial integration and
is done to simplify the algebra below. 

We must first check the invariance of the Lagrangian under the supersymmetry
transformations discussed in the previous section. 
\ba
\delta {\cal L} & = & - \sqrt{2} \xi  \partial^\mu \psi \partial_\mu \phi^*
-{i \over \sqrt{2}}  \xi^\dagger F^* \overline\sigma^\mu \partial_\mu \psi
-{1 \over \sqrt{2}} \xi \sigma^\nu \overline\sigma^\mu \partial_\mu \psi
\partial_\nu \phi^* \nonumber \\
& & \qquad \qquad  + {i \over \sqrt{2}} \xi^\dagger \partial_\mu
F^* \overline\sigma^\mu \psi + {1 \over \sqrt{2}} \xi \sigma^\nu 
\overline\sigma^\mu \psi
\partial_\mu \partial_\nu \phi^* + h.c.
\ea
Now with the help of still one more identity, $(\sigma^\mu
\overline\sigma^\nu + \sigma^\nu \overline\sigma^\mu)^\beta_\alpha = -2
\eta^{\mu\nu} \delta^\beta_\alpha$, we can expand the above expression
\ba
\delta {\cal L} & = & - \sqrt{2} \xi  \partial^\mu \psi \partial_\mu \phi^*
\nonumber \\
& &  + {1 \over \sqrt{2}} \xi \sigma^\nu  \overline\sigma^\mu \psi
\partial_\mu
\partial_\nu \phi^* + \sqrt{2} \xi \partial^\mu \psi \partial_\mu \phi^*
+ {1 \over \sqrt{2}} \xi \sigma^\mu  \overline\sigma^\nu 
\partial_\mu \psi \partial_\nu \phi^* \nonumber \\
& & - {i \over \sqrt{2}}  \xi^\dagger \bigl( F^* \overline\sigma^\mu
\partial_\mu
\psi - \partial_\mu F^* \overline\sigma^\mu \psi\bigr) + h.c.
\label{varl}
\ea
Fortunately, we now have some cancellations. The first and third terms in
(\ref{varl}) trivially cancel. Using the commutivity of the partial
derivative and performing a simple integration by parts we see that the
second and fourth terms also cancel. We left with (again after an integration
by parts) 
\beq
\delta {\cal L} = -i \sqrt{2} \xi^\dagger F^* \overline\sigma^\mu
\partial_\mu \psi + h.c.
\label{notinv}
\eeq
indicating the lack of invariance of the Lagrangian (\ref{lwz}). 

We can recover the invariance under the supersymmetry transformations by
considering in addition to the Lagrangian (\ref{lwz}) the following,
\beq
{\cal L}_{aux} = F^* F
\eeq
and its variation
\beq
\delta{\cal L}_{aux} = \delta F^* F + F^* \delta F
\eeq
The variation of the auxiliary field, $F$, was determined in (\ref{varf}) and
gives
\beq
\delta{\cal L}_{aux} = i \sqrt{2} \xi^\dagger F^* \overline\sigma^\mu
\partial_\mu \psi + h.c.
\eeq
and exactly cancels the piece left over in (\ref{notinv}).
Therefore the Lagrangian
\beq
{\cal L}  =  -\partial_\mu \phi^* \partial^\mu \phi - i \psi^\dagger
\overline\sigma^\mu\partial_\mu \psi + F^* F
\label{lss}
\eeq
is fully invariant under the set of supersymmetry transformations.

\subsection{Interactions for Chiral Multiplets}

Our next task is to include interactions for chiral multiplets which are
also consistent with supersymmetry. We will therefore consider a set of
chiral multiplets, $(\phi_i, \psi_i, F_i)$ and a renormalizable Lagrangian,
${\cal L}_{int}$. Renormalizability limits the mass dimension of any term in
the Lagrangian to be less than or equal to 4. Since the interaction Lagrangian
must also be invariant under the supersymmetry transformations, we do not
expect any terms which are cubic or quartic in the scalar fields $\phi_i$. 
Clearly no term can be linear in the fermion fields either. 
This leaves us with only the following possibilities
\beq
{\cal L}_{int} = {1 \over 2} A^{ij} \psi_i \psi_j + B^i F_i + h.c.
\label{lint}
\eeq
where $A^{ij}$ is some linear function of the $\phi_i$ and ${\phi^i}^*$
and $B^i$ is some function which is at most quadratic in the scalars and
their conjugates. Here, and in all that follows, it will be assumed
that repeated indices such as $ii$ are summed. Furthermore, since
$\psi_i
\psi_j =
\psi_j
\psi_i$ (spinor indices are suppressed), the function $A^{ij}$ must be
symmetric in
$ij$. As we will see, the functions $A$ and
$B$ will be related by insisting on the invariance of (\ref{lint}). 

We begin therefore with the variation of ${\cal L}_{int}$
\ba
\delta {\cal L}_{int} & = & {1 \over 2}{\partial A^{ij} \over \partial
\phi_k} (\sqrt{2}
\xi \psi_k) (\psi_i \psi_j) + {1 \over 2}{\partial A^{ij} \over \partial
{\phi^k}^*} (\sqrt{2} \xi^\dagger {\psi^k}^\dagger) (\psi_i \psi_j) \nonumber
\\ & & +{1 \over 2} A^{ij} (\sqrt{2} \xi F_i + \sqrt{2} i \sigma^\mu
\xi^\dagger
\partial_\mu \phi_i) \psi_j \nonumber \\
& & + {1 \over 2} A^{ij} \psi_i (\sqrt{2} \xi F_j + \sqrt{2} i \sigma^\mu
\xi^\dagger \partial_\mu \phi_j) \nonumber \\
& & + {\partial B^i \over \partial \phi_j} (\sqrt{2} \xi \psi_j) F_i +
{\partial B^i \over \partial {\phi^j}^*} (\sqrt{2} \xi^\dagger
{\psi^j}^\dagger) F_i \nonumber \\
& & + B^i \sqrt{2} i \xi^\dagger \overline\sigma^\mu \partial_\mu \psi_i +
h.c.
\label{varint}
\ea
where the supersymmetry transformations of the previous section have
already been performed.  The notation $(\psi_i \psi_j)$ refers to
$\psi_i^\alpha {\psi_j}_\alpha$ as clearly spinor indices have everywhere been
suppressed.  The Fierz
identity $(\xi \psi_k) (\psi_i \psi_j) + (\xi \psi_i) (\psi_j \psi_k) + (\xi
\psi_j) (\psi_k \psi_i) = 0$ implies that the derivative of the function
$A^{ij}$ with respect to $\phi_k$ (as in the first term of (\ref{varint}))
must be symmetric in ${ijk}$. Because there is no such identity for the
second term with derivative with respect to ${\phi^k}^*$, this term must
vanish.  Therefore, the function $A^{ij}$ is a holomorphic function of the
$\phi_i$ only. Given these constraints, we can write 
\beq
A^{ij} = - M^{ij} - y^{ijk} \phi_k
\eeq
where by (\ref{lint}) we interpret $M^{ij}$ as a symmetric fermion mass
matrix, and $y^{ijk}$ as a set of (symmetric) Yukawa couplings. In fact, it
will be convenient to write 
\beq
A^{ij} = - {\partial^2 W \over \partial \phi_i \partial \phi_j}
\eeq
where
\beq
W =  {1\over 2} M^{ij} \phi_i \phi_j + {1 \over 6} y^{ijk}\phi_i \phi_j
\phi_k
\label{superp}
\eeq
and is called the superpotential.

Noting that the 2nd and 3rd lines of (\ref{varint}) are equal due to the
symmetry of $A^{ij}$, we can rewrite the remaining terms as
\ba
\delta {\cal L}_{int} & = &  A^{ij} \psi_i (\sqrt{2} \xi F_j + \sqrt{2} i
\sigma^\mu
\xi^\dagger \partial_\mu \phi_j) \nonumber \\
& & + {\partial B^i \over \partial \phi_j} (\sqrt{2} \xi \psi_j) F_i +
{\partial B^i \over \partial {\phi^j}^*} (\sqrt{2} \xi^\dagger
{\psi^j}^\dagger) F_i \nonumber \\
& & - B^i \sqrt{2} i \partial_\mu \psi_i \sigma^\mu \xi^\dagger +
h.c.
\label{varint2}
\ea
using in addition one of our previous spinor identities on the last line.
Further noting that because of our definition of the superpotential in terms
of $A^{ij}$, we can write $A^{ij} \partial_\mu \phi_j =  - \partial_\mu
(\partial W/\partial \phi_i)$. Then the 2nd and last terms of (\ref{varint2})
can be combined as a total derivative if
\beq 
B^i = {\partial W \over \partial \phi_i}
\label{bdef}
\eeq
and thus is also related to the superpotential $W$. Then the 4th term
proportional to $\partial B^i / \partial {\phi^j}^*$ is absent due to the
holomorphic property of $W$, and the definition of $B$ (\ref{bdef})
allows for a trivial cancellation of the 1st and 3rd terms in (\ref{varint2}).
Thus our interaction Lagrangian (\ref{lint}) is in fact supersymmetric with
the imposed relationships between the functions $A^{ij}, B^i$, and the
superpotential $W$.

After all of this, what is the auxiliary field $F$? It has been designated as
an ``auxiliary" field, because it has no proper kinetic term in (\ref{lss}).
It can therefore be removed via the equations of motion. Combining the
Lagrangians (\ref{lss}) and (\ref{lint}) we see that the variation of the
Lagrangian with respect to $F$ is
\beq
{\delta {\cal L} \over \delta F} = {F^i}^* + W^i
\label{varlf}
\eeq
where we can now use the convenient notation that $W^i = \partial W / \partial
\phi_i$, $W_i^* = \partial W / \partial {\phi^i}^*$, and $W^{ij} = \partial^2
W / \partial \phi_i \partial \phi_j$, etc. The vanishing of (\ref{varlf}) then
implies that 
\beq
F_i = - W_i^*
\eeq

Putting everything together we have 
\ba
{\cal L} & = & -\partial_\mu \phi^{i*} \partial^\mu \phi_i - i \psi^{i\dagger}
\overline\sigma^\mu\partial_\mu \psi_i \nonumber \\
& & - {1 \over 2} (W^{ij} \psi_i \psi_j + W_{ij}^* {\psi^i}^\dagger
{\psi^j}^\dagger) - W^i W_i^*
\label{totl}
\ea
As one can see the last term plays the role of the scalar potential
\beq
V(\phi_i,\phi^{i*}) =  W^i W_i^*
\eeq

\subsection{Gauge Multiplets}

In principle, we should next determine the invariance of the Lagrangian
including a vector or gauge multiplet.  To repeat the above exercise performed
for chiral multiplets, while necessary, is very tedious.  Here, I will only
list some of the more important ingredients.  

Similar to the definition in (\ref{gcov}), the gauge covariant derivative
acting on scalars  and chiral fermions is 
\beq
D_\mu = \partial_\mu -i g T \cdot A_\mu
\label{ginocov}
\eeq  
where $T$ is the relevant representation of the gauge group. For SU(2), we
have simply that $T^i = \sigma^i/2$.  
In the gaugino kinetic term, the covariant derivative becomes
\beq
(D_\mu \lambda)^a = \partial_\mu \lambda^a + g f^{abc} A^b \lambda^c
\eeq
where the
$f^{abc}$ are the (antisymmetric) structure constants of the gauge group
under consideration ($[T^a,T^b] = if^{abc} T^c$).
Gauge invariance for a vector field,
$A^a_\mu$, is manifest through a gauge transformation of the form
\beq
\delta_{gauge} A^a_\mu = -\partial_\mu \Lambda^a + g f^{abc} \Lambda^b
A^c_\mu 
\eeq
where $\Lambda$ is an infinitesimal gauge transformation parameter.  To this, we must add the
gauge transformation of the spin 1/2 fermion partner, or gaugino, in the
vector multiplet
\beq
\delta_{gauge} \lambda^a = g f^{abc} \Lambda^b \lambda^c
\eeq

Given our experience with chiral multiplets, it is not too difficult to
construct the supersymmetry transformations for the the vector multiplet.
Starting with $A_\mu$, which is real, and taking $Q A_\mu = 0$, one finds
that 
\beq
\delta_\xi A^a_\mu = i \bigl[ \xi^\dagger \overline\sigma_\mu
\lambda^a - {\lambda^a}^\dagger \overline\sigma_\mu \xi \bigr]
\eeq
Similarly, applying the supersymmetry transformation to $\lambda^a$, 
leads to a derivative of $A^a_\mu$ (to account for the mass dimension) which
must be in the form of the field strength $F^a_{\mu\nu}$, and an auxiliary
field, which is conventionally named $D^a$.
Thus,
\beq
\delta_\xi \lambda^a = {1 \over {2}}  \bigl( \sigma^\mu
\overline\sigma^\nu \xi \bigr)
F^a_{\mu\nu} + i \xi D^a
\eeq
As before, we can determine the transformation properties of $D^a$ by
applying successive supersymmetry transformations as in (\ref{genop}) with the
substitution $\partial_\mu \to D_\mu$ using (\ref{ginocov}) above. The result
is,
\beq
\delta_\xi D^a =   \bigl[ \xi^\dagger \overline\sigma^\mu D_\mu
\lambda^a + D_\mu {\lambda^a}^\dagger \overline\sigma^\mu \xi \bigr]
\label{varD}
\eeq

Also in analogy with the chiral model, 
the simplest Lagrangian for the vector multiplet is
\beq
{\cal L}_{gauge} = -{1 \over 4} F^a_{\mu\nu} {F^a}^{\mu\nu} - 
i{\lambda^a}^\dagger \overline\sigma^\mu D_\mu \lambda^a 
+ {1 \over 2} D^a D^a
\label{lagauge}
\eeq
In (\ref{lagauge}), the gauge kinetic terms are given in general by
$F^a_{\mu\nu} = \partial_\mu A^a_\nu - \partial_\nu A^a_\mu +
g f^{abc}A^b_\mu A^c_\nu$.

If we have both chiral and gauge multiplets in the theory (as we must) then
we must make simple modifications to the supersymmetry transformations
discussed in the previous section and add new gauge invariant interactions
between the chiral and gauge multiplets which also respect supersymmetry. 
To (\ref{cftrans}), we must only change $\partial_\mu \to D_\mu$ using
(\ref{ginocov}).  To (\ref{varf}), it will be necessary to add a term
proportional to
$(T^a \phi) \xi^\dagger {\lambda^a}^\dagger$ so that, 
\beq
\delta_\xi F = \sqrt{2} i (\xi^\dagger \overline\sigma^\mu D_\mu \psi)
+ 2 i g (T^a \phi) \xi^\dagger {\lambda^a}^\dagger
\label{varf1}
\eeq
The new interaction terms take the form
\beq
{\cal L}_{int} =  \sqrt{2} g i \left [
(\phi^* T^a \psi)\lambda^a - \lambda^{\dagger a} (\psi^\dagger T^a \phi)
\right ]
\nonumber\\
 + g  (\phi^* T^a \phi) D^a
\eeq
Furthermore, invariance under supersymmetry requires not only the additional
term in (\ref{varf1}), but also the condition
\beq
W^i (T^a)_i^j \phi_j = 0
\label{forgold}
\eeq

Finally, we must eliminate the auxiliary field $D^a$ using the equations of
motion which yield
\beq
D^a = -g (\phi^* T^a \phi )
\eeq
Thus the ``D-term" is also seen to be a part of the scalar potential which in
full is now,
\beq
V(\phi,\phi^*) = |F^i|^2 +  {1 \over 2} |D^a|^2 = |W^i|^2 +
{1 \over 2} g^2 (\phi^* T^a \phi)^2.
\label{FD}
\eeq
Notice a very important property of the scalar potential in supersymmetric
theories: the potential is positive semi-definite, $V \ge 0$. 

\subsection{Interactions}

The types of gauge invariant interactions allowed by supersymmetry are
scattered throughout the pieces of the supersymmetric Lagrangian.  Here, we
will simply identify the origin of a particular interaction term, its
coupling, and associated Feynmann diagram. In the diagrams below, the arrows
denote the flow of chirality.  Here, $\psi$ will represent an incoming
fermion, and $\psi^\dagger$ an outgoing one. While there is no true chirality
associated with the scalars, we can still make the association as the scalars
are partnered with the fermions in a given supersymmetric multiplet. We will
indicate $\phi$ with an incoming scalar state, and $\phi^*$ with an outgoing
one. 

Starting from the superpotential (\ref{superp}) and the Lagrangian
(\ref{totl}),  we can identify several interaction terms and their associated 
diagrams:
\begin{itemize}
\item a fermion mass insertion from $W^{ij}  \psi_i \psi_j$
\end{itemize}
\vskip .5in

$\qquad \qquad \qquad M^{ij} \psi_i \psi_j \qquad$

\begin{figure}[h]
\vspace*{-.45in} 
 \hspace*{2.5in}
\begin{minipage}{7.5in}
\includegraphics[width=4.5truecm]{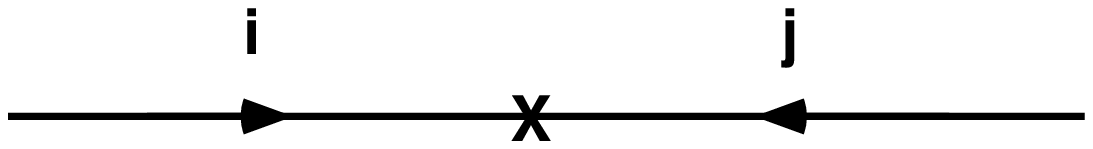}
\end{minipage}
\end{figure}

\vskip .5in

\begin{itemize}
\item a scalar-fermion Yukawa coupling, also from $W^{ij}  \psi_i \psi_j$
\end{itemize}
\vskip .7in

$\qquad \qquad \qquad y^{ijk} \phi_k  \psi_i \psi_j \qquad $

\begin{figure}[h]
\vspace*{-0.75in} 
 \hspace*{2.5in}
\begin{minipage}{7.5in}
\includegraphics[width=4.5truecm]{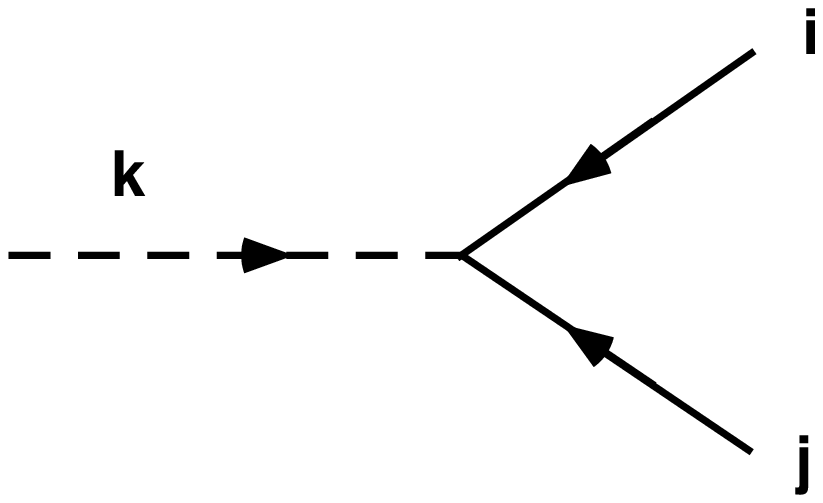}
\end{minipage}
\end{figure}

\vskip .4in

\begin{itemize}
\item a scalar mass insertion from $|W^{i}|^2$
\end{itemize}
\vskip .5in

$\qquad \qquad \qquad M^{il} {M_{jl}}^* \phi_i {\phi^j}^* \qquad$

\begin{figure}[h]
\vspace*{-0.45in} 
 \hspace*{2.5in}
\begin{minipage}{7.5in}
\includegraphics[width=4.5truecm]{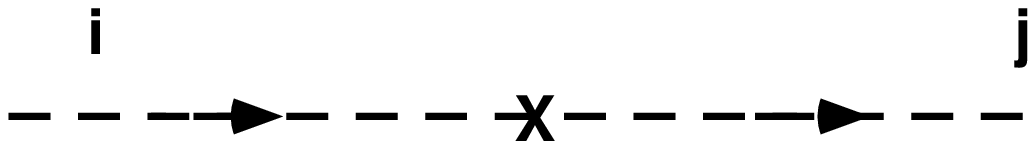}
\end{minipage}
\end{figure}

\newpage

\begin{itemize}
\item a scalar cubic interaction from $|W^{i}|^2$ (plus its complex conjugate
which is not shown)
\end{itemize}
\vskip .9in

$\qquad \qquad \qquad {M_{li}}^* y^{ljk} {\phi^i}^* \phi_j \phi_k \qquad $

\begin{figure}[h]
\vspace*{-0.8in} 
 \hspace*{2.5in}
\begin{minipage}{7.5in}
\includegraphics[width=4.5truecm]{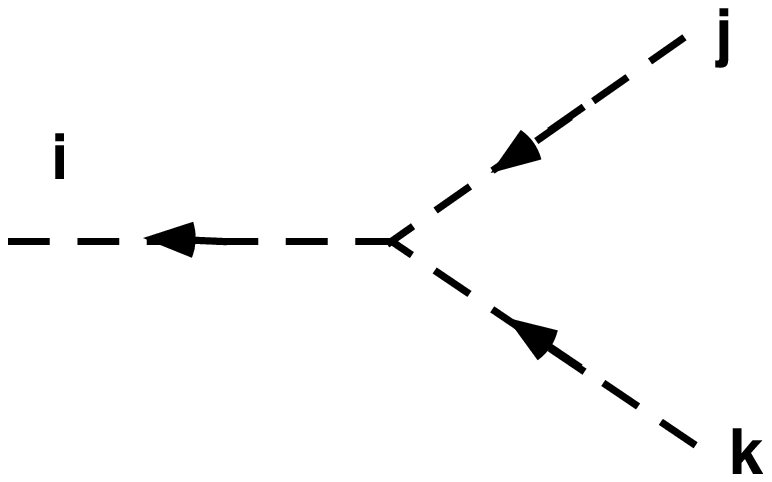}
\end{minipage}
\end{figure}

\vskip 0.5in

\begin{itemize}
\item and finally a scalar quartic interaction from $|W^{i}|^2$ 
\end{itemize}
\vskip 1.1in

$\qquad \qquad ~~ {y^{ijn}} y_{kln}^* \phi_i \phi_j {\phi^k}^* {\phi^l}^*
\qquad
$

\begin{figure}[h]
\vspace*{-1in} 
 \hspace*{2.5in}
\begin{minipage}{7.5in}
\includegraphics[width=4.5truecm]{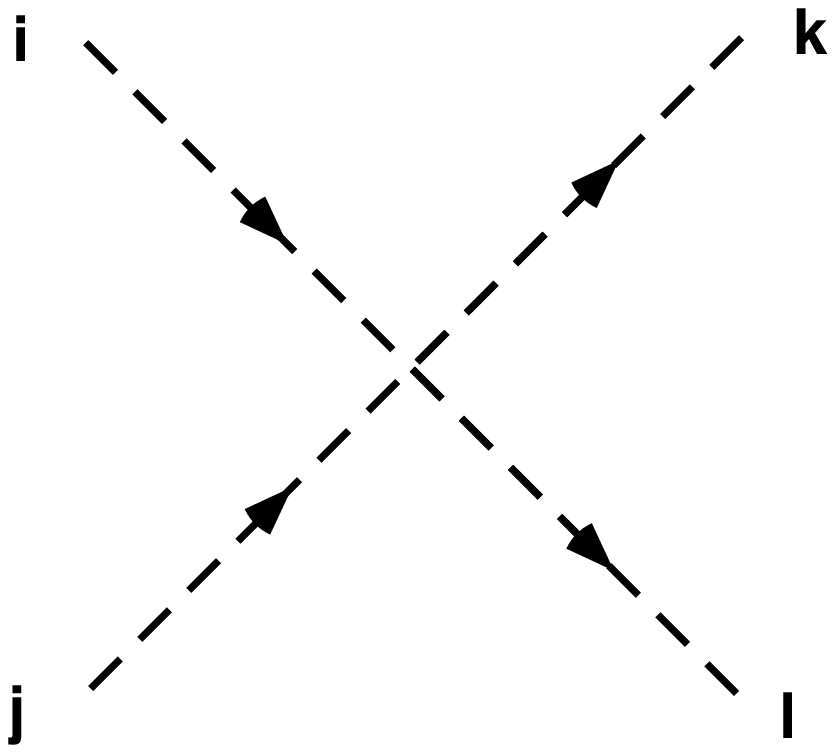}
\end{minipage}
\end{figure}

\vskip 0.5in

Next we must write down the interactions and associated diagrams for the gauge
multiplets. The first two are standard gauge field interaction terms in
any non-abelian gauge theory (so that $f^{abc} \ne 0$) and arise from the
gauge kinetic term $F_{\mu\nu}^2$, the third is an interaction between the
vector and its fermionic partner, and arises from the gaugino kinetic term
$\lambda^\dagger \overline\sigma^\mu D_\mu \lambda$.

\newpage
\begin{itemize}
\item The quartic gauge interaction from $F_{\mu\nu}^2$ (to be summed over
the repeated gauge indices)
\end{itemize}
\vskip 0.6in

$\qquad ~~ ~~ g^2 f^{abc} f^{ade} {A^b}^\mu {A^c}^\nu {A^d}_\mu {A^e}_\nu
\qquad
$

\begin{figure}[h]
\vspace*{-1.2in} 
 \hspace*{2.5in}
\begin{minipage}{7.5in}
\includegraphics[width=5.5truecm]{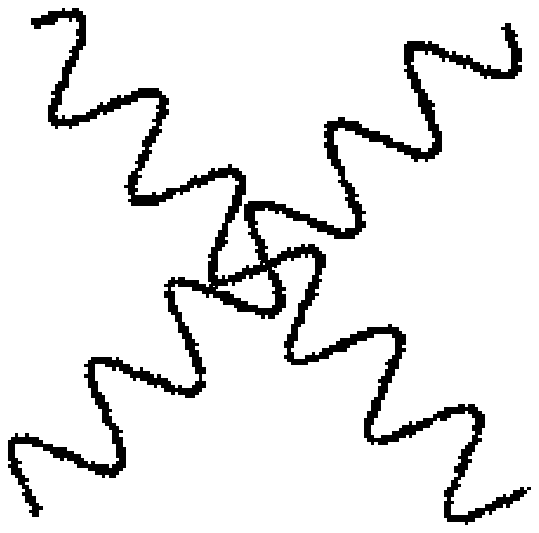}
\end{minipage}
\end{figure}

\vskip -0.1in

\begin{itemize}
\item The trilinear gauge interaction also from $F_{\mu\nu}^2$ 
\end{itemize}
\vskip 1.1in

$\qquad ~~  g f^{abc} {A^b}^\mu {A^c}^\nu (\partial_\mu {A^a}_\nu -
\partial_\nu {A^a}_\mu) \qquad
$

\begin{figure}[h]
\vspace*{-1.0in} 
 \hspace*{2.5in}
\begin{minipage}{7.5in}
\includegraphics[width=5.5truecm]{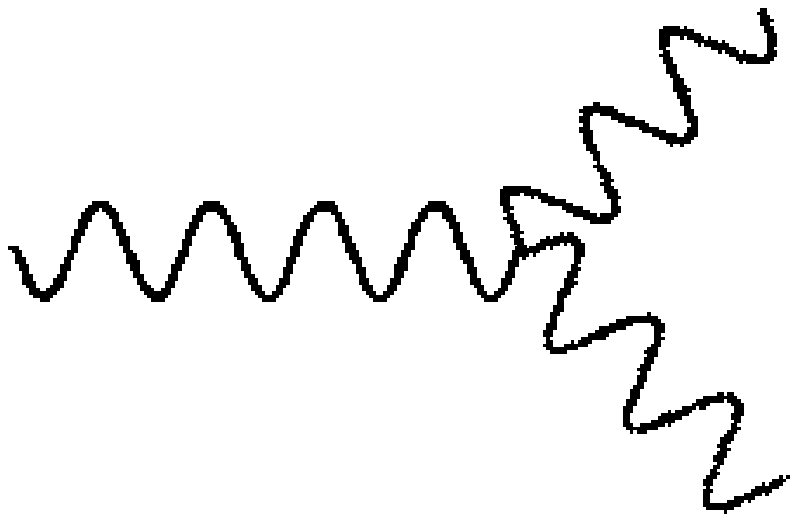}
\end{minipage}
\end{figure}

\vskip 0.1in

\begin{itemize}
\item The gauge-gaugino interaction from $\lambda^\dagger \overline\sigma^\mu
D_\mu \lambda$
\end{itemize}
\vskip 0.7in

$\qquad ~~ ~~ g f^{abc} {A^b}_\mu {\lambda^a}^\dagger \overline\sigma^\mu
 \lambda^c \qquad
$

\begin{figure}[h]
\vspace*{-1.05in} 
 \hspace*{2.5in}
\begin{minipage}{7.5in}
\includegraphics[width=5.5truecm]{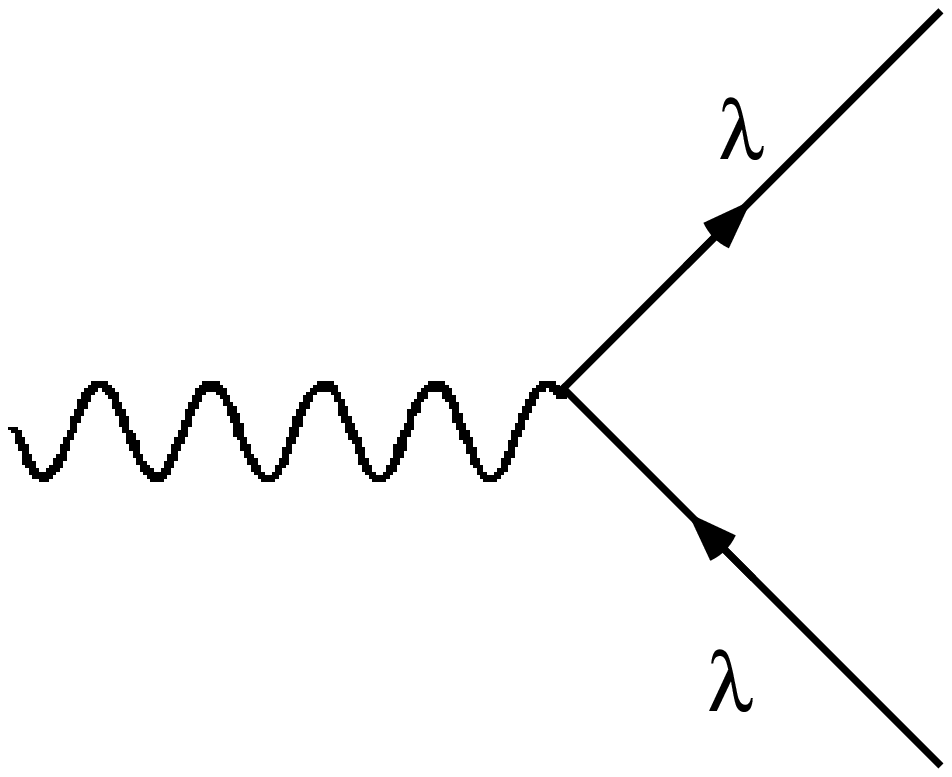}
\end{minipage}
\end{figure}

\vskip 0.5in

If our chiral multiplets are not gauge singlets, then we also have
interaction terms between the vectors and the fermions and scalars of the
chiral multiplet arising from the chiral kinetic terms. Recalling that the
kinetic terms for the chiral multiplets must be expressed in terms of the
gauge covariant derivative (\ref{ginocov}), we find the following
interactions, from$|D_\mu \phi|^2$ and $\psi^\dagger \overline\sigma^\mu
D_\mu \psi$,

\begin{itemize}
\item a quartic interaction interaction involving two gauge bosons and two
scalars,  
\end{itemize}
\vskip 0.7in

$\qquad \qquad ~~ g^2 {A^a}^\mu {A^b}_\mu (T^a \phi) (\phi^* T^b)
\qquad
$

\begin{figure}[h]
\vspace*{-1in} 
 \hspace*{2.5in}
\begin{minipage}{7.5in}
\includegraphics[width=4.5truecm]{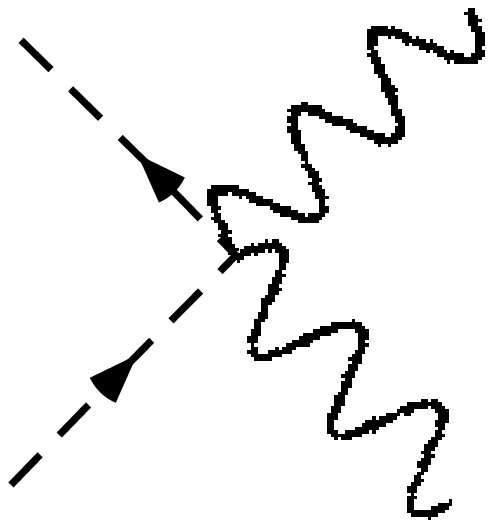}
\end{minipage}
\end{figure}

\vskip -.2in

\begin{itemize}
\item a cubic interaction involving one gauge boson and two
scalars,  
\end{itemize}
\vskip 0.9in

$\qquad \qquad ~~ g ({A^a}^\mu  (T^a\phi) \partial_\mu \phi^* + h.c.)
\qquad
$

\begin{figure}[h]
\vspace*{-0.95in} 
 \hspace*{2.5in}
\begin{minipage}{7.5in}
\includegraphics[width=4.5truecm]{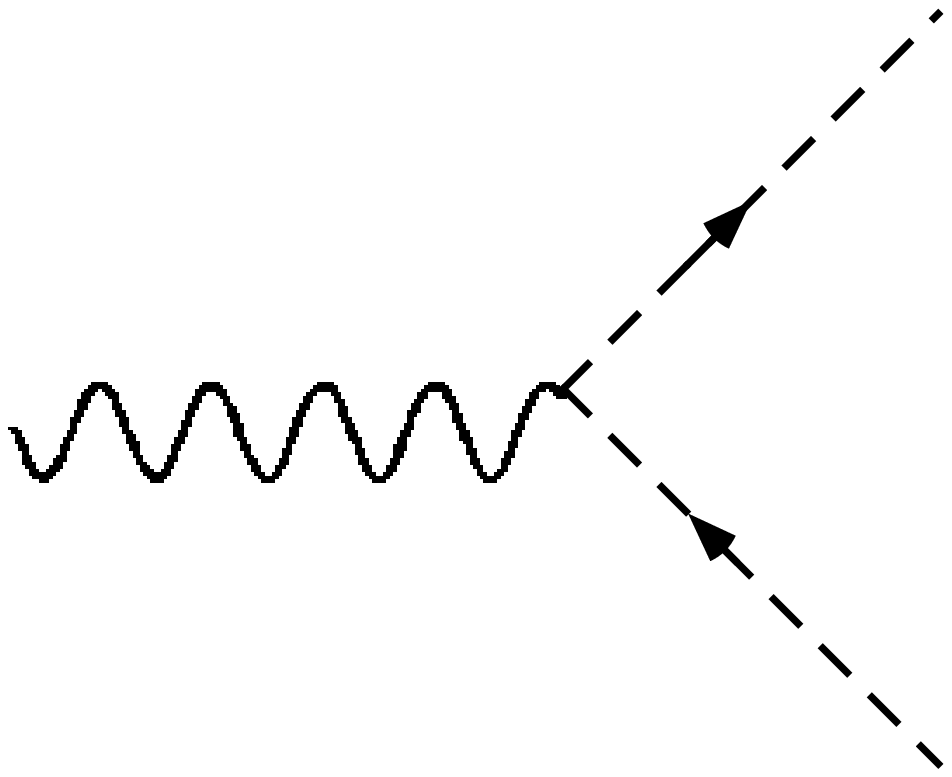}
\end{minipage}
\end{figure}

\vskip 0.4in

\begin{itemize}
\item a cubic interaction involving one gauge boson and two
fermions,  
\end{itemize}
\vskip .7in

$\qquad \qquad ~~ g ({A^a}_\mu \psi^\dagger \overline\sigma^\mu  (T^a\psi)
 + h.c.)
\qquad
$

\begin{figure}[h]
\vspace*{-.95in} 
 \hspace*{2.5in}
\begin{minipage}{7.5in}
\includegraphics[width=4.5truecm]{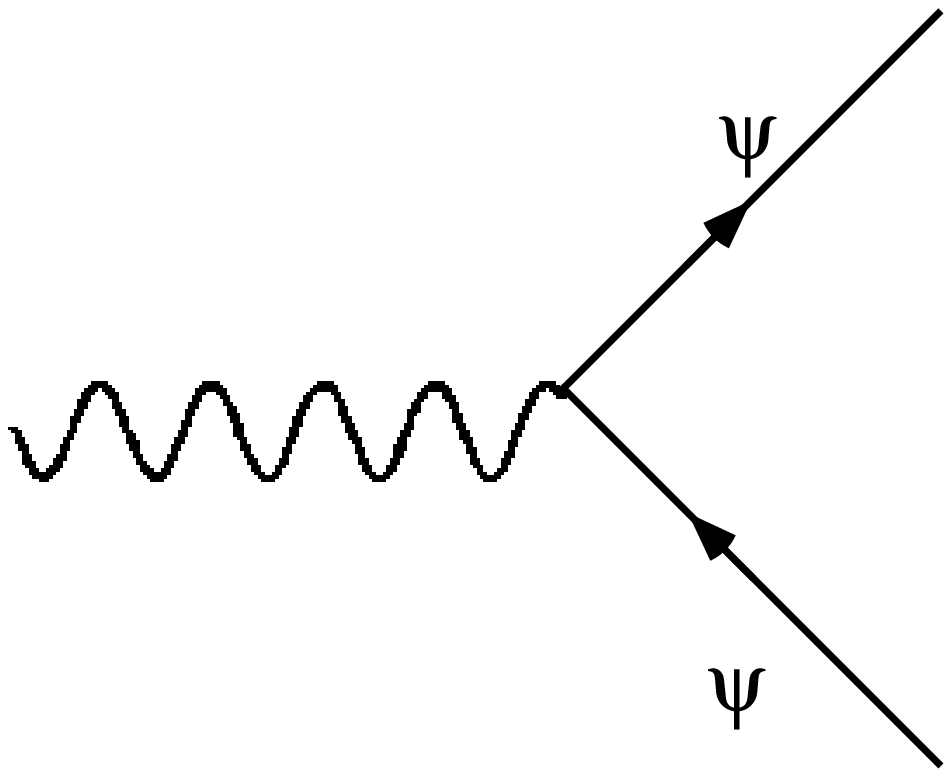}
\end{minipage}
\end{figure}

\vskip 0.2in

Finally, there will be two additional diagrams. One coming from the
interaction term involving both the chiral and gauge multiplet, and one
coming from $D^2$, 

\newpage
\begin{itemize}
\item a cubic interaction  involving a gaugino, and 
a chiral scalar and fermion pair, 
\end{itemize}
\vskip 0.5in

$\qquad  ~~ g ((\phi^*T^a\psi)\lambda^a + h.c.)
\qquad
$

\begin{figure}[h]
\vspace*{-0.7in} 
 \hspace*{2.5in}
\begin{minipage}{7.5in}
\includegraphics[width=4.5truecm]{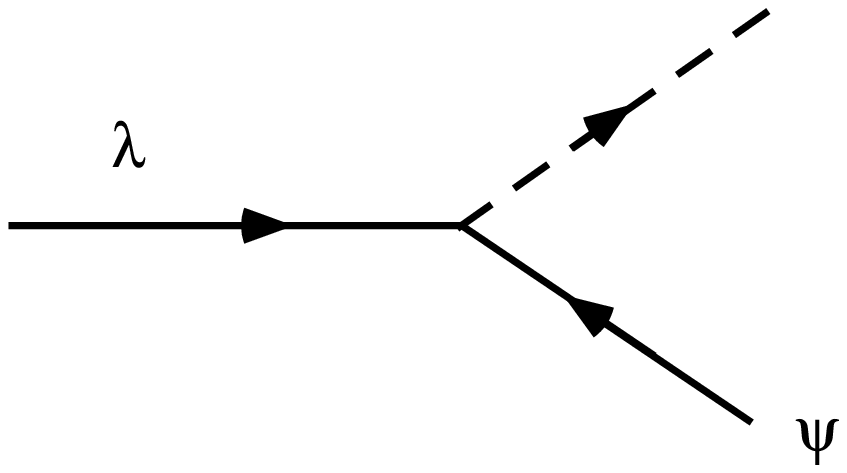}
\end{minipage}
\end{figure}

\vskip 0.4in

\begin{itemize}
\item another quartic interaction interaction involving a gaugino, and 
a chiral scalar and fermion pair, 
\end{itemize}
\vskip .7in

$\qquad  \qquad ~~ g^2 (\phi^*T^a\phi)^2
\qquad
$

\begin{figure}[h]
\vspace*{-1.1in} 
 \hspace*{2.5in}
\begin{minipage}{7.5in}
\includegraphics[width=4.0truecm]{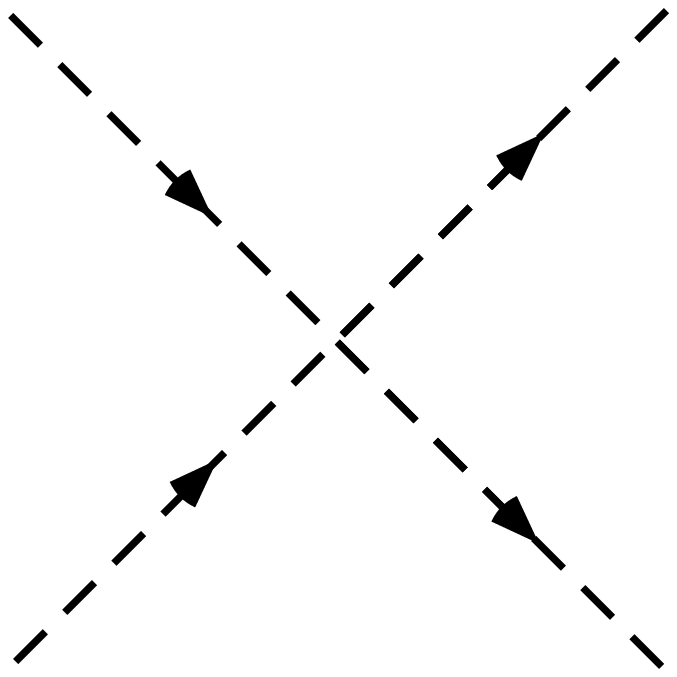}
\end{minipage}
\end{figure}

\vskip 0.5in

\subsection{Supersymmetry Breaking}

The world, as we know it, is clearly {\em not} supersymmetric.  Without much
ado, it is clear from the diagrams above, that for every chiral fermion of
mass $M$, we expect to have a scalar superpartner of equal mass. This is,
however, not the case, and as such we must be able to incorporate some degree
of supersymmetry breaking into the theory. At the same time, we would like to
maintain the positive benefits of supersymmetry such as the resolution of the
hierarchy problem.

To begin, we must be able to quantify what we mean by supersymmetry
breaking. From the anti-commutation relations (\ref{anticom}), we see that we
can write an expression for the Hamiltonian or $P^0$ using the explicit forms
of the Pauli matrices as
\beq
P^0 = {1 \over 4} \sum_{\alpha = 1}^2 \{ Q_\alpha , Q_{\dot \alpha}^\dagger \}
\eeq
A supersymmetric vacuum must be invariant under the supersymmetry
transformations and therefore would require $Q|0\rangle = 0$ and
$Q^\dagger|0\rangle = 0$ and therefore corresponds to $H=0$ and also
$V=0$.  Thus, the supersymmetric vacuum must have $|F|=|D|=0$.
Conversely, if supersymmetry is spontaneously broken, the vacuum is not
annihilated by the supersymmetry charge $Q$ so that $Q|0\rangle = \chi$
and $\langle \chi|Q|0\rangle = f_\chi^2$, where $\chi$ is a fermionic field
associated with the breakdown of supersymmetry and in analogy with the
breakdown of a global symmetry, is called the Goldstino.
For $f_\chi \ne 0$, $\langle 0|H|0\rangle = V_0 \ne 0$, and requires
therefore either (or both) $|F|\ne 0$ or $|D| \ne 0$. Mechanisms for the
spontaneous breaking of supersymmetry will be discussed in the next lecture.

It is also possible that to a certain degree, supersymmetry is explicitly
broken in the Lagrangian. In order to preserve the hierarchy between the
electroweak and GUT or Planck scales, it is necessary that the explicit
breaking of supersymmetry be done softly, i.e., by the insertion of 
weak scale mass terms in the Lagrangian. This ensures that the theory remain
free of quadratic divergences \cite{gg}. The possible forms for such terms are
\ba
{\cal L}_{soft} & = & -{1\over 2} M^a_\lambda \lambda^a \lambda^a
-{1\over 2} ({m^2})^{i}_{j} \phi_i {\phi^j}^* \nonumber \\
& & -{1\over 2} {(BM)}^{ij} \phi_i \phi_j - {1\over 6} {(Ay)}^{ijk} \phi_i
\phi_j \phi_k + h.c.
\ea 
\noindent where the $M^a_\lambda$ are gaugino masses, $m^2$ are soft scalar
masses, 
$B$ is a bilinear mass term, and $A$ is a trilinear mass term. 
Masses for the gauge bosons are of course forbidden by gauge invariance and
masses for chiral fermions are redundant as such terms are explicitly present
in
$M^{ij}$ already. The diagrams corresponding to these terms are

\begin{itemize}
\item a soft supersymmetry breaking gaugino mass insertion 
\end{itemize}
\vskip .3in

$\qquad \qquad \qquad  M^a_\lambda \lambda^a \lambda^a \qquad$

\begin{figure}[h]
\vspace*{-0.45in} 
 \hspace*{2.5in}
\begin{minipage}{7.5in}
\includegraphics[width=4.5truecm]{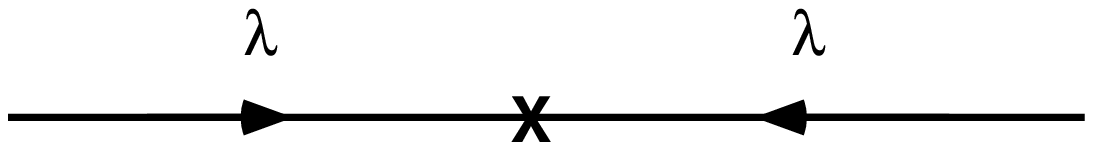}
\end{minipage}
\end{figure}

\vskip .3in

\begin{itemize}
\item a soft supersymmetry breaking scalar mass insertion
\end{itemize}
\vskip .3in

$\qquad \qquad \qquad ({m^2})^{i}_{j} \phi_i {\phi^j}^* \qquad$

\begin{figure}[h]
\vspace*{-0.45in} 
 \hspace*{2.5in}
\begin{minipage}{7.5in}
\includegraphics[width=4.5truecm]{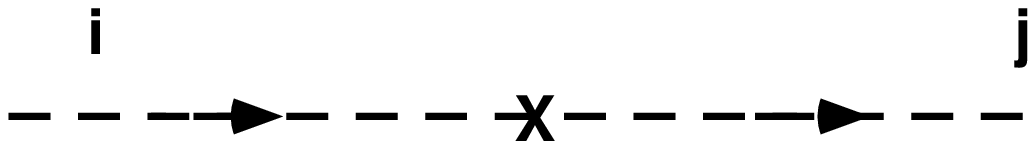}
\end{minipage}
\end{figure}

\vskip .3in 

\begin{itemize}
\item a soft supersymmetry breaking bilinear mass insertion
\end{itemize}
\vskip .3in

$\qquad \qquad \qquad {(BM)}^{ij} \phi_i \phi_j \qquad$

\begin{figure}[h]
\vspace*{-0.45in} 
 \hspace*{2.5in}
\begin{minipage}{7.5in}
\includegraphics[width=4.5truecm]{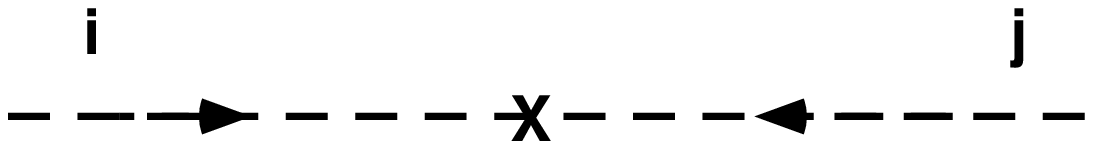}
\end{minipage}
\end{figure}

\vskip .5in
\newpage
\begin{itemize}
\item a soft supersymmetry breaking trilinear scalar interaction
\end{itemize}
\vskip .7in

$\qquad \qquad \qquad {(Ay)}^{ijk} \phi_i
\phi_j \phi_k \qquad$
\begin{figure}[h]
\vspace*{-.8in} 
 \hspace*{2.5in}
\begin{minipage}{7.5in}
\includegraphics[width=4.5truecm]{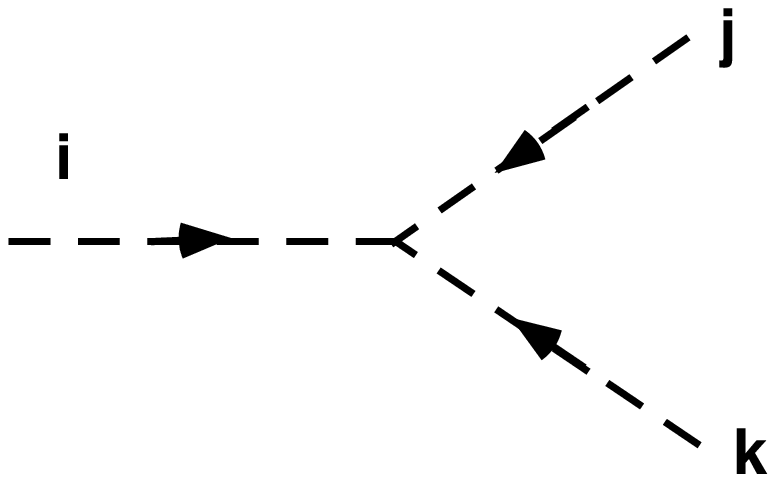}
\end{minipage}
\end{figure}

\vskip 0.4in

We are now finally in a position to put all of these pieces together and
discuss realistic supersymmetric models.

\section{The Minimal Supersymmetric Standard Model}

To construct the supersymmetric standard model \cite{fay} we start with the
complete set of chiral fermions in (\ref{chiralf}), and add a scalar
superpartner to each Weyl fermion so that each of the fields in
(\ref{chiralf}) represents a chiral multiplet. Similarly we must add a
gaugino for each of the gauge bosons in the standard model making up the
gauge multiplets. The minimal supersymmetric standard model (MSSM) \cite{MSSM}
is defined by its minimal field content (which accounts for the known
standard model fields) and minimal superpotential necessary to account for
the known  Yukawa mass terms. As such we define the MSSM by the superpotential
\beq
W = \epsilon_{ij} \bigl[ y_e H_1^j  L^i e^c + y_d H_1^j Q^i d^c + y_u H_2^i
Q^j u^c \bigr] + W_\mu 
\label{WMSSM}
\eeq 
where
\beq
W_\mu = \epsilon_{ij} \mu H_1^i H_2^j
\label{Wmu}
\eeq
In (\ref{WMSSM}), the indices, $\{ij\}$, are SU(2)$_L$ doublet indices.
The Yukawa couplings, $y$, are all $3 \times 3$ matrices in generation space.
Note that there is no generation index for the Higgs multiplets. Color and
generation indices have been suppressed in the above expression. There are
two Higgs doublets in the MSSM. This is a necessary addition to the standard
model which can be seen as arising from the holomorphic property of the
superpotential.  That is, there would be no way to account for all of the
Yukawa terms for both up-type and down-type multiplets with a single Higgs
doublet.  To avoid a massless Higgs state, a mixing term $W_\mu$ must be
added to the superpotential.

From the rules governing the interactions in supersymmetry discussed in the
previous section, it is easy to see that the terms in (\ref{WMSSM}) are
easily identifiable as fermion masses if the Higgses obtain vacuum expectation
values (vevs). For example, the first term will contain an interaction which
we can write as
\ba
&  & - {1 \over 2} {\partial^2 W \over \partial L \partial e^c} \bigl( \psi_L
\psi_{e^c} + \psi_L^\dagger \psi_{e^c}^\dagger \bigr)\nonumber \\
& = & - {1 \over 2} y_e H^0_1 \bigl( \ e {e^c} + e^\dagger {e^c}^\dagger
\bigr)
\label{emassl}
\ea
where it is to be understood that in (\ref{emassl}) that $H_1$ refers to the
scalar component of the Higgs $H_1$ and $\psi_L$ and $\psi_{e^c}$ represents
the fermionic component of the left-handed lepton doublet and right-handed
singlet respectively.  Gauge invariance  requires that as defined in
(\ref{WMSSM}), $H_1$ has hypercharge $Y_{H_1} = -1$ (and $Y_{H_2} = +1$).
Therefore if the two doublets obtain expectation values of the form
\beq
\langle H_1 \rangle = \pmatrix{v_1 \cr 0 \cr} \qquad
\langle H_2 \rangle = \pmatrix{0 \cr v_2 \cr}
\eeq
then (\ref{emassl}) contains a term which 
corresponds to an electron mass term with 
\beq
m_e = y_e v_1
\label{emass}
\eeq
Similar expressions are easily obtained for all of the other massive fermions
in the standard model. Clearly as there is no $\nu^c$ state in the minimal
model, neutrinos remain massless.  Both Higgs doublets must obtain vacuum
values and it is convenient to express their ratio as a parameter of the model,
\beq
\tan \beta = {v_2 \over v_1}
\eeq

\subsection{The Higgs sector}

Of course if the vevs for $H_1$ and $H_2$ exist, they must be derivable from
the scalar potential which in turn is derivable from the superpotential 
and any soft terms which are included. The part of the scalar potential which
involves only the Higgs bosons is
\ba
V & = & |\mu|^2 (H_1^* H_1 + H_2^* H_2)  +  {1 \over 8} {g^\prime}^2
(H_2^* H_2 - H_1^* H_1)^2 \nonumber \\
 & & + {1 \over 8} {g}^2 \left(4 |H_1^* H_2|^2 - 2(H_1^* H_1) (H_2^* H_2) +
(H_1^* H_1)^2 + (H_2^* H_2)^2\right) \nonumber \\
& & + m_1^2 H_1^* H_1 + m_2^2 H_2^* H_2 + (B \mu \epsilon_{ij} H_1^i H_2^j +
h.c. )
\label{shiggspot}
\ea
In (\ref{shiggspot}), the first term is a so-called $F$-term, derived from
$|(\partial W / \partial H_1)|^2$ and $|(\partial W / \partial H_2)|^2$
setting all sfermion vevs equal to 0. The next two terms are $D$-terms, the
first a U(1)-$D$-term, recalling that the hypercharges for the Higgses are
$Y_{H_1} = -1$ and $Y_{H_2} = 1$, and the second is an SU(2)-$D$-term, taking
$T^a =
\sigma^a /2$ where $\sigma^a$ are the three Pauli matrices. 
Finally, the last three terms are soft supersymmetry breaking masses $m_1$
and $m_2$, and the bilinear term $B\mu$. The Higgs doublets can be written as
\beq
\langle H_1 \rangle = \pmatrix{H_1^0 \cr H_1^- \cr} \qquad
\langle H_2 \rangle = \pmatrix{H_2^+ \cr H_2^0 \cr}
\eeq
and by $(H_1^* H_1)$, we mean ${H_1^0}^* {H_1^0} + {H_1^-}^* {H_1^-}$ etc. 

The neutral portion of (\ref{shiggspot}) can be expressed more simply as 
\ba
V & = &   {g^2 + {g^\prime}^2 \over 8} 
\left(|H_1^0|^2 - |H_2^0|^2 \right)^2 +  (m_1^2 + |\mu|^2)  |H_1^0|^2 
\nonumber \\
 & & + (m_2^2 + |\mu|^2)  |H_2^0|^2 + (B \mu
 H_1^0 H_2^0 + h.c. )
\label{neuthiggspot}
\ea
For electroweak symmetry breaking, it will be required that either one (or
both) of the soft masses ($m_1^2, m_2^2$) be negative (as in the standard
model).

In the standard model, the single Higgs doublet leads to one real scalar
field, as the other three components are eaten by the massive electroweak
gauge bosons.  In the supersymmetric version, the eight components result in
2 real scalars ($h,H$); 1 pseudo-scalar ($A$); and one charged Higgs
($H^\pm$); the other three components are eaten.
Also as in the standard model, one must expand the Higgs doublets about their
vevs, and we can express the components of the Higgses in terms of the mass
eigenstates
\ba
\langle H_1 \rangle = \pmatrix{v_1 + {1\over \sqrt{2}}\left[ H \cos \alpha -
h \sin \alpha + iA \sin \beta \right]
\cr H^- \sin \beta
\cr} \nonumber \\  \nonumber \\ \nonumber \\
\langle H_2 \rangle = \pmatrix{H^+ \cos \beta \cr  v_2 + {1\over
\sqrt{2}}\left[ H \sin \alpha + h \cos \alpha + i A \cos \beta \right] \cr}
\ea
From the two vevs, we can define $v^2 = v_1^2 + v_2^2$ so that
$M_w^2 = {1\over 2} g^2 v^2$ as in the standard model.

In addition, electroweak symmetry breaking places restrictions on the set of
parameters appearing in the Higgs potential (\ref{neuthiggspot}).
If we use the two conditions
\beq
{\partial V \over \partial |H_1^0|} = 0 \qquad {\partial V \over \partial
|H_2^0|} = 0
\eeq
with a little algebra, we can obtain the following two conditions
\beq
-2B\mu = (m_1^2 + m_2^2 + 2\mu^2) \sin 2\beta
\label{cond1}
\eeq
and
\beq
v^2 = {4 \left( m_1^2 + \mu^2 - (m_2^2 + \mu^2)\tan^2 \beta \right)
\over (g^2 + {g^\prime}^2) (\tan^2 \beta -1)}
\label{cond2}
\eeq
From the potential and these two conditions, the masses of the physical
scalars can be obtained. At the tree level,
\beq
m_{H^\pm}^2 = m_A^2 + m_W^2
\eeq
\beq
m_A^2 = m_1^2 + m_2^2 + 2\mu^2 = -B\mu (\tan \beta + \cot \beta)
\eeq
\beq
m_{H,h}^2 = {1\over2} \left[ m_A^2 + m_Z^2 \pm \sqrt{( m_A^2 + m_Z^2)^2 - 4
m_A^2 m_Z^2 \cos^2 2\beta}
~\right]
\eeq
The Higgs mixing angle is defined by
\beq
\tan 2\alpha = \tan 2\beta \left[ {m_H^2 + m_h^2 \over  m_A^2 - m_Z^2} \right]
\eeq
Notice that these expressions and the above constraints limit the number of
free inputs in the MSSM.  First, from the mass of the pseudoscalar, we see
that $B\mu$ is not independent and can be expressed in terms of $m_A$ and
$\tan \beta$. Furthermore from the conditions (\ref{cond1}) and (\ref{cond2}),
we see that if we keep $\tan \beta$, we can either either choose $m_A$ and
$\mu$ as free inputs thereby determining the two soft masses, $m_1$ and $m_2$,
or we can choose the soft masses as inputs, and fix $m_A$ and $\mu$ by the
conditions for electroweak symmetry breaking. Both choices of parameter
fixing are widely used in the literature. 

The tree level expressions for the Higgs masses make some very definite
predictions.  The charged Higgs is heavier than $M_W$, and the light Higgs
$h$, is necessarily lighter than $M_Z$.  Note if uncorrected, the MSSM would
already be excluded (see discussion on current accelerator limits in section
6). However, radiative corrections to the Higgs masses are not negligible in
the MSSM, particularly for a heavy top mass
$m_t \sim$ 175 GeV. The leading one-loop
corrections to $m^2_h$ depend quartically on $m_t$ and can be expressed as
\cite{susyHiggs}
\ba
\Delta m^2_h & = & {3m^4_t\over 4\pi^2v^2}~~\ln~~\left({m_{\tilde t_1}
m_{\tilde t_2}\over m^2_t}\right) \nonumber \\
& & + {3m^4_t \hat A^2_t\over
8\pi^2 v^2}~~\left[2h(m^2_{\tilde t_1}, m^2_{\tilde t_2})+ \hat A^2_t
~~g(m^2_{\tilde t_1}, m^2_{\tilde t_2})\right] + \ldots
\label{higgs1loop}
\ea
where $m_{\tilde t_{1,2}}$ are the physical masses of the two stop
squarks $\tilde t_{1,2}$ to be discussed in more detail shortly, $\hat A_t
\equiv A_t + \mu \cot\beta$, ($A_t$ is supersymmetry breaking trilinear term
associated with the top quark Yukawa coupling).  The functions $h$ and $f$ are
\beq
h(a,b) \equiv {1\over a-b}~\ln \left({a\over b}\right)~,~~g(a,b) =
{1\over (a-b)^2}~\left[2 - {a+b\over a-b}~\ln\left({a\over
b}\right)\right]
\label{funcs}
\eeq
Additional corrections to coupling vertices, two-loop corrections and
\linebreak renormalization-group resummations have also been computed
in the MSSM \cite{moreradcorr}. With these corrections one can allow
\beq
m_h \la 130~{\rm GeV}
\label{higgslimit}
\eeq
within the MSSM.
While certainly higher than the tree level limit of $M_Z$, the limit still
predicts a relatively light Higgs boson, and allows the MSSM to be
experimentally excluded (or verified!) at the LHC.

Recalling the expression for the electron mass in the MSSM (\ref{emass}), we
can now express the Yukawa coupling $y_e$ in terms of masses and $(\tan)
\beta$,
\beq
y_e = {g m_e \over \sqrt{2} M_W \cos \beta}
\eeq
There are similar expressions for the other fermion masses, with the
replacement $\cos \beta \to \sin \beta$ for up-type fermions. 

\subsection{The sfermions}

We turn next to the question of the sfermion masses \cite{ER}.  As an example,
let us consider the $\tilde u$ mass$^2$ matrix. Perhaps the most complicated
entry in the mass$^2$ matrix is the $L-L$ component. To begin with, there is
a soft supersymmetry breaking mass term, $m_Q^2$. In addition, from the
superpotential term, $y_u H_2 Q u^c$, we obtain an $F$-term contribution by
taking $\partial W / \partial u^c = y_u H_2 Q$.  Inserting the vev for $H_2$,
we have in the $F$-term,
\beq
|\partial W / \partial u^c|^2 = |y_u v_2 \t u|^2 =
m_u^2 |{\t u}|^2
\eeq
This is generally negligible for all but third generation sfermion
masses. Next we have the $D$-term contributions. Clearly to generate a $\t u^*
\t u$ term, we need only consider the $D$-term contributions from diagonal
generators, i.e., $T^3$ and $Y$, that is from 
\ba
D^3 & = & -{1 \over 2} g \left[ |H_1^0|^2  - |H_2^0|^2 + |\t u|^2 + \cdots
\right] \\
D^\prime & = & -{1 \over 2} g^\prime \left[ |H_2^0|^2  - |H_1^0|^2 + Y_Q
|\t u|^2 +\cdots
\right]
\ea
where $Y_Q = 2q_u -1 = 1/3$ is the  quark doublet hypercharge. Once again,
inserting vevs for $H_1$ and $H_2$ and keeping only relevant terms, we have
for the $D$-term contribution
\ba
{1 \over 2} |D^3|^2 + {1 \over 2} |D^\prime|^2 & = &
{1 \over 4} \left( g^2 v^2 \cos 2\beta - {g^\prime}^2 v^2 \cos 2 \beta (2q_u -
1) \right) {\tilde u}^* {\tilde u} \nonumber \\
& = & M_Z^2 \cos 2\beta \left( {1\over 2} - q_u \sin^2 \theta_W \right)
{\tilde u}^* {\tilde u}
\ea
Thus the total contribution to the $L-L$ component of the up-squark mass$^2$
matrix is 
\beq
m_{u_L}^2 = m_Q^2 + m_u^2 + M_Z^2 \cos 2\beta \left( {1\over 2} - q_u \sin^2
\theta_W \right)
\eeq
Similarly it is easy to see that the $R-R$ component can be found from the
above expressions by discarding the SU(2)$_L$ $D$-term contribution
and recalling that $Y_{u^c} = 2q_u$.
Then,
\beq
m_{u_R}^2 = m_{u^c}^2 + m_u^2 + M_Z^2 \cos 2\beta \left( q_u \sin^2
\theta_W \right)
\eeq
There are, however, in addition to the above diagonal entries, off-diagonal
terms coming from a supersymmetry breaking $A$-term, and an $F$-term. The
$A$-term is quickly found from $A_Q y_u H_2 Q u^c$ when setting the vev for
$H_2 = v_2$ and yields a term $A_Q m_u$. The $F$-term contribution comes from
$\partial W / \partial H_2 = \mu H_1 + y_u Q u^c$. When inserting the vev
and taking the square of the latter term, and keeping the relevant mass
term, we find for the total off-diagonal element
\beq
m_{u_L u_R}^2 = m_u ( A_Q + \mu \cot \beta ) = m_u {\hat A_Q}
\eeq 
Note that for a down-type sfermion, the definition of ${\hat A}$ is modified
by taking $\cot \beta \to \tan \beta$. Also note that the off-diagonal term
is negligible for all but the third generation sfermions. 

Finally to determine the physical sfermion states and their masses we must
diagonalize the sfermion mass matrix. This mass matrix is easily diagonalized
by writing the diagonal sfermion eigenstates as
\ba
\tilde f_1 &=& \tilde f_L \cos\theta_f + \tilde f_R \sin\theta_f\;,\nonumber \\
\noalign{\medskip}
\tilde f_2 &=& - \tilde f_L \sin \theta_f + \tilde f_R \cos \theta_f\;.
\ea
With these conventions we have the diagonalizing angle and mass eigenvalues
\[
\theta_f = {\rm sign}[-m_{LR}^2]\left\{ \begin{array}{ll}
{\pi\over 2} -{ 1 \over 2 } \arctan |2 m_{LR}^2/
(m_L^2 - m_R^2)|,  & m_L^2 > m_R^2\;,\\
\noalign{\medskip}
{ 1 \over 2 } \arctan |2 m_{LR}^2/
(m_L^2 - m_R^2)|,  & m_L^2 < m_R^2\;,
                   \end{array}
           \right.
\]
\beq
m^2_{ 1,2 } =  { 1 \over 2 }  \left[ ( m_R^2 + m_L^2 ) \mp \sqrt {
( m_R^2 - m_L^2 )^2 +
4 m_{LR}^4 } \right]\;.
\eeq
Here $\theta_f$ is chosen so that $m_1$ is always lighter that $m_2$.
Note that in the special case $m_L = m_R = m$, we have
$\theta_f= $ sign[$-m_{LR}^2$]$(\pi/4)$ and $ m^2_{ 1,2 } = m^2 \mp
|m_{LR}^2|$.

\subsection{Neutralinos}

There are four new neutral fermions in the MSSM which not only receive mass
but mix as well. These are the gauge fermion partners of the neutral $B$ and
$W^3$ gauge bosons, and the partners of the Higgs.  The two gauginos are
called the bino, ${\widetilde B}$, and wino, ${\widetilde W^3}$ respectively.
The latter two are the Higgsinos, ${\widetilde H_1}$ and ${\widetilde H_2}$.
In addition to the supersymmetry breaking gaugino mass terms, $- {1 \over
2} M_1 {\widetilde B} {\wt B}$, and $-{1\over 2} M_2 {\wt W}^3 \wt W^3$, there
are supersymmetric mass contributions of the type $W^{ij}  \psi_i \psi_j$,
giving a mixing term between $\wt H_1$ and $\wt H_2$, ${1\over 2} \mu \wt H_1
\wt H_2$, as well as terms of the form $g (\phi^*T^a\psi)\lambda^a$ giving the
following mass terms after the appropriate Higgs vevs have been inserted,
${1 \over \sqrt{2}} g^\prime v_1 \wt H_1 \wt B$, 
$-{1 \over \sqrt{2}} g^\prime v_2 \wt H_2 \wt B$,
$-{1 \over \sqrt{2}} g v_1 \wt H_1 \wt W^3$, and
${1 \over \sqrt{2}} g v_2 \wt H_2 \wt W^3$.
These latter terms can be written in a simpler form noting that for example,
$g^\prime v_1 / \sqrt{2} = M_Z \sin \theta_W  \cos \beta$. Thus we can write
the neutralino mass matrix as (in the $({\wt B}, {\wt W}^3, {{\wt
H}^0}_1,{{\wt H}^0}_2 )$ basis) \cite{ehnos}
\beq     
  \left( \begin{array}{cccc}
M_1 & 0 & {-M_Z s_{\theta_W} \cos \beta} &  {M_Z s_{\theta_W} \sin \beta} \\
0 & M_2 & {M_Z c_{\theta_W} \cos \beta} & -{M_Z c_{\theta_W} \sin \beta} \\
{-M_Z s_{\theta_W} \cos \beta} & {M_Z c_{\theta_W} \cos \beta} & 0 & -\mu \\
{M_Z s_{\theta_W} \sin \beta} & -{M_Z c_{\theta_W} \sin \beta} & -\mu & 0 
\end{array} \right) 
\label{neutmass}
\eeq
where $s_{\theta_W} = \sin \theta_W$ and $c_{\theta_W} = \cos \theta_W$.
The mass eigenstates (a linear combination of the four neutralino states) and
the mass eigenvalues are found by diagonalizing the mass matrix
(\ref{neutmass}). However, by a change of basis involving two new states
\cite{ehnos}
\begin{equation}
		\wt S^0 = \wt H_1 \sin\beta + \wt H_2 \cos\beta
\end{equation}
\beq
	{\wt A}^0 = -{{{\wt H}^0}_1 \cos \beta + {{\wt H}^0}_2} \sin \beta
\eeq
  the mass matrix simplifies and becomes ( in the $(\wt B, \wt W^3, \wt A,
\wt S)$ basis)
\beq     
  \left( \begin{array}{cccc}
M_1 & 0 & {M_Z s_{\theta_W}} &  0 \\
0 & M_2 & -{M_Z c_{\theta_W}}  & 0 \\
{M_Z s_{\theta_W}} & -{M_Z c_{\theta_W}} & \mu \sin 2\beta & \mu \cos 2\beta
\\ 0 & 0 & \mu \cos 2\beta & -\mu \sin 2\beta
\end{array} \right) 
\label{neutmassS}
\eeq
In this basis, the eigenstates (which as one can see depend only the the
three input mass, $M_1, M_2$, and $\mu$) can be solved for analytically.

Before moving on to discuss the chargino mass matrix, it will be useful for
the later discussion to identify a few other neutralino states. These are the
photino, 
\begin{equation}
		\wt {\gamma} = \wt W^3\sin\theta_W + \wt B \cos\theta_W
\end {equation}
and a symmetric and antisymmetric combination of Higgs bosons,
\begin{equation}
		\wt H_{(12)} = \frac{1}{\sqrt{2}}  (\wt H_1 + \wt H_2)
\end{equation}
\begin{equation}
		\wt H_{[12]} = \frac{1}{\sqrt{2}}  ( \wt H_1 - \wt H_2)
\end{equation}

\subsection{Charginos}

There are two new charged fermionic states which are the partners of the
$W^\pm$ gauge bosons and the charged Higgs scalars, $H^\pm$, which are the
charged gauginos, $\wt W^\pm$ and charged Higgsinos, $\wt H^\pm$, or
collectively charginos. The chargino mass matrix is composed similarly to the
neutralino mass matrix. The result for the mass term is
\beq
-{1\over 2} ~(\wt W^-, \wt H^-)~~ \left(\matrix{M_2 & \sqrt{2}
m_W\sin\beta \cr \sqrt{2} m_W\cos\beta & \mu}\right) ~~\left(\matrix{\wt
W^+\cr\wt H^+}\right)~~+~~h.c.
\label{chmass}
\eeq
Note that unlike the case for neutralinos, two unitary matrices must be
constructed to diagonalize (\ref{chmass}).  The result for the mass
eigenstates of the two charginos is
\ba
m_{\wt c_1}^2,m_{\wt c_2}^2 & = & {1 \over 2} \Bigl[ M_2^2 + \mu^2 + 2M_W^2 
\nonumber \\
& \mp & \sqrt{(M_2^2 + \mu^2 + 2M_W^2 )^2 - 4(\mu M_2 - M_W^2 \sin 2\beta)^2}
~\Bigr]
\ea

\subsection{More Supersymmetry Breaking}
As was noted earlier, supersymmetry breaking can be associated with a
positive vacuum energy density, $V > 0$.  Clearly from the definition of the
scalar potential, this can be achieved if either (or both) the $F$-terms or
the $D$-terms are non-zero.  Each of the these two possibilities is called
$F$-breaking and $D$-breaking respectively (for obvious reasons).

\subsubsection{$D$-Breaking}

One of the simplest mechanisms for the spontaneous breaking of supersymmetry,
proposed by Fayet and Illiopoulos \cite{fi}, involves merely adding to the
Lagrangian a term proportional to $D$,
\beq
{\cal L}_{\rm FI} = \kappa D
\label{lagfi}
\eeq 
It is easy to see by examining (\ref{varD}) that this is possible only for a
U(1) gauge symmetry.  For a U(1), the variation of (\ref{lagfi}) under
supersymmetry is simply a total derivative.  The scalar potential is now
modified
\beq
V(D) = - {1\over 2} |D|^2 - \kappa D - g (q_i {\phi^i}^* \phi_i) D
\eeq
where $q_i$ is the U(1) charge of the scalar $\phi_i$. 
As before, the equations of motion are used to eliminate the auxiliary field
$D$ to give
\beq
D = - \kappa - g (q_i {\phi^i}^* \phi_i)
\eeq
So long as the U(1) itself remains unbroken (and the scalar fields $\phi_i$
do not pick up expectation values, we can set $\langle \phi_i \rangle = 0$,
and hence 
\beq
\langle D \rangle = - \kappa
\eeq
with
\beq
V = {1\over 2} \kappa^2
\eeq
and supersymmetry is broken.
Unfortunately, it is not possible that $D$-breaking of this type occurs on
the basis of the known U(1) in the standard model, i.e., U(1)$_Y$, as the
absence of the necessary mass terms in the superpotential would not prevent
the known sfermions from acquiring vevs. It may be possible that some other
U(1) is responsible for supersymmetry breaking via $D$-terms, but this is
most certainly beyond the context of MSSM.

\subsubsection{$F$-Breaking}

Although $F$-type breaking also requires going beyond the standard model, it
does not do so in the gauge sector of the theory.  $F$-type breaking can be
achieved relatively easily by adding a few gauge singlet chiral multiplets,
and one the simplest mechanisms was proposed by O'Raifertaigh \cite{oraif}.
In one version of this model, we add three chiral supermultiplets, $A,B$, and
$C$, which are coupled through the superpotential
\beq
W = \alpha AB^2 + \beta C (B^2 - m^2)
\label{woraif}
\eeq
The scalar potential is easily determined from (\ref{woraif})
\ba
V & = & |F_A|^2 + |F_B|^2 + |F_C|^2 \nonumber \\
& = & |\alpha B^2|^2 + |2B (\alpha A + \beta C)|^2 + 
|\beta (B^2 - m^2)|^2
\ea
Clearly, the first and third terms of this potential can not be made to
vanish simultaneously, and so for example,  if $B = 0$, $F_C \ne 0$, $V > 0$,
and supersymmetry is broken.

It is interesting to examine the fermion mass matrix for the above toy model. 
The mass matrix is determined from the superpotential through $W^{ij}
\psi_i \psi_j$ and in the $(A,B,C)$ basis gives
\beq
\pmatrix{ 0 & 2 \alpha B & 0 \cr
2 \alpha B & 2 ( \alpha A + \beta C) & 2 \beta B \cr
0 & 2 \beta B&  0 \cr}
\eeq
The fact that the determinant of this matrix is zero, indicates that there is
at least one massless fermion state, the Goldstino. 

The existence of the Goldstino as a signal of supersymmetry breaking was
already mentioned in the previous section. It is relatively straightforward
to see that the Goldstino can be directly constructed from the $F$- and
$D$-terms which break supersymmetry. Consider the mass matrix for a gaugino
$\lambda^a$, and chiral fermion $\psi_i$
\beq
\pmatrix{
0   &
\sqrt{2} g  (\langle \phi^{*}\rangle T^a)^i    \cr
\sqrt{2} g  (\langle \phi^{*}\rangle T^a)^j  &
\langle W^{ij} \rangle
}
\label{gmatrix}
\eeq
where we do not assume any supersymmetry breaking gaugino mass.
Consider further, the fermion 
\beq
{\wt G} = 
\pmatrix{{\langle D^a \rangle /\sqrt{2}} , \langle F_i
\rangle \cr}
\eeq
in the ($\lambda,\psi$) basis. Now from the condition (\ref{forgold})
and the requirement that we are sitting at the minimum of the potential so
that
\beq
{\partial V\over \partial \phi_j} = 0 \leftrightarrow 
g  (\langle \phi^{*}\rangle T^a)^i + F_j W^{ij} = 0
\eeq
we see that the fermion $\wt G$ is massless, that is, it is annihilated by
the mass matrix (\ref{gmatrix}). The Goldstino state $\wt G$ is physical so
long as one or both $\langle D \rangle \ne 0$, or  $\langle F \rangle \ne 0$.
This is the analog of the Goldstone mechanism for the breaking of global
symmetries. 

\subsection{R-Parity}

In defining the supersymmetric standard model, and in particular the
minimal model or MSSM, we have limited the model to contain a minimal field
content.  That is, the only new fields are those which are {\em required} by
supersymmetry.  In effect, this means that other than superpartners, only the
Higgs sector was enlarged from one doublet to two. However, in writing the
superpotential (\ref{WMSSM}), we have also made a minimal choice regarding
interactions.  We have limited the types of interactions to include only
those required in the standard model and its supersymmetric
generalization.

However, even if we stick to the minimal field content, there are several
other superpotential terms which we can envision adding to (\ref{WMSSM}) which
are consistent with all of the symmetries (in particular the gauge symmetries)
of the theory.  For example, we could consider
\beq
W_{R} = {1\over 2} \lambda^{ijk} L_iL_j {e^c_k}
+ \lambda^{\prime ijk} L_i Q_j {d^c_k} + {1\over 2} \lambda^{\prime\prime ijk}
{u^c_i}{d^c_j}{d^c_k}
+ \mu^{\prime i} L_i H_u
\label{wr}
\eeq 
In (\ref{wr}), the terms proportional to $\lambda, \lambda^\prime$, and
$\mu^\prime$, all violate lepton number by one unit.  The term proportional
to $\lambda^{\prime\prime}$ violates baryon number by one unit. 

Each of
the terms in (\ref{wr}) predicts new particle interactions and can be to some
extent constrained by the lack of observed exotic phenomena. However, the
combination of terms which violate both baryon and lepton number can be
disastrous. For example, consider the possibility that both $\lambda^\prime$
and $\lambda^{\prime\prime}$ were non-zero.  This would lead to the following
diagram which mediates proton decay, $p \to e^+ \pi^0, \mu^+ \pi^0, \nu \pi^+,
\nu K^+$ etc.
\begin{figure}[h]
\vspace*{0.25in} 
\hspace*{1.5in}
\centering
\begin{minipage}{7.5in}
\includegraphics[width=8.5truecm]{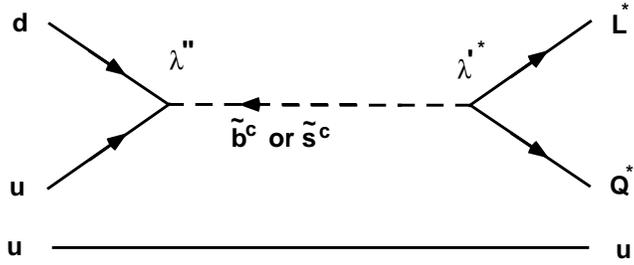}
\end{minipage}
\caption{R-parity violating contribution to proton decay.}
\end{figure}
Because of the necessary antisymmetry of the final two flavor indices in
$\lambda^{\prime\prime}$, there can be no $\t d^c$ exchange in this diagram.
The rate of proton decay as calculated from this diagram will be enormous due
to the lack of any suppression by superheavy masses. There is no GUT or
Planck scale physics which enters in, this is a purely (supersymmetric)
standard model interaction. The (inverse) rate can be easily estimated to be 
\beq
\Gamma^{-1}_p \sim {\t m^4 \over m_p^5} \sim 10^8 {\rm GeV}^{-1}
\eeq
assuming a supersymmetry breaking scale of $\t m$ of order 100 GeV. 
This should be compared with current limits to the proton life-time of $\ga
10^{63}$ GeV$^{-1}$. 

It is possible to eliminate the unwanted superpotential terms by imposing a
discrete symmetry on the theory.  This symmetry has been called $R$-parity
\cite{Rparity}, and can be defined as
\beq
R = (-1)^{3B + L + 2s}
\eeq
where $B,L$, and $s$ are the baryon number, lepton number, and spin
respectively. With this definition, it turns out that all of the known
standard model particles have $R$-parity +1.  For example, the electron has
$B=0$, $L=-1$, and $s=1/2$, the photon as $B=L=0$ and $s=1$.  In both cases,
$R=1$. Similarly, it is clear that all superpartners of the known standard
model particles have $R=-1$, since they must have the same value of $B$ and
$L$ but differ by 1/2 unit of spin. If $R$-parity is exactly conserved, then
all four superpotential terms in (\ref{wr}) must be absent.  But perhaps an
even more important consequence of $R$-parity is the prediction that the
lightest supersymmetric particle or LSP is stable. In much the same way that
baryon number conservation predicts proton stability, $R$-parity predicts
that the lightest $R = -1$ state is stable.  This makes supersymmetry an
extremely interesting theory from the astrophysical point of view, as the LSP
naturally becomes a viable dark matter candidate \cite{gold,ehnos}.  This will
be discussed in detail in the 6th lecture.

\section{The Constrained MSSM and Supergravity}

As a phenomenological model, while the MSSM has all of the ingredients which
are necessary, plus a large number of testable predictions, it contains far
too many parameters to pin down a unique theory.  Fortunately, there are a
great many constraints on these parameters due to the possibility of exotic
interactions as was the case for additional $R$-violating superpotential
terms. The supersymmetry breaking sector of the theory contains a huge number
of potentially independent masses. However, complete arbitrariness in the
soft sfermion masses would be a phenomenological disaster.  For example,
mixing in the squark sector, would lead to a completely unacceptable
preponderance of flavor changing neutral currents \cite{EN}.

Fortunately, there are some guiding principles that we can use to relate the
various soft parameters which not only greatly simplifies the theory, but also
removes the phenomenological problems as well. Indeed, among the motivations
for supersymmetry was a resolution of the hierarchy problem \cite{hier}. We
can therefore look to unification (grand unification or even unification with
gravity) to establish these relations \cite{DG}.

The simplest such hypothesis is to assume that all of the soft supersymmetry
breaking masses have their origin at some high energy scale, such as the GUT
scale.  We can further assume that these masses obey some unification
principle. For example, in the context of grand unification, one would expect
all of the members of a given GUT multiplet to receive a common supersymmetry
breaking mass. For example, it would be natural to assume that at the GUT
scale, all of the gaugino masses were equal, $M_1=M_2=M_3$ (the latter is the
gluino mass).  While one is tempted to make a similar assumption in the case
of the sfermion masses (and we will do so), it is not as well justified.
While one can easily argue that sfermions in a given GUT multiplet should
obtain a common soft mass, it is not as obvious why all $R=-1$ scalars should
receive a common mass. 

Having made the assumption that the input parameters are fixed at the GUT
scale, one must still calculate their values at the relevant low energy scale.
This is accomplished by ``running" the renormalization group equations
\cite{Inoue}.  Indeed, in standard (non-supersymmetric) GUTs, the gauge
couplings are fixed at the unification scale and run down to the electroweak
scale.  Conversely, one can use the known values of the gauge couplings and
run them up  to determine the unification scale (assuming that the couplings
meet at  a specific renormalization scale). 

\subsection{RG evolution}

To check the prospects of unification in detail  requires using the two-loop
renormalization equations
\beq
{d\alpha_i\over dt} = -{1\over 4\pi}~~\left( b_i +
{b_{ij}\over 4\pi}\alpha_j\right)\alpha_i^2
\label{runalpha}
\eeq
where $t = \ln (M_{GUT}^2/Q^2)$, and the $b_i$ are given by
\beq
b_i = \left(\matrix{ 0 \cr -{22\over 3} \cr -11}\right) + N_g
\left(\matrix{{4\over 3} \cr\cr {4\over 3} \cr\cr {4\over 3}}\right) + N_H
\left(\matrix{{1\over 10} \cr\cr {1\over 6} \cr \cr 0}\right)
\label{foureight}
\eeq
from gauge bosons, $N_g$ matter generations and $N_H$ Higgs doublets,
respectively, and at two loops
\beq
b_{ij} = \left(\matrix{0&0&0\cr\cr 0&-{136\over 3} & 0 \cr\cr
0&0&-102}\right) + N_g
\left(\matrix{{19\over 15} & {3\over 5} & {44\over 15} \cr\cr {1\over 5} & {49\over 3} & 4
\cr\cr {11\over 30} & {3\over 2} & {76\over 3}}\right) +  N_H \left(
\matrix{{9\over 50} &
{9\over 10} & 0 \cr\cr {3\over 10} & {13\over 6} & 0 \cr\cr 0 & 0 & 0}\right)
\label{bsm}
\eeq
These coefficients 
depend only on the light particle content of the theory. 

\begin{figure}[hbtp]
	\centering
	\includegraphics[width=9.truecm]{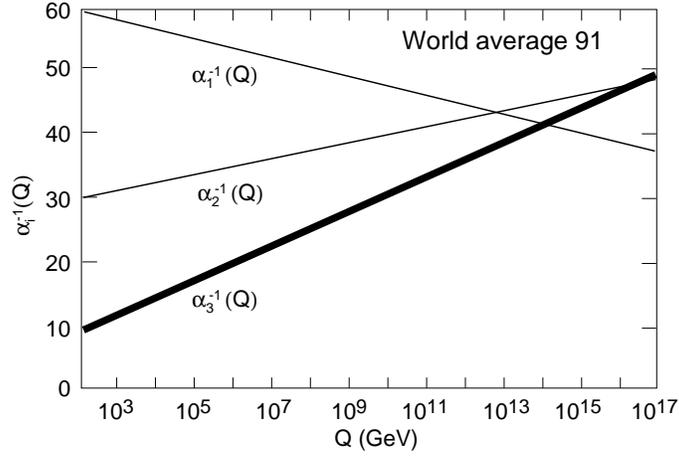}
	\caption{RG evolution of the inverse gauge couplings in the standard model
\cite{EKN,ellis}.}
	\label{smgauge}
\end{figure}

However, using the known inputs at the electroweak scale, one finds
\cite{EKN} that the couplings of the standard model are not unified at any
high energy scale.  This is shown in Figure \ref{smgauge}.

In the MSSM, the additional particle content changes the slopes in
the RGE evolution equations. Including supersymmetric particles, one
finds~\cite{DRW}
\beq
b_i = \left(\matrix{0 \cr\cr -6 \cr\cr -9}\right) + N_g
\left(\matrix{2\cr\cr 2 \cr\cr 2}\right) + N_H \left(\matrix{{3\over 10}
\cr\cr {1\over 2}\cr\cr 0}\right)
\label{fourten}
\eeq
and
\beq
b_{ij} = \left(\matrix{0&0&0\cr\cr 0&-24 & 0 \cr\cr 0&0&-54}\right) + N_g
\left(\matrix{{38\over 15} & {6\over 5} & {88\over 15} \cr\cr {2\over 5} & 14 & 8
\cr\cr {11\over 15} & 3 & {68\over 3}}\right) +  N_H \left( \matrix{{9\over
50} & {9\over 10} & 0 \cr\cr {3\over 10} & {7\over 2} & 0 \cr\cr 0 & 0 &
0}\right)
\label{fbmssm}
\eeq

In this case, it is either a coincidence, or it is rather remarkable that the
RG evolution is altered in just such a way as to make the MSSM consistent
with unification.  The MSSM evolution is shown in Figure \ref{mssmgauge}
below. 
\begin{figure}[hbtp]
	\centering
	\includegraphics[width=9.truecm]{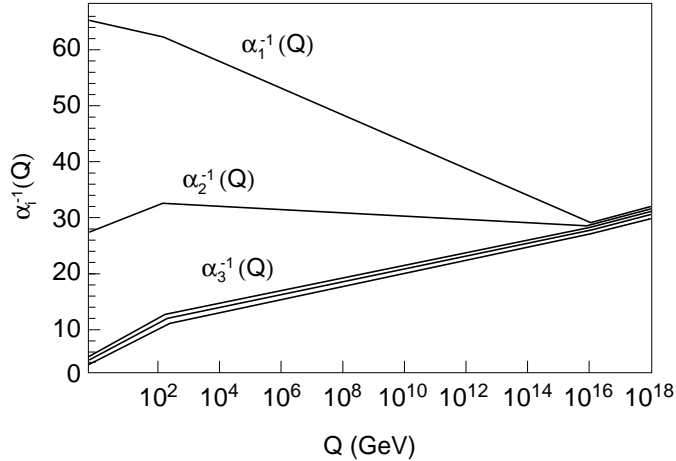}
	\caption{RG evolution of the inverse gauge couplings in the MSSM
\cite{EKN,ellis}.}
	\label{mssmgauge}
\end{figure}
For many, this concordance of the gauge couplings and GUTs offers strong
motivation for considering supersymmetry. 

As was noted earlier, most of the parameters in the MSSM are also subject to
RG evolution. For example, all of the Yukawa couplings also run.
Here, I just list the RG equation for the top quark Yukawa,
\beq
{d \alpha_t \over dt} = {\alpha_t \over 4\pi} \left( {16 \over 3} \alpha_3 + 
{3} \alpha_2 + {13 \over 15} \alpha_1 - 6\alpha_t - \alpha_b +\cdots \right)
\eeq
where $\alpha_t = y_t^2/4\pi$. This is the leading part of the 1-loop
correction.  For a more complete list of these equations see \cite{rge}.
These expressions are also known to higher order \cite{spm}. Note that the
scaling of the supersymmetric couplings are all proportional to the couplings
themselves. That means that if the coupling is not present at the tree level,
it will not be generated by radiative corrections either.  This is a general
consequence of supersymmetric nonrenormalization theorems \cite{nrt}.

The supersymmetry breaking mass parameters also run. Starting with the
gaugino masses, we have
\beq
{dM_i \over dt} = -b_i \alpha_i M_i /4 \pi
\eeq
Assuming a common gaugino mass, $m_{1/2}$ at the GUT scale as was discussed
earlier, these equations are easily solved in terms of the fine structure
constants, 
\beq
M_i (t)  = {\alpha_i (t) \over \alpha_i(M_{GUT})} m_{1/2}
\eeq
This implies that 
\beq
{M_1 \over g_1^2}={M_2 \over g_2^2}={M_3 \over g_3^2}
\eeq
(Actually, in a GUT, one must modify the relation due to the difference
between the U(1) factors in the GUT and the standard model, so that we have
$M_1 = {5\over 3} {\alpha_1\over \alpha_2} M_2$.)

Finally, we have the running of the remaining mass parameters. 
A few examples are:
\ba
{d \mu^2 \over dt} & = & {3 \mu^2 \over 4 \pi} \left( \alpha_2 + {1 \over 5}
\alpha_1 - \alpha_t - \alpha_b +\cdots \right) \\
{d m_{e^c}^2 \over dt} & = & {12\over 5} {\alpha_1 \over 4 \pi} M_1^2 + \cdots
\\ {d A_i \over dt} & = & {1 \over 4 \pi} \left({16 \over 3} \alpha_3 M_3 +
3 \alpha_2 M_2 + {13 \over 15} \alpha_1 M_1 - 6 \alpha_t A_t + \cdots
\right)\\ {dB
\over dt} & = & {3 \over 4 \pi} \left(
 \alpha_2 M_2 + {1 \over 5} \alpha_1 M_1 - 3 \alpha_t A_t + \cdots
\right)
\ea

\subsection{The Constrained MSSM}

As the name implies, the constrained MSSM or CMSSM, is a subset of the
possible parameter sets in the MSSM. In the CMSSM \cite{rewsb,kane}, we try to
make as many reasonable and well motivated assumptions as possible.  
To begin with gaugino mass unification is assumed. (This is actually a common
assumption in the MSSM as well).  Furthermore soft scalar mass unification or
universality is also assumed.  This implies that {\em all} soft scalar masses
are assumed equal at the GUT input scale, so that
\beq
{\t m}^2(M_{GUT}) = m_0^2
\eeq
This condition is applied not only to the sfermion masses, but also to the
soft Higgs masses, $m_{1,2}^2$ as well. By virtue of the conditions
(\ref{cond1}) and (\ref{cond2}), we see that in the CMSSM, $\mu$, and
$B\mu$, (or $m_A^2$), are no longer free parameters since these conditions
amount to $m_A^2(m_1^2, m_2^2, \mu^2)$ and $v^2((m_1^2, m_2^2, \mu^2)$. 
Thus we are either free to pick $m_A, \mu$ as free parameters (this fixes
$m_{1,2}$, though we are usually not interested in those quantities) as in
the MSSM, or choose $m_{1,2}$ (say at the GUT scale) and $m_A$ and $\mu$
become predictions of the model. Universality of the soft trilinears, $A_i$,
is also assumed.

In the CMSSM therefore, we have only the following free input parameters:
$m_{1/2}, m_0$, $\tan \beta$, $A_0$, and the sign of $\mu$. We could of course
choose phases for some these parameters.  In the MSSM and CMSSM, there are
two physical phases which can be non-vanishing, $\theta_\mu$, and $\theta_A$.
If non-zero, they lead to various CP violating effects such as inducing
electric dipole moments in the neutron and electron. For some references
regarding these phases see \cite{phases,fkos,fko}, but we will not discuss
them further in these lectures. 

In the figure below, an example of the running of the mass parameters
in the CMSSM is shown.  Here, we have chosen $m_{1/2} = 250$ GeV, $m_0 = 100$
GeV, $\tan \beta = 3$, $A_0 = 0$, and $\mu < 0$.
\begin{figure}[t]
	\centering
	\includegraphics[width=9.truecm]{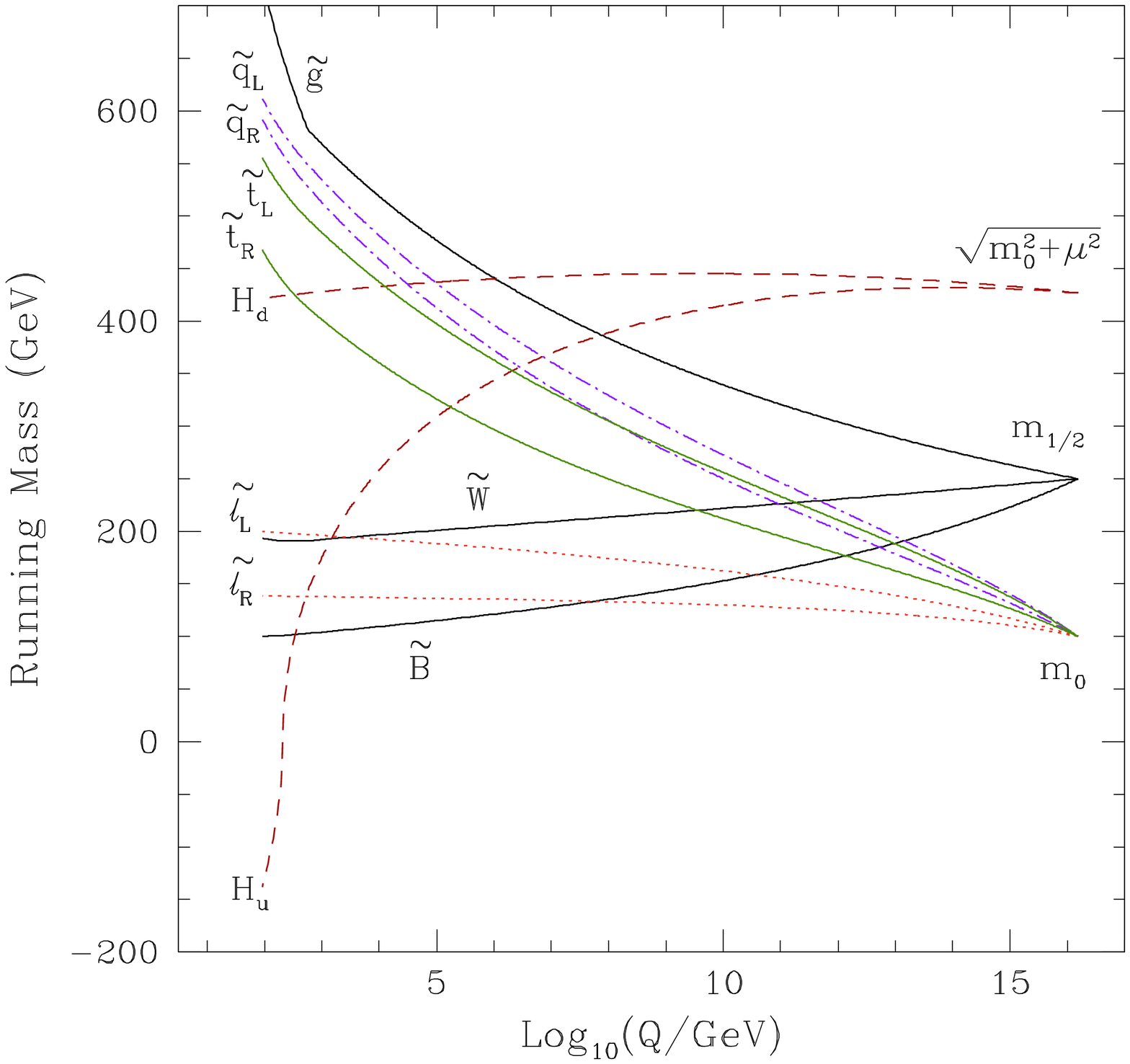}
\vskip -.2in
	\caption{RG evolution of the mass parameters in the CMSSM.
I thank Toby Falk for providing this figure.}
	\label{running}
\end{figure}
Indeed, it is rather amazing that from so few input parameters, all of the
masses of the supersymmetric particles can be determined. 
The characteristic features that one sees in the figure, are for example, that
the colored sparticles are typically the heaviest in the spectrum.  This is
due to the large positive correction to the masses due to $\alpha_3$ in the
RGE's.  Also, one finds that the $\wt B$, is typically the lightest
sparticle.  But most importantly, notice that one of the Higgs mass$^2$, goes
negative triggering electroweak symmetry breaking \cite{rewsb}. (The negative
sign in the figure refers to the sign of the mass$^2$, even though it is the
mass of the sparticles which are depicted.) In the Table below, I list some
of the resultant electroweak scale masses, for the choice of parameters used
in the figure. 

\begin{table}[htb]
\caption{Physical mass parameters for $m_{1/2} = 250$ GeV, $m_0 = 100$
GeV, $\tan \beta = 3$, $A_0 = 0$, and $\mu < 0$. (These are related but not
equal to the running mass parameters shown in Figure
\protect{\ref{running}}.)}
\begin{center}
\begin{tabular}{cccc}
\hline\hline
 particle & mass & parameter & value
\\
\hline
$m_{\t l}$ & 203 & $\mu$ & -391  \\
$m_{\t e^c}$ & 144 & $M_1$ & 100  \\
$m_{\t \nu}$ & 190 & $M_2$ & 193  \\
$m_{\t q}$ & 412--554 & $M_3$ & 726  \\
$m_{\t \chi_1}$ & 104 & $\alpha_3(M_Z)$ & .123  \\
$m_{\t \chi^\pm_1}$ & 203 & $A's$ & 163--878  \\
$m_{h}$ & 93 &  &   \\
$m_{A}$ & 466 & &   \\
\hline\hline
\end{tabular}
\end{center}
\label{tab:data}
\end{table}

\subsection{Supergravity}

Up until now, we have only considered global supersymmetry.  Recall, our
generator of supersymmetry transformations, the spinor $\xi$.
In all of the derivations of the transformation and invariance properties in
the first two sections, we had implicitly assumed that $\partial_\mu \xi =
0$.  By allowing
$\xi(x)$ and
$\partial_\mu
\xi(x)
\ne 0 $, we obtain local supersymmetry or supergravity \cite{sugr}.
It is well beyond the means of these lectures to give a thorough treatment
of local supersymmetry.  We will therefore have to content ourself with some
general remarks from which we can glimpse at some features for which we can
expect will have some phenomenological relevance.

First, it is important to recognize that our original Lagrangian for the
Wess-Zumino model involving a single noninteracting chiral multiplet will no
longer be invariant under local supersymmetry transformations.  New terms,
proportional to $\partial_\mu \xi(x)$ must be canceled.  In analogy with
gauge, theories which contain similar terms and are canceled by the
introduction of vector bosons, here the terms must be canceled by
introducing a new spin 3/2 particle called the gravitino.
The supersymmetry transformation property of the gravitino must be
\beq
\delta_\xi \Psi_\mu^\alpha  \propto \partial_\mu \xi^\alpha
\eeq
Notice that the gravitino carries both a gauge and a spinor index. 
The gravitino is part of an $N=1$ multiplet
which contains the spin two graviton. In unbroken supergravity, the massless
gravitino has two spin components ($\pm$ 3/2) to match the two polarization
states of the graviton.

The scalar potential is now determined by a analytic function of the scalar
fields, called the K\"ahler potential, $K(\phi,\phi^*)$.  The  K\"ahler
potential can be thought of as a metric in field space, 
\beq
{\cal L}_{kin} = K^i_j \partial_\mu \phi_i \partial^\mu {\phi^j}^*
\eeq 
where $K^i = \partial K / \partial \phi_i$ and $K_i = \partial K / \partial
{\phi^i}^*$. In what is known as minimal $N=1$ supergravity, the K\"ahler
potential is given by
\beq
K = \kappa^2 \phi_i {\phi^i}^*  + \ln (\kappa^6 |W|^2)
\label{min}
\eeq
where $W(\phi)$ is the superpotential, $\kappa^{-1} = M_P/\sqrt{8\pi}$ and the
Planck mass is $M_P = 1.2\times 10^{19}$ GeV. The scalar potential (neglecting
any gauge contributions) is \cite{sugr2}
\beq
V(\phi,\phi^*) = e^K \kappa^{-4} \left[ K^i (K^{-1})^j_i K_j -3 \right]
\eeq
For minimal supergravity, we have $K^i = \kappa^2 {\phi^i}^* + {W^i}/W$, $K_i
= \kappa^2 \phi_i + W_i^*/W^*$, and $({K^{-1}})^j_i = \delta ^j_i/\kappa^2$.
Thus the resulting scalar potential is
\beq
V(\phi,\phi^*)  = e^{\kappa^2 \phi_i {\phi^i}^*}  \left[ |W^i + {\phi^i}^* W
|^2 - 3\kappa^2|W|^2
\right]
\label{sgpot}
\eeq

As we will now see, one of the primary motivations for the CMSSM, and scalar
mass universality comes from the simplest model for local supersymmetry
breaking.  The model \cite{polonyi} involves one additional chiral multiplet
$z$, (above the normal matter fields $\phi_i$). Let us consider therefore, a
superpotential which is separable in the so-called Polonyi field and matter
so that 
\beq
W(z,\phi_i) = f(z) + g(\phi_i)
\label{sep}
\eeq
and in particular let us choose 
\beq
f(z) = \mu(z + \beta)
\label{polonyi}
\eeq
and for reasons to be
clear shortly, $\beta = 2 - \sqrt{3}$. I will from now on work in units such
that $\kappa = 1$. 
If we ignore for the moment the matter fields $\phi$, the potential for $z$
becomes
\beq
V(z,z^*) = e^{zz^*} \mu^2 \left[ |1 + z^* (z + \beta)|^2 - 3|(z+\beta)|^2
\right]
\label{VZ}
\eeq
It is not difficult to verify that with the above choice of $\beta$, the
minimum of $V$ occurs at $\langle z \rangle = \sqrt{3} - 1$, with
$V(\langle z \rangle) = 0$.

Note that by expanding this potential, one can define two real scalar fields
$A$ and $B$, with mass eigenvalues, 
\beq
m_A^2 = 2\sqrt{3} m_{3/2}^2 \qquad m_B^2 = 2(2-\sqrt{3}) m_{3/2}^2
\eeq
where the gravitino mass is
\beq
m_{3/2} = e^{K/2} = e^{2-\sqrt{3}} \mu
\eeq
Note also that there is a mass relation, $m_A^2 + m_B^2  = 4 m_{3/2}^2$,
which is a guaranteed consequence of supertrace formulae in supergravity
\cite{sugr2}.  Had we considered the fermionic sector of the theory, we would
now find that the massive gravitino has four helicity states $\pm 1/2$ and
$\pm 3/2$. The ``longitudinal" states arise from the disappearance of the
goldstino (the fermionic partner of $z$ in this case) by the superHiggs
mechanism, again in analogy with the spontaneous breakdown of a gauge symmetry
\cite{superh,sugr2,polonyi}. 

We next consider the matter potential from eqs. (\ref{sep}) and
(\ref{sgpot}). In full, this takes the form \cite{carlos}
\ba
V & = & e^{(|z|^2 + |\phi|^2)} \left[ |{\partial f \over \partial z} + z^* (
f(z) + g(\phi) )|^2 \right. \nonumber \\
& & + \left. |{\partial g \over \partial \phi} + \phi^* (
f(z) + g(\phi) )|^2 - 3 | f(z) + g(\phi) |^2 \right]
\ea
Here again, I have left out the explicit powers of $M_P$. Expanding this
expression, and at the same time dropping terms which are suppressed by
inverse powers of the Planck scale (this can be done by simply dropping terms
of mass dimension greater than four), we have, after inserting the vev for
$z$ \cite{carlos}, 
\ba
V & = & e^{(4 - 2\sqrt{3})} \left[ |\mu + (\sqrt{3} - 1) (\mu +
g(\phi)) |^2 \right. \nonumber \\
& & \left. +|{\partial g \over \partial \phi} + \phi^* (
\mu + g(\phi) )|^2 - 3 | \mu + g(\phi) |^2 \right] \nonumber \\
& = & e^{(4 - 2\sqrt{3})} \left[ -\sqrt{3} \mu (g(\phi) + g^*(\phi^*)) + 
|{\partial g \over \partial \phi}|^2 \right. \nonumber \\
& & + \left. ( \mu ( \phi {\partial g \over \partial \phi} + \phi^* {\partial
g^* \over \partial \phi^*}) + \mu^2 \phi \phi^* \right] \nonumber \\
& = & e^{(4 - 2\sqrt{3})} |{\partial g \over \partial \phi}|^2 \nonumber \\
& & + 
m_{3/2} e^{(2 - \sqrt{3})}(\phi {\partial g \over \partial \phi} - \sqrt{3} g
+ h.c.) )  + m_{3/2}^2 \phi \phi^*
\label{cpot}
\ea

This last expression deserves some discussion. First, up to an overall
rescaling of the superpotential, $g \to e^{\sqrt{3}-2} g$, the first term is
the ordinary $F$-term part of the scalar potential of global supersymmetry. 
The next term, proportional to $m_{3/2}$ represents a universal trilinear
$A$-term. This can be seen by noting that $\sum \phi \partial g / \partial
\phi = 3 g$, so that in this model of supersymmetry breaking, $A = (3 -
\sqrt{3}) m_{3/2}$.  Note that if the superpotential contains bilinear
terms, we would find $B = (2 - \sqrt{3}) m_{3/2}$. The last term represents a
universal scalar mass of the type advocated in the CMSSM, with
$m_0^2 = m_{3/2}^2$.  The generation of such soft terms is a rather generic
property of low energy supergravity models \cite{mark}. 

Before concluding this section, it worth noting one other class of
supergravity models, namely the so-called no-scale supergravity model
\cite{noscale}. No-scale supergravity is based on the K\"ahler potential of
the form
\beq
K = -\ln (S + S^*) - 3 \ln (T + T^* - \phi_i {\phi^i}^*) + \ln |W|^2
\label{ns}
\eeq
where the $S$ and $T$ fields are related to the dilaton and moduli fields in
string theory \cite{sns}.  If only the $T$ field is kept in (\ref{ns}), the
resulting scalar potential is exactly flat, i.e., $V = 0$ identically.  In
such a model, the gravitino mass is undetermined at the tree level, and up to
some field redefinitions, there is a surviving global supersymmetry. 
No-scale supergravity has been
used heavily in constructing supergravity models in which all mass
scales below the Planck scale are determined radiatively
\cite{nsguts},\cite{nshid}.

\section{Cosmology}

Supersymmetry has had a broad impact on cosmology.  In these last two
lectures, I will try to highlight these.  In this lecture, I will briefly
review the problems induced by supersymmetry, such as the Polonyi or moduli
problem, and the gravitino problem.  I will also discuss the question of
cosmological inflation in the context of supersymmetry. Finally, I will
describe a mechanism for baryogenesis which is purely supersymmetric in
origin.  I will leave the question of dark matter and the accelerator
constraints to the last lecture. 

Before proceeding to the problems, it will be useful to establish some of the
basic quantities and equations in cosmology. 
The standard big bang model assumes homogeneity and
isotropy, so that space-time can be described by the 
Friedmann-Robertson-Walker metric which in co-moving coordinates is given by
\beq
	ds^2  = -dt^2  + R^2(t)\left[ {dr^2 \over \left(1-kr^2\right) }
      + r^2 \left(d\theta^2  + \sin^2 \theta d\phi^2 \right)\right]	
\label{met}
\eeq
where $R(t)$ is the cosmological scale factor and $k$ is the three-space
curvature constant ($k = 0, +1, -1$ for a spatially flat, closed or open
Universe). $k$ and $R$ are the only two quantities in the
metric which distinguish it from flat Minkowski space.
It is  also common to assume
 the perfect fluid form for the energy-momentum
tensor
\beq
	T_{\mu\nu}   = pg_{\mu\nu}   + (p + \rho)u_\mu u_\nu 		
\eeq
where $g_{\mu\nu}$   is the space-time metric described by (\ref{met}),
 $p$ is the isotropic
pressure, $\rho$ is the energy density and $u^\mu  = (1,0,0,0)$
 is the velocity vector
for the isotropic fluid.  Einstein's equation  yield the
Friedmann equation,
\beq
	H^2  \equiv \left({\dot{R} \over R}\right)^2  = {1 \over 3} \kappa^2 \rho
 - { k \over R^2}  + {1 \over 3} \Lambda
\label{H}
\eeq
and
\beq
	\left({\ddot{R} \over R}\right) = {1 \over 3} \Lambda -
 {1 \over 6} \kappa^2 ( \rho + 3p)
\eeq
where $\Lambda$ is the cosmological constant,
or equivalently from $	{T^{\mu\nu}}_{;\nu}   =  0$
\beq
	\dot{\rho} = -3H(\rho + p)	
\label{rhod}	
\eeq
These equations form the basis of the standard big bang model.

If our Lagrangian contains scalar fields, then from the scalar field
contribution to the energy-momentum tensor
\beq
	T_{\mu\nu}  
	    = \partial_{\mu} \phi \partial_{\nu} \phi 
- {1 \over 2} g_{\mu\nu}  \partial_{\rho} \phi
\partial^{\rho} \phi - g_{\mu\nu}V(\phi)	
\eeq
we can identify 
the energy density and pressure due to a scalar $\phi$,
\begin{eqnarray}
	\rho =  {1 \over 2} {\dot{\phi}}^2  + {1 \over 2} R^{-2}(t)
(\nabla\phi)^2  + V(\phi)		\\
	p =  {1 \over 2} {\dot{\phi}}^2  - {1 \over 6} R^{-2}(t)
(\nabla\phi)^2  - V(\phi)	
\label{prho}
\end{eqnarray}
In addition, we have the equation of motion,
\beq
{\ddot \phi} + 3 H {\dot \phi} + {\partial V \over \partial \phi} = 0
\label{sevol}
\eeq

Finally, I remind the reader that in the early radiation dominated Universe,
the energy density (as well as the Hubble parameter) is determined by the
temperature,
\beq
\rho = N {\pi^2 \over 30} T^4 \qquad H = \sqrt{\pi^2 N \over 90} \kappa {T^2}
\eeq 
The critical energy density (corresponding to $k= 1$, is 
\beq
\rho_c = {3 H^2 \kappa^{-2}} = 1.88 \times 10^{-29} {\rm g cm}^{-3} h_0^2
\eeq
where the scaled Hubble parameter is $h_0 = H_0/100$km Mpc$^{-1}$s$^{-1}$.
The cosmological density parameter is defined as $\Omega = \rho/\rho_c$.

\subsection{The Polonyi Problem}

The Polonyi problem, although based on the specific choice for breaking
supergravity discussed in the previous lecture (eq. \ref{polonyi}), is a
generic problem in broken supergravity models. In fact, the problem is
compounded in string theories, where there are in general many such fields
called moduli. 
Here, attention will be restricted to the simple example of the Polonyi
potential.

The potential in eq. (\ref{VZ}) has the property that at its minimum occurs
at $\langle z \rangle = (\sqrt{3} -1) M$, where $M = \kappa^{-1}$ is the
reduced Planck mass. Recall that the constant $\beta$ was chosen so
that at the minimum, $V(\langle z \rangle) = 0$. In contrast to the value of
the expectation value, the curvature of the potential at the minimum, is
$m_z^2
\sim \mu^2$, which as argued earlier is related to the gravitino mass and
therefore must be of order the weak scale. In addition the value of the
potential at the origin is of order $V(0)\sim \mu^2 M^2$, i.e., an
intermediate scale.   Thus, we have  a long and very flat potential.

Without too much difficulty, it is straightforward to show that such a
potential can be very problematic cosmologically \cite{pprob}.
The evolution of the Polonyi field $z$, is governed by eq. (\ref{sevol}) with
potential (\ref{VZ}). There is no reason to expect that the field $z$ is
initially at its minimum. This is particularly true if there was a prior
inflationary epoch, since quantum fluctuations for such a light field would be
large, displacing the field by an amount of order $M$ from its minimum. If $z
\ne \langle z \rangle$, the evolution of
$z$ must be traced.
When the Hubble parameter $H > \mu$ , $z$ is approximately constant.
That is, the potential term (proportional to $\mu^2$) can be neglected. 
At later times, as $H$ drops, 
$z$ begins to oscillate about the minimum when $H \la \mu$. 
Generally, oscillations begin when $H \sim m_z \sim \mu$ as can be seen from
the solution for the evolution of a non-interacting massive field with $V =
m_z^2 z^2/2$. This solution would take the form of $z \sim \sin (m_z t) /t$
with $H = 2/3t$.

At the time that the $z$-oscillations begin, the Universe becomes dominated
by the potential $V(z)$, since $H^2 \sim \rho /M^2$. Therefore all other
contributions to $\rho$ will redshift away, leaving the potential as the
dominant component to the energy density. Since the oscillations evolve as
non-relativistic matter (recall that in the above solution for $z$, $H =
2/3t$ as in a matter dominated Universe).   As the Universe evolves, we can
express the energy density as $\rho \sim \mu^2 M^2 (R_z/R)^3$, where $R_z$
is the value of the scale factor when the oscillations begin. 
Oscillations continue, until the $z$-fields can decay. Since they are only
gravitationally coupled to matter, their decay rate is given by $\Gamma_z
\sim \mu^3/M^2$. Therefore oscillations continue until $H\sim \Gamma_z$ or
when $R= R_{dz} \sim (M/\mu)^{4/3}$. The energy density at this time is
only $\mu^6/M^2$.  Even if the the thermalization of the decay products
occurs rapidly, the Universe reheats only to a temperature of order $T_R \sim
\rho^{1/4} \sim \mu^{3/2}/M^{1/2}$.  For $\mu \sim 100$ GeV, we have $T_R
\sim 100$ keV! There are two grave problems with this scenario.  The first is
is that big bang nucleosynthesis would have taken place during the
oscillations which is characteristic of a matter dominated expansion rather
than a radiation dominated one.  Because of the difference in the expansion
rate the abundances of the light elements would be greatly altered (see e.g.
\cite{ks}).  Even more problematic is the entropy release due to the decay of
these oscillations.  The entropy increase \cite{pprob} is related to the
ratio of the reheat temperature to the temperature of the radiation in the
Universe when the oscillations decay, $T_d \sim T_i (R_z/R_{dz})$ where
$T_i$ is the temperature when oscillations began $T_i \sim (\mu M)^{1/2}$.
Therefore, the entropy increase is given by
\beq
	S_f /S_i \sim (T_R /T_d  )^3 \sim (M /\mu)  \sim 10^{16}  	
\eeq
This is far too much to understand the present value of the baryon-to-entropy
ratio, of order $10^{-11} - 10^{-10}$ as required by nucleosynthesis and
allowed by baryosynthesis. That is, even if a baryon asymmetry of order one
could be produced, a dilution by a factor of $10^{16}$ could not be
accommodated.

\subsection{The Gravitino Problem}

Another problem which results from the breaking of supergravity is the
gravitino problem \cite{grprob}. If gravitinos are present with equilibrium
number  densities, we can write  their energy density as
\beq
\rho_{3/2} = m_{3/2} n_{3/2} = m_{3/2} \left( 3\zeta(3) \over \pi^2\right)
T_{3/2}^2
\eeq
where today one expects that the gravitino temperature $T_{3/2}$ is reduced
relative to the photon temperature due to the annihilations of particles
dating back to the Planck time \cite{oss}.  Typically one can expect
$Y_{3/2} = (T_{3/2}/T_\gamma)^3 \sim 10^{-2}$. Then for $\Omega_{3/2} h^2 \la
1$, we obtain the limit that $m_{3/2} \la 1$ keV.

Of course, the above mass limits assumes a stable gravitino, the problem
persists however, even if the gravitino decays, since its gravitational decay
rate is very slow.  Gravitinos decay when their decay rate, $\Gamma_{3/2}
\simeq m_{3/2}^3/M_P^2$,  becomes comparable to the expansion rate of the
Universe (which becomes dominated by the mass density of gravitinos), $H
\simeq m_{3/2}^{1/2} T_{3/2}^{3/2}/M_P$.  Thus decays occur at $T_d \simeq
m_{3/2}^{5/3}/M_P^{2/3}$. After the decay, the Universe is ``reheated" to 
a temperature 
\beq
T_R \simeq \rho(T_d)^{1/4} \simeq m_{3/2}^{3/2}/M_P^{1/2}
\eeq
As in the case of the decay of the Polonyi fields, the Universe must reheat
sufficiently so that big bang nucleosynthesis occurs in a standard radiation
dominated Universe. For $T_R \ga 1$ MeV, we must require $m_{3/2} \ga 20$ TeV.
This large value threatens the solution of the hierarchy problem.
In addition, one must still be concerned about the
dilution of the baryon-to-entropy ratio \cite{grinf}, in this case by a factor
$\Delta = (T_R/T_D)^3 \sim Y (M_P/m_{3/2})^{1/2}$.
Dilution may not be a problem if the baryon-to-entropy ratio
is initially large.

Inflation (discussed below) could alleviate the gravitino problem by diluting
the gravitino abundance to safe levels \cite{grinf}. 
If gravitinos satisfy the noninflationary bounds,
then their reproduction after inflation is never a problem.
For gravitinos with mass of order 100 GeV,
dilution without over-regeneration will also solve the problem,
but there are several
factors one must contend with in order to be cosmologically safe.
Gravitino decay products can also upset
the successful predictions of Big Bang nucleosynthesis,
 and decays into LSPs (if R-parity is conserved) can also yield
too large a mass density in the now-decoupled LSPs \cite{ehnos}.
 For unstable
gravitinos, the most restrictive bound on their number density comes form the
photo-destruction of the light elements produced during nucleosynthesis
\cite{grnuc}
\beq
	n_{3/2}/n_\gamma  \la 10^{-13}   (100 {\rm GeV}/m_{3/2}   )		
\eeq
for lifetimes $ > 10^4$  sec.  Gravitinos are regenerated after inflation and
one can estimate \cite{ehnos,grinf,grnuc}
\beq
n_{3/2}/n_\gamma \sim (\Gamma /H)(T_{3/2}/T_\gamma)^3 \sim \alpha N(T_R) (T_R
/M_P )(T_{3/2}/T_\gamma)^3	
\eeq
where $\Gamma \sim \alpha N(T_R) (T_R^3/M_P^2)$  is the production rate of
gravitinos.  Combining these last two equations one can derive bounds on $T_R$
\beq
	T_R \la 4 \times 10^9~{\rm GeV} (100~{\rm GeV}/m_{3/2})		
\label{grlimit}
\eeq
using a more recent calculation of the gravitino regeneration rate
\cite{enor}. A slightly stronger bound (by an order of magnitude in $T_R$)
was found in \cite{km}.

\subsection{Inflation}

It would be impossible in the context of these lectures to give any kind of
comprehensive review of inflation whether supersymmetric or not.  I refer the
reader to several reviews \cite{infl}.  Here I will mention only the most
salient features of inflation as it connects with supersymmetry.

Supersymmetry was first introduced \cite{pri} in inflationary models as a
means to cure some of the problems associated with the fine-tuning of new
inflation \cite{new}.
New inflationary models based on a Coleman-Weinberg type of $SU(5)$
 breaking produced density fluctuations \cite{pert} with magnitude
$\delta\rho/\rho \sim O(10^2)$ rather 
than $\delta\rho/\rho \sim 10^{-5}$
 as needed to remain consistent with microwave background
 anisotropies. Other more technical problems\cite{lin2}
 concerning slow rollover and
the effects of quantum fluctuations also passed doom on this original model.

The problems associated with new inflation, had to with the interactions
of the scalar field driving inflation, namely the $SU(5)$ adjoint. One
cure is to (nearly) completely decouple the field driving inflation, the
inflaton, from the gauge sector. As gravity becomes the primary interaction
to be associated with the inflaton it seemed only natural to take all scales
to be the Planck scale \cite{pri}. Supersymmetry was also
employed to give flatter potentials and acceptable density
perturbations\cite{dens}.
These models were then placed in the context of N=1
supergravity\cite{nost1,hrg}. 

The simplest such model for the inflaton $\eta$, is based on a superpotential
of the form
\beq
	W(\eta) = \mu^2(1 - \eta/M_P)^2 M_P
\label{minin}
\eeq
or
\beq	
	W(\eta) = \mu^2 (\eta - \eta^4 / 4 {M_P}^3)	
\label{1,1}
\eeq
where Eq.(\ref{minin})\cite{hrg} is to be used in minimal
 supergravity while Eq.(\ref{1,1})\cite{eenos} is to be used
 in no-scale supergravity. Of course the real goal is to determine the
identity of the inflaton.  Presumably string theory holds the answer to this
question, but a fully string theoretic inflationary model has yet to be
realized \cite{bg}.

For the remainder of the discussion, it will be
sufficient to  consider only a generic model of inflation whose potential is
of the form:
\begin{equation}
 V( \eta ) = {{\mu}^4} P( \eta )
\label{a}
\end{equation}
where $\eta$ is the scalar field driving inflation, the inflaton,  
$\mu$ is an as yet unspecified mass parameter, and $P(\eta)$ is a  
function of $\eta$ which possesses the features necessary for  
inflation, but contains no small parameters, i.e., 
where all of the couplings in $P$ are $O(1)$ but may contain
non-renormalizable terms. 

The requirements for successful inflation can be expressed in terms of two
conditions:
\linebreak 1) enough inflation;
\beq {\partial^2 V \over \partial \eta^2}\mid_{\eta \sim {\eta_i} \pm H}
< {3 H^2 \over 65} = {8 \pi V(0) \over 65 {M_P}^2}
\eeq
2) density perturbations of the right magnitude\cite{pert};
\beq
{\delta \rho \over \rho} \simeq {H^2 \over 10 \pi^{3/2} \dot{\eta}}
\simeq O(100) {\mu^2 \over {M_P}^2}
\label{perts}
\eeq
given here for scales which ``re-enter" the
 horizon during the matter dominated
era. 
       For large scale fluctuations of the type measured by COBE\cite{cobe},
we can use Eq. (\ref{perts}) to fix the inflationary scale $\mu$ \cite{cdo}:
\begin{equation}
{\frac{\mu^2}{M_P^2} = {\rm few} \times{10^{-8}}}
\label{cobemu}
\end{equation}

       Fixing $({\mu^2}/{M_P^2})$ has immediate general consequences  
for inflation\cite{eeno}. For example, the Hubble parameter during inflation,  
${{H^2} \simeq (8\pi/3)({\mu^4}/{M_P^2})}$ so that $H \sim  
10^{-7}M_P$. The duration of inflation is $\tau \simeq  
{M_P^3}/{\mu^4}$, and the number of e-foldings of expansion is $H\tau  
\sim 8\pi({M_P^2}/{\mu^2}) \sim 10^{9}$. If the inflaton decay rate  
goes as $\Gamma \sim {m_{\eta}^3}/{M_P^2} \sim {\mu^6}/{M_P^5}$, the  
universe recovers at a temperature $T_R \sim (\Gamma{M_P})^{1/2} \sim  
{\mu^3}/{M_P^2} \sim 10^{-11} {M_P} \sim 10^8 GeV$. However, it
was noted in \cite{eeno} that in fact the Universe is not immediately
thermalized subsequent to inflaton decays,
and the process of thermalization actually leads to a smaller
reheating temperature,
\begin{equation}
T_R \sim \alpha^2 \mu^3/M_P^2 \sim 10^5 GeV~,
\end{equation}
 where $\alpha^2 \sim 10^{-3}$ characterizes the strength of the
interactions leading to thermalization.
This low reheating temperature
is certainly safe with regards to the gravitino limit (\ref{grlimit})
discussed above.  

\subsection{Baryogenesis}

The production of a net baryon asymmetry requires baryon number violating
interactions, C and CP violation and a departure 
from thermal equilibrium\cite{sak}.
The first two of these ingredients are contained in GUTs, 
the third can be realized in an expanding universe
 where it is not uncommon that interactions 
come in and out of equilibrium.  

In the original and simplest model of baryogenesis \cite{ww}, a GUT gauge or
Higgs boson decays out of equilibrium producing a net baryon asymmetry. 
While the masses of the gauge bosons is fixed to the rather high GUT scale
$10^{15-16}$ GeV, the masses of the triplets could be lighter $O(10^{10})$
GeV and still remain compatible with proton decay because of the Yukawa
suppression in the proton decay rate when mediated by a heavy Higgs. This
reduced mass allows the simple out-out-equilibrium decay scenario to proceed
after inflation so long as the Higgs is among the inflaton decay products
\cite{dlnos}. From the arguments above, an inflaton mass of $10^{11}$ GeV is
sufficient to realize this mechanism.  Higgs decays in this mechanism 
would be well out of equilibrium as at reheating $T \ll m_H$ and $n_H \sim
n_\gamma$.  In this case, the baryon asymmetry is given simply by
\beq
{n_B \over s} \sim \epsilon {n_H \over {T_R}^3} 
\sim \epsilon {n_\eta \over {T_R}^3}
\sim \epsilon {T_R \over m_\eta} \sim \epsilon 
\left( {m_\eta \over M_P} \right)^{1/2}
\sim \epsilon {\mu \over M_P}\sim 10^{-4} \epsilon
\eeq
where $\epsilon$ is the CP violation in the decay, $T_R$ is the reheat
temperature after inflation,  and I have substituted for $n_\eta =
\rho_\eta/m_\eta 
\sim \Gamma^2{M_P}^2/m_\eta$.

\begin{figure}[hbtp]
	\centering
	\includegraphics[width=10.truecm]{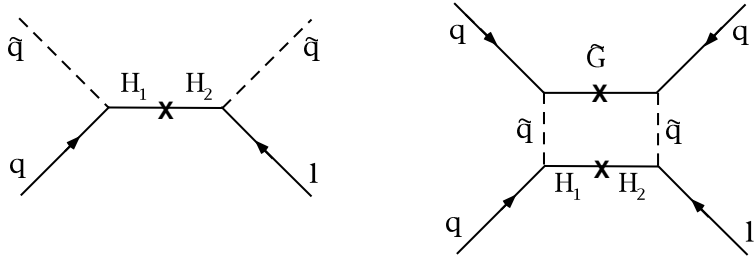}
\caption{Dimension 5 and induced dimension 6 graphs violating baryon number.}
\label{dim56}
\end{figure}

In a supersymmetric grand unified SU(5)
 theory, the superpotential $F_Y$ must be expressed in terms
of SU(5) multiplets
\beq
	F_Y  = h_d {\bf H_2 ~{\bar 5}~10}  +  h_u {\bf H_1~10~10}
\eeq
where $10, {\bar 5}, H_1$ and $H_2$ are chiral
supermultiplets for the 10 and ${\bar 5}$-plets of
SU(5) matter fields and the Higgs
5 and ${\bar 5}$ multiplets respectively.  There are now new dimension 5 
operators \cite{DG,sy} which violate baryon number and lead to proton decay 
as shown in Figure \ref{dim56}.
The first of these diagrams leads to effective dimension 5 Lagrangian terms
such as
\beq
 {\cal L}_{\rm eff}^{(5)} = {h_u h_d \over M_{H}} ( \tilde q
\tilde q q l)
\eeq
and the resulting dimension 6 operator for proton decay \cite{enr}
\beq
{\cal L}_{\rm eff} = {h_u h_d \over M_{H}} \left( {g^2 \over M_{\tilde G}}
 \right) (  q q q l)
\eeq
As a result of diagrams such as these, the proton decay rate scales as $\Gamma
\sim h^4 g^4/M_H^2 M_{\tilde G}^2$ where $M_H$ is the triplet mass, and 
$M_{\tilde G}$ is a typical gaugino mass of order $\la$ 1 TeV.  This rate
however is much too large unless $M_H \ga 10^{16}$ GeV.

It is however possible to have a lighter ($O(10^{10}-10^{11})$ GeV)
Higgs triplet needed for baryogenesis in the out-of-equilibrium decay
scenario with inflation.  One needs  two pairs of Higgs five-plets 
($H_1, H_2$ and  $H_1^\prime, H_2^\prime$ which 
is anyway necessary to have sufficient 
C and CP violation in the decays.
By coupling one pair $(H_2$ and $H_1^\prime)$ 
only to the third generation of fermions
via \cite{nt}
\beq
a {\bf H_1 10 10} + b {\bf H_1^\prime 10_3 10_3} + c {\bf H_2 10_3 {\bar 5}_3}
+ d {\bf H_2^\prime 10 {\bar 5}}
\eeq
proton decay can not be induced by the dimension five operators.

\subsubsection{The Affleck-Dine Mechanism}

Another mechanism for generating the  cosmological baryon asymmetry
is the decay of scalar condensates as first 
proposed by Affleck and Dine\cite{ad}.
This mechanism is truly a product of supersymmetry.
It is straightforward though tedious to show that
  there are many directions in field space such that the scalar potential 
given in eq. (\ref{FD}) vanishes identically
 when SUSY is unbroken. That is, with a particular
assignment of scalar vacuum expectation values, $V=0$ in both the
$F-$ and $D-$ terms.  An example of such a direction
is 
\beq
u_3^c = a \qquad s_2^c = a \qquad
-u_1 = v \qquad \mu^- = v \qquad b_1^c = e^{i\phi} \sqrt{v^2 + a^2}
\label{flat}
\eeq
where $a,v$ are arbitrary complex vacuum expectation values.
 SUSY breaking lifts this degeneracy so that
\begin{equation}
	V  \simeq \tilde{m}^2 \phi^2
\eeq
where $\tilde{m}$ is the SUSY breaking scale and $\phi$ is the direction
 in field space corresponding to the flat direction.
  For large initial values of $\phi$, \ $\phi_o \sim M_{gut}$,
 a large baryon asymmetry can be generated\cite{ad,lin}. This requires
the presence of baryon number violating operators such as $O=qqql$ such that
$\langle O \rangle \neq 0$.  The decay of these
 condensates through such an operator
can lead to a net baryon asymmetry.

In a supersymmetric gut, as we have seen above, there are 
precisely these types of operators. In Figure \ref{adop}, a 4-scalar diagram
involving the fields of the flat direction (\ref{flat}) is shown. Again,
$\tilde G$ is a  (light) gaugino, and $\wt X$ is a superheavy gaugino. The two
supersymmetry breaking insertions are of order 
$\tilde m$, so that the diagram produces an effective quartic coupling
of order ${\tilde m}^2/(\phi_o^2 + M_X^2)$.

\begin{figure}[hbtp]
	\centering
	\includegraphics[width=5.truecm]{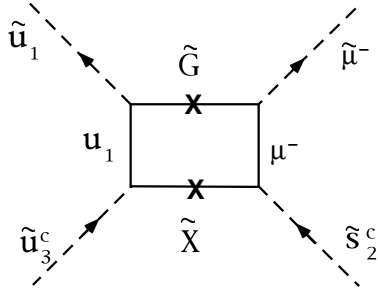}
\caption{Baryon number violating diagram involving flat direction fields.}
\label{adop}
\end{figure}

The baryon asymmetry is computed by tracking the evolution of the
sfermion condensate, which is determined by
\begin{equation}
\ddot{\phi} + 3H\dot{\phi} = - {\tilde m}^2 \phi
\end{equation}
To see how this works, it is instructive to consider 
a toy model with potential
\cite{lin}
\beq
V(\phi,\phi^*)  = \tilde{m}^2 \phi \phi^* + {1 \over 2} i \lambda [\phi^4
-{\phi^*}^4 ]
\label{toy}
\eeq
The equation of motion becomes
\begin{eqnarray}
\ddot{\phi_1} + 3H\dot{\phi_1} = - {\tilde m}^2 \phi_1 + 
3 \lambda \phi_1^2 \phi_2
-\lambda \phi_2^3 \\
\ddot{\phi_2} + 3H\dot{\phi_2} = - {\tilde m}^2 \phi_2 -
3 \lambda \phi_2^2 \phi_1
+ \lambda \phi_1^3
\end{eqnarray}
with $\phi = (\phi_1 + i \phi_2)/\sqrt{2}$.
Initially, when the expansion rate of the Universe, $H$, is large, we 
can neglect $\ddot \phi$ and $\tilde m$. As one can see from (\ref{toy})
the flat direction lies along $\phi \simeq \phi_1 \simeq \phi_o$
with $\phi_2 \simeq 0$. In this case, $\dot{\phi_1} \simeq 0$ and
$\dot{\phi_2} \simeq {\lambda \over 3H} \phi_o^3$.  Since the baryon density
can be written as $n_B = j_o = {1 \over 2} ( \phi_1 \dot{\phi_2} - 
\phi_2 \dot{\phi_1} ) \simeq {\lambda \over 6H} \phi_o^4$, by generating
some motion in the imaginary $\phi$ direction, we have generated a net baryon
density.

When $H$ has fallen to order $\tilde m$ (when $t^{-1} \sim \tilde m$),
$\phi_1$ begins to oscillate about the origin with $\phi_1 \simeq
\phi_o \sin (\tilde {m} t)/\tilde{m}t$. At this
 point the baryon number generated is 
conserved and the baryon density, $n_B$ falls as $R^{-3}$.  Thus,
\beq
n_B \sim {\lambda \over \tilde m} \phi_o^2  \phi^2 \propto R^{-3}
\eeq
and relative to the number density of $\phi$'s 
($n_\phi = \rho_\phi / \tilde m
= \tilde{m} \phi^2$)
\beq
{n_B \over n_\phi} \simeq {\lambda \phi_o^2 \over {\tilde m}^2}
\eeq

If it is assumed that the energy density of the Universe is dominated
by $\phi$, then the oscillations will cease, when
\beq
\Gamma_\phi \simeq {{\tilde m}^3 \over \phi^2} \simeq H
\simeq {\rho_\phi^{1/2} \over M_P} \simeq {{\tilde m} \phi \over M_P}
\eeq
or when the amplitude of oscillations has dropped to $\phi_D \simeq (M_P 
{\tilde m}^2 )^{1/3}$. Note that the decay rate is suppressed as 
fields coupled directly to $\phi$ gain masses $\propto \phi$.
It is now straightforward to compute the baryon to entropy ratio,
\beq
{n_B \over s} = {n_B \over \rho_\phi^{3/4}} \simeq {\lambda \phi_o^2
\phi_D^2 \over {\tilde m}^{5/2} \phi_D^{3/2}} =
{\lambda \phi_o^2 \over {\tilde m}^2} \left({M_P \over \tilde m}\right)^{1/6}
\eeq
and after inserting the quartic coupling
\beq
{n_B \over s} \simeq \epsilon {\phi_o^2 \over (M_X^2 + \phi_o^2)}
\left({M_P \over \tilde m}\right)^{1/6}
\eeq
which could be $O(1)$.

In the context of inflation, a couple of significant changes to the scenario
take place. First, it is more likely that the energy density
is dominated by the inflaton rather than the sfermion condensate.
The sequence of events leading to a baryon
 asymmetry is then as follows \cite{eeno}:
After inflation, oscillations of the inflaton begin at $R=R_\eta$
when $H \sim m_\eta$ and oscillations of the sfermions begin at
$R=R_\phi$ when $H\sim {\tilde m}$. If the Universe is inflaton dominated,
$H \sim m_\eta (R_\eta / R)^{3/2}$ since $H \sim \rho_\eta^{1/2}$ and
$\rho_\eta \sim \eta^2 \sim R^{-3}$ Thus one can relate $R_\eta$ and $R_\phi$,
$R_\phi \simeq (m_\eta / {\tilde m})^{2/3} R_\eta$. As discussed earlier, 
inflatons decay when $\Gamma_\eta = m_\eta^3/M_P^2 = H$ or when
$R=R_{d\eta} \simeq (M_p/m_\eta)^{4/3} R_\eta$.  
The Universe then becomes dominated
by the relativistic decay products of the inflaton,
$\rho_{r\eta} = m_\eta^{2/3} M_P^{10/3} (R_\eta/R)^4$ and 
$H = m_\eta^{1/3} M_P^{2/3} (R_\eta/R)^2$.
Sfermion decays still occur when $\Gamma_\phi =H$ which now
corresponds to a value of the scale factor $R_{d\phi}
=(m_\eta^{7/15} \phi_o^{2/5} M_P^{2/15}/{\tilde m}) R_\eta$.
The final baryon asymmetry in the Affleck-Dine 
 scenario with inflation becomes \cite{eeno}
\begin{equation}
	\frac{n_B}{s} \sim  \frac{\epsilon {\phi_o}^4 {m_\eta}^{3/2}}
{{M_X}^2 {M_P}^{5/2} \tilde{m}} \sim 
 \frac{\epsilon m_\eta^{7/2}}{{M_X}^2 {M_P}^{1/2} \tilde{m}}
  \sim  (10^{-6}-1) \epsilon
\end{equation}
for  
 $\tilde{m} \sim (10^{-17}-10^{-16}) M_P$,
 and $M_X \sim (10^{-4}-10^{-3}) M_P$
and $m_\eta \sim (10^{-8} - 10^{-7} ) M_P$.

When combined with inflation, it is important to verify that the AD
flat directions remain flat. In general, during inflation,
supersymmetry is broken. The gravitino mass is related to the vacuum
energy and $m_{3/2}^2 \sim V/M_P^2 \sim H^2$, thus lifting the flat
directions and potentially preventing the  realization of
the AD scenario as argued in \cite{drt}.
To see this, recall the minimal supergravity model defined in eqs.
(\ref{min}) - (\ref{sep}).  Recall also, the last term in eq. (\ref{cpot}),
which gives a mass to all scalars (in the
matter sector), including flat directions of order the gravitino mass which
during inflation is large. This argument can be generalized to
supergravity models with non-minimal K\"ahler potentials.

However, in no-scale
supergravity models, or more generally in models which possess a
Heisenberg symmetry \cite{heis}, the K\"ahler potential
can be written as (cf. eq. (\ref{ns}))
\begin{equation}
G = f(z + z^* - \phi_i^* \phi^i) + \ln |W(\phi)|^2  \label{kpot}
\end{equation}
Now, one can write 
\begin{equation}
V = e^{f(\eta) }\left[
	\left( \frac{f^{\prime 2}}{f^{\prime\prime}} - 3 \right)
		|W|^2
	- \frac{1}{f'} |W_i|^2 \right] 
\label{hepot}
\end{equation}
It is important to notice that the cross term $|\phi_i^* W|^2$ has 
disappeared in the
scalar potential.  Because of the absence of the cross term, flat
directions remain flat even during inflation \cite{gmo}.  
 The no-scale model corresponds to $f = -3 \ln \eta$,
$f^{\prime 2} = 3 f^{\prime\prime}$ and the first term in (\ref{hepot})
vanishes. The potential then takes the form
\beq
V = \left[\frac{1}{3} e^{{2 \over 3}f} |W_i|^2 
	\right],
\label{nspot}
\eeq
which is positive definite.  The requirement that the vacuum energy vanishes
implies
$\langle W_i\rangle = \langle g_a\rangle = 0 $ at the minimum. As a
consequence $\eta$ is undetermined 
and so is the gravitino mass $m_{3/2}(\eta)$.  

The above argument is only valid at the tree level.
An explicit one-loop calculation~\cite{tr} shows  
that the
effective potential along the flat direction has the form 
\begin{equation}
V_{eff} \sim \frac{g^2}{(4\pi)^2}
\langle V \rangle \left( 
	-2 \phi^2 \log \left(\frac{\Lambda^2}{g^2 \phi^2}\right) 
	+ \phi^2 \right)
	+ {\cal O}(\langle V \rangle)^2 ,
\label{finalV} \end{equation}
where $\Lambda$ is the cutoff of the effective supergravity theory,  
and
has a minimum around $\phi \simeq 0.5 \Lambda$.  
Thus, $\phi_0 \sim M_P$ will be generated and 
in this case the subsequent sfermion oscillations will
dominate the energy density and a baryon asymmetry will result 
which is independent of inflationary parameters as originally  
discussed in
\cite{ad,lin} and will produce $n_B/s \sim O(1)$.
Thus we are left with the problem that the baryon asymmetry in
no-scale type models is too large \cite{eno,gmo,cgmo}. 

In \cite{cgmo}, several possible solutions were presented to dilute
the baryon asymmetry.  These included 1) entropy production from moduli decay,
2) the presence of non-renormalizable interactions, and 3) electroweak
effects. Moduli decay in this context, turns out to be insufficient to bring
an initial asymmetry of order $n_B/s \sim 1$ down to acceptable levels.
However, as a by-product one can show that there is no moduli problem
either. In contrast, adding non-renormalizable Planck scale
operators of the form $\phi^{2n-2}/M_P^{2n-6}$ leads to a smaller
initial value for $\phi_o$ and hence a smaller value for $n_B/s$.
For dimension 6 operators ($n=4$), a baryon asymmetry of order $n_B/s
\sim 10^{-10}$ is produced. Finally, another possible suppression
mechanism is to  employ the smallness of the
fermion masses.  The baryon asymmetry is  known to be wiped out if the
net $B-L$ asymmetry vanishes because of the sphaleron transitions at
high temperature.   However, Kuzmin, Rubakov and
Shaposhnikov \cite{KRS} pointed out that this  erasure can be
partially circumvented if the individual $(B-3L_{i})$  asymmetries,
where $i=1,2,3$ refers to three generations, do not vanish even when
the total asymmetry vanishes.  Even though there is  still a tendency
that the baryon asymmetry is erased by the chemical equilibrium due to
the sphaleron transitions, the finite mass of the tau lepton shifts
the chemical equilibrium between $B$ and 
$L_{3}$ towards the $B$ side and leaves a finite asymmetry in the 
end.  Their estimate is
\begin{equation}
	B = - \frac{4}{13} \sum_{i} \left(L_{i} - \frac{1}{3}B\right)
		\left( 1 + \frac{1}{\pi^{2}} \frac{m_{l_{i}}^{2}}{T^{2}}\right)
\end{equation}
where the temperature $T \sim T_C \sim 200$~GeV is when the sphaleron 
transition freezes out (similar to the temperature of the electroweak phase
transition)  and 
$m_{\tau}(T)$ is expected to be somewhat smaller than $m_{\tau}(0) = 
1.777$~GeV. Overall, the sphaleron transition suppresses the baryon 
asymmetry by a factor of $\sim 10^{-6}$.  This suppression factor is
sufficient to keep the total baryon asymmetry at a reasonable order of
magnitude in many of the cases discussed above.

\section{Dark Matter and Accelerator Constraints}

 There is considerable evidence  
for dark matter in the Universe 
 \cite{dm}.
 The best observational evidence 
is found on the scale of galactic halos and comes from the 
observed flat rotation curves of galaxies. There is also good
evidence for dark matter in elliptical galaxies, as well as clusters
of galaxies coming from X-ray observations of these objects.
In theory, we 
expect dark matter because 1) inflation predicts $\Omega = 1$, and the upper
limit on the baryon (visible) density of the Universe from big bang
nucleosynthesis is $\Omega_B < 0.1$ \cite{osw2}; 2) Even in the absence of
inflation (which does not distinguish between matter and a cosmological
constant), the large scale power spectrum is consistent with a cosmological
matter density of $\Omega \sim 0.3$, still far above the limit from
nucleosynthesis;
 and 3) our current understanding of galaxy 
formation is inconsistent with observations if the Universe is dominated by
baryons.  

	It is also evident that not only must there be dark matter, the bulk of the
dark matter must be non-baryonic.  In addition to the problems with
baryonic dark matter associated with nucleosynthesis or the growth of
density perturbations, it is very difficult to hide baryons.
There are indeed very good constraints on the possible forms for baryonic
dark matter in our galaxy.  Strong cases can be made against hot gas, dust,
jupiter size objects, and stellar remnants such as white dwarfs and neutron
stars \cite{hio}.  

In what follows, I will focus on the region of the parameter space in which
the relic abundance of dark matter contributes a significant though not
excessive amount to the overall energy density. Denoting by $\Omega_\chi$ the
fraction of the critical energy density provided by
the dark matter, the density of interest falls in the range
\beq 
0.1 \le \Omega_\chi h^2 \le 0.3
\label{orange}
\eeq
The lower limit in eq.(\ref{orange}) is motivated by astrophysical
relevance. For lower values of $\Omega_\chi h^2$, there is not enough
dark matter to play a significant role in structure
formation, or constitute a large fraction of the critical density.
The upper bound in (\ref{orange}), on the other hand, is an absolute
constraint,
derivable from the age of the Universe, which can be expressed as
\beq
H_0 t_0 = \int_0^1 dx \left(1 - \Omega - \Omega_\Lambda + \Omega_\Lambda
x^2
 + \Omega /x\right)^{-1/2}
\label{ht}
\eeq
In (\ref{ht}), $\Omega$ is the density of matter relative to 
critical density, while $\Omega_\Lambda$ is the equivalent contribution
due a cosmological constant.
Given a lower bound on the age of the Universe, one can establish an
upper bound on $\Omega h^2$ from eq.(\ref{ht}).
A safe lower bound to the age of the Universe
is $t_0 \ga 12$ Gyr, which translates into the upper bound given in
(\ref{orange}).
Adding a cosmological constant does not relax the upper bound on $\Omega h^2$,
so long as $\Omega + \Omega_\Lambda \le 1$.  If indeed, the indications for a
cosmological constant from recent supernovae observations \cite{snov} turn out
to be correct, the density of dark matter will be constrained to the lower
end of the range in (\ref{orange}).

As these lectures are focused on supersymmetry, I will not dwell on the
detailed evidence for dark matter, nor other potential dark matter candidates.
Instead, I will focus on the role of supersymmetry and possible
supersymmetric candidates. As was discussed at the end of section 3, one way
to insure the absence of unwanted, $B$ and $L$-violating superpotential
terms, is to impose the conservation of $R$-parity. In doing so, we have the
prediction that the lightest supersymmetric particle (LSP) will be stable. 
It is worth noting that $R$-parity conservation is consistent with certain
mechanisms for generating neutrino masses in supersymmetric models. For
example, by adding a new chiral multiplet $\nu^c$, along with 
superpotential terms of the form, $H_2 L \nu^c + M \nu^c \nu^c$, although
lepton number is violated (by two units), $R$-parity is conserved. In this
way a standard see-saw mechanism for neutrino masses can be recovered. 

The stability of the LSP almost certainly renders it a neutral weakly
interacting particle \cite{ehnos}. Strong and electromagnetically interacting LSPs
would become bound with normal matter forming anomalously heavy isotopes.
Indeed, there are very strong upper limits on the abundances, relative of
hydrogen, of nuclear isotopes~\cite{isotopes},
$n/n_H \la 10^{-15}~~{\rm to}~~10^{-29}
$
for 1 GeV $\la m \la$ 1 TeV. A strongly interacting stable relics is expected
to have an abundance $n/n_H \la 10^{-10}$
with a higher abundance for charged particles.

There are relatively few supersymmetric candidates which are not colored and
are electrically neutral.  The sneutrino \cite{snu} is one possibility,
but in the MSSM, it has been excluded as a dark matter candidate by direct
\cite{dir} searches, indirect \cite{indir} and accelerator\cite{accel}
searches.  In fact, one can set an accelerator based limit on the sneutrino
mass from neutrino counting, 
$m_{\tilde\nu}\ga$ 43 GeV~\cite{EFOS}. In this case, the direct relic
searches in
underground low-background experiments require  
$m_{\tilde\nu}\ga$ 1 TeV~\cite{uground}. Another possibility is the
gravitino which is probably the most difficult to exclude. 
I will concentrate on the remaining possibility in the MSSM, namely the
neutralinos.

The neutralino mass matrix was discussed in section 3.3 along with some
particular neutralino states. In general, neutralinos can  be expressed as a
linear combination
\begin{equation}
	\chi = \alpha \tilde B + \beta \tilde W^3 + \gamma \tilde H_1 +
\delta
\tilde H_2
\end{equation}
and the coefficients $\alpha, \beta, \gamma,$ and $\delta$ depend only on
$M_2$, $\mu$, and $\tan \beta$ (assuming gaugino mass unification at the GUT
scale so that $	M_1 = {5 \over 3}  {\alpha_1 \over \alpha_2}  M_2	$).

 There are some limiting cases in which the LSP 
is nearly a pure state \cite{ehnos}.  When $\mu  \rightarrow 0$, ${\tilde
S}^0$ 
 is the LSP with
\beq
	m_{\tilde S}  \rightarrow  {2 v_1 v_2 \over v^2} \mu = \mu \sin 2\beta
\eeq
When $M_2 \rightarrow 0$, the photino is the LSP with \cite{gold}
\beq
	m_{\tilde \gamma} \rightarrow {8 \over 3}
{ {g_1}^2 \over  ( {g_1}^2 + {g_2}^2) } M_2
\eeq
When $M_2$  is large and $M_2 \ll \mu$  then the bino ${\tilde B}$
 is the LSP \cite{osi3}    and
\beq
	m_{\tilde B}  \simeq M_1 	
\eeq
and finally when $\mu$ is large and $\mu  \ll M_2$
the Higgsino states 
$	{\tilde H}_{(12)}$  with mass $m_{{\tilde H}_{(12)}} = - \mu$ for $\mu <
0$,  or ${\tilde H}_{[12]}$ 
with mass $m_{{\tilde H}_{[12]}} =  \mu$ for $\mu > 0$ are the LSPs depending
on the sign of $\mu $ \cite{osi3}.

In Figure \ref{osi399} \cite{osi3}, regions in
the $M_2, \mu$  plane with $\tan\beta = 2$ are shown in which the LSP
is one of several nearly pure states, the photino, $\tilde \gamma$, the
bino,
$\tilde B$, a symmetric combination of the Higgsinos, 
$\tilde{H}_{(12)}$, or the Higgsino, 
$\tilde{S}$. The dashed lines show the LSP mass contours.
 The cross hatched regions correspond to parameters giving
  a chargino ($\tilde W^{\pm}, \tilde H^{\pm}$) state 
with mass $m_{\tilde \chi} \leq 45 GeV$ and as such are 
excluded by LEP\cite{lep2}.
This constraint has been extended by LEP1.5 \cite{lep15}, 
and LEP2 \cite{LEP2} and is shown by the 
light shaded region and corresponds to regions where the chargino mass is $\la
95$ GeV. The newer limit does not extend deep into the Higgsino region
because of the degeneracy between the chargino and neutralino.
 Notice that the parameter space is dominated by the  
$\tilde B$ or $\tilde H_{12}$
 pure states and that the photino (often quoted as the LSP in the past
\cite{gold,phot})
 only occupies a small fraction of the parameter space,
 as does the Higgsino combination $\tilde S^0$. Both of these light states are
now experimentally excluded.

\begin{figure}[hbtp]
	\centering
	\includegraphics[width=9.truecm]{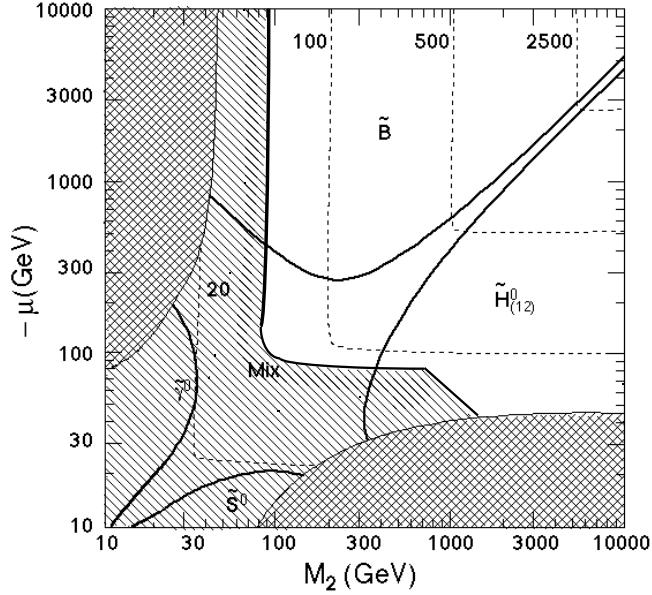}
	\caption{Mass contours and composition of nearly pure LSP states in the MSSM
\cite{osi3}.}
	\label{osi399}
\end{figure}

The relic abundance of LSP's is 
determined by solving
the Boltzmann
 equation for the LSP number density in an expanding Universe.
 The technique\cite{wso} used is similar to that for computing
 the relic abundance of massive neutrinos\cite{lw}.
The relic density depends on additional parameters in the MSSM beyond $M_2,
\mu$, and $\tan \beta$. These include the sfermion masses, $m_{\tilde f}$, the
Higgs pseudo-scalar mass, $m_A$, and the tri-linear masses $A$ as well as two
phases $\theta_\mu$ and $\theta_A$.
To determine, the relic density it is necessary to obtain the general
annihilation cross-section for neutralinos.  This has been done in
\cite{mcdos,dn,jkg,bb}. In much of the parameter space of interest, the LSP is
a bino and the annihilation proceeds mainly through sfermion exchange as shown in
Figure \ref{ann}.
 For binos, as was the case for photinos \cite{gold,phot}, it is possible
 to adjust the sfermion masses to obtain closure density in a wide mass range.
Adjusting the sfermion mixing parameters \cite{fkmos} or CP violating phases
\cite{fkos,fko} allows even greater freedom.

\begin{figure}[hbtp]
	\centering
	\includegraphics[width=8.truecm]{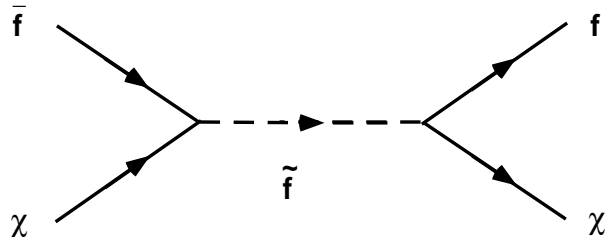}
	\caption{Typical annihilation diagram for neutralinos through sfermion
exchange.}
	\label{ann}
\end{figure}

Because of the p-wave suppression associated with Majorana fermions, the s-wave
part of the annihilation cross-section is suppressed by the outgoing fermion masses. 
This means that it is necessary to expand the cross-section to include p-wave
corrections which can be expressed as a term proportional to the temperature if
neutralinos are in equilibrium. Unless the ${\tilde B}$ mass happens to lie near
$m_Z/2$ or $m_h/2$, in which case there are large contributions to the
annihilation through direct $s$-channel resonance exchange, the dominant
contribution to
the $\tilde{B} \tilde{B}$ annihilation cross section comes from crossed
$t$-channel sfermion exchange. 
In the absence of such a resonance, the thermally-averaged cross section
for $\tilde{B} \tilde{B} \to f \bar{f}$ takes the generic form
\ba
\langle \sigma v \rangle & = & (1 - {m_f^2 \over m_{\t B}^2})^{1/2} {g_1^4 \over
128 \pi} \left[ (Y_L^2 + Y_R^2)^2 ({m_f^2 \over \Delta_f^2}) \right. \nonumber \\
& & \qquad +\, \left. (Y_L^4
+ Y_R^4)  ({4 m_{\t B}^2 \over \Delta_f^2}) (1 + ...) \,x \,\right] \nonumber
\\
& \equiv & a + b x
\label{eqn:sigv}
\ea
where $Y_{L(R)}$ is the hypercharge of $f_{L(R)}$, $\Delta_f \equiv
m_{\t f}^2 + m_{\t B}^2 - m_f^2$, and we have shown only the leading $P$-wave
contribution proportional to $x \equiv T/m_{\t B}$.
Annihilations in the early Universe continue until the annihilation rate $\Gamma
\simeq \sigma v n_\chi$ drops below the expansion rate, $H$.  For particles which
annihilate through approximate weak scale interactions, this occurs when $T \sim
m_\chi /20$. Subsequently, the relic density of neutralinos is fixed relative to
the number of relativistic particles. 

As noted above, the number density of neutralinos is tracked by a
Boltzmann-like equation,
\beq
{dn \over dt} = -3{{\dot R} \over R} n - \langle \sigma v \rangle (n^2 -
n_0^2)
\eeq
where $n_0$ is the equilibrium number density of neutralinos.
By defining the quantity $f = n/T^3$, we can rewrite this equation in terms of $x$,
as
\beq
{df \over dx} = m_\chi \left( {1 \over 90} \pi^2 \kappa^2 N \right)^{1/2}
(f^2 - f_0^2)
\eeq
The solution to this equation at late times (small $x$) yields a constant value of
$f$, so that $n \propto T^3$. 
The final relic density expressed as a fraction of the critical energy density 
can be written as \cite{ehnos}
\beq
\Omega_\chi h^2 \simeq 1.9 \times 10^{-11} \left({T_\chi \over
T_\gamma}\right)^3 N_f^{1/2} \left({{\rm GeV} \over ax_f + {1\over 2} b
x_f^2}\right)
\label{relic}
\eeq 
where $(T_\chi/T_\gamma)^3$ accounts for the subsequent reheating of the
photon temperature with respect to $\chi$, due to the annihilations of
particles with mass $m < x_f m_\chi$ \cite{oss}. The subscript $f$ refers to
values at freeze-out, i.e., when annihilations cease. 

 In Figure \ref{figbga} \cite{efgos}, regions in the
$M_2-\mu$
 plane (rotated with respect to Figure \ref{osi399}) with $\tan\beta = 2$,
and with a relic abundance $0.1 \le \Omega h^2 \le 0.3$ are shaded. In Figure
\ref{figbga}, the sfermion masses have been fixed such that $m_0 = 100$ GeV
(the dashed curves border the region when $m_0 = 1000$ GeV). 
 Clearly the MSSM offers sufficient room to solve the dark matter problem.
In the higgsino sector ${\tilde H}_{12}$, additional types of
annihilation processes known as co-annihilations
\cite{gs,dnr,co2}
 between ${\tilde H}_{(12)}$ and the next lightest 
neutralino (${\tilde H}_{[12]}$)
must be included. These tend to significantly lower 
the relic abundance in much
of this sector and as one can see there is little room left for Higgsino dark
matter \cite{efgos}.

As should be clear from Figures \ref{osi399} and \ref{figbga}, binos are a good and
likely choice for dark matter in the MSSM.  For fixed $m_{\tilde f}$,
$\Omega h^2 \ga 0.1$, for all $m_{\tilde B} = 20 - 250$ GeV
largely independent of $\tan \beta$ and the sign of $\mu$.
In addition, the requirement that $m_{\tilde f} > m_{\tilde B}$
translates into an upper bound of about 250 GeV on the bino mass
\cite{osi3,gkt}.  By further adjusting the trilinear $A$ and accounting for
sfermion mixing this upper bound can be relaxed \cite{fkmos} and by allowing for
non-zero phases in the MSSM, the upper limit can be extended to about 600 GeV
\cite{fkos}.
 For fixed $\Omega h^2 = 1/4$, we would require
sfermion masses of order 120 -- 250 GeV for binos with masses in the range
20 -- 250 GeV.  The Higgsino relic density, on the other hand, is largely 
independent of $m_{\tilde f}$.  For large $\mu$, annihilations into $W$
and $Z$ pairs dominate, while for lower $\mu$, it is the annihilations
via Higgs scalars which dominate.  Aside from a narrow region with 
$m_{\tilde H_{12}} < m_W$ and very massive Higgsinos with 
$m_{\tilde H_{12}} \ga 500$ GeV, the relic density of ${\tilde H_{12}}$
is very low. Above about 1 TeV, these Higgsinos are also excluded. 

\begin{figure}[ht]
	\centering
\vskip -1in
	\includegraphics[width=10.truecm]{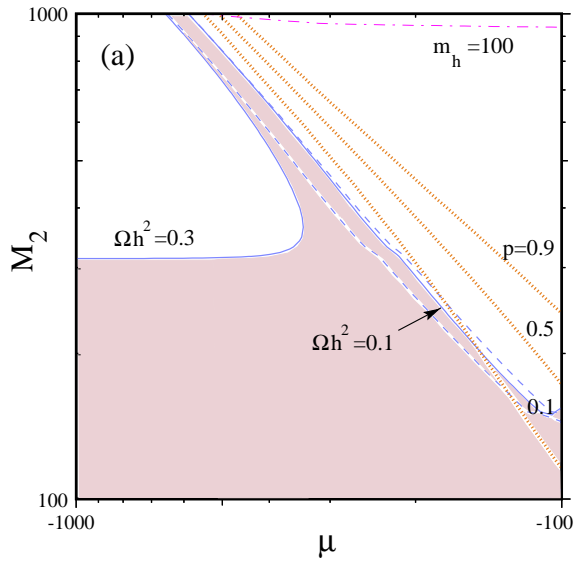}
\vskip -.6in
	\caption{Regions in the $M_2$--$\mu$ plane where $0.1 \le \Omega h^2\le
0.3$ \cite{efgos}. Also shown are the Higgsino purity contours (labeled
0.1, 0.5, and 0.9). As one can see, the shaded region is mostly gaugino (low
Higgsino purity).
 Masses are in GeV.}
	\label{figbga}
\end{figure}

As discussed in section 4, one can make a further reduction in the number of
parameters by setting all of the soft scalar masses equal at the GUT scale, thereby
considering the CMSSM.
For a given value of $\tan \beta$, the parameter space is best described
in terms of the common gaugino masses $m_{1/2}$ and scalar masses $m_0$
at the GUT scale.  In Figure \ref{rd2c} \cite{fko},
this parameter space is shown for $\tan
\beta = 2$.  The light shaded region corresponds to the portion of parameter
space where the relic density $\Omega_\chi h^2$ is between 0.1 and 0.3. The darker
shaded region corresponds to the parameters where the LSP is not a neutralino
but rather a ${\t \tau}_R$. In the $m_0-m_{1/2}$ plane, the upper limit to
$m_0$ is due to the upper limit $\Omega_\chi h^2 < 0.3$.  For larger $m_0$, 
the large sfermion masses suppress the annihilation cross-section which is
then too small to bring the relic density down to acceptable levels. In this
region, the LSP is mostly $\wt B$, and the value of
$\mu$ can not be adjusted to make the LSP a Higgsino which would allow an enhanced
annihilation particularly at large $m_0$. The cosmologically interesting region at
the left of the figure is rather jagged, due to the appearance of pole effects. 
There, the LSP can annihilate through s-channel $Z$ and $h$ (the light Higgs)
exchange, thereby allowing a very large value of $m_0$. Because the
$\Omega_\chi h^2 = 0.3$ contour runs into the ${\t \tau}_R$-LSP region at a
given value of
$m_{1/2}$, it was thought \cite{kane} that this point corresponded to an upper
limit to the LSP mass (since $m_{\t B}$ is approximately 0.4 $m_{1/2}$. As we
will see, this limit has been extended due to co-annihilations of the ${\wt
B}$ and ${\t \tau}_R$
\cite{efo}.

\begin{figure}[ht]
	\centering
\vskip -1in
	\includegraphics[width=10.truecm]{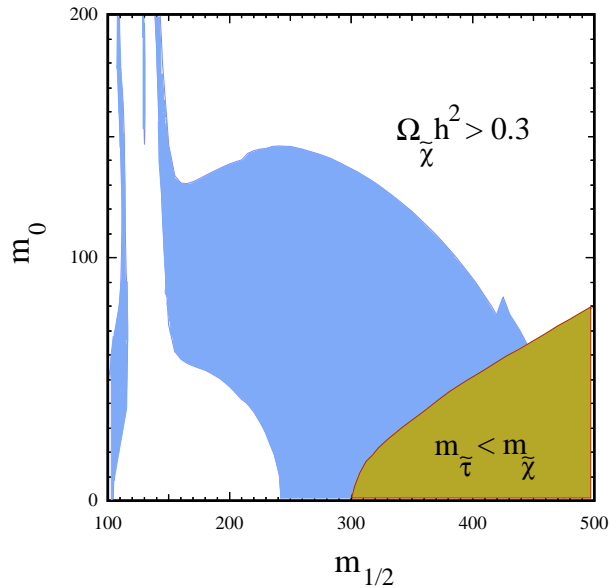}
\vskip -.6in
	\caption{Region in the $m_{1/2}$--$m_0$ plane where $0.1 \le \Omega h^2 \le
0.3$
\cite{fko}. Masses are in GeV.}
	\label{rd2c}
\end{figure}

The $m_{1/2} - m_0$ parameter space is further constrained by the recent 
runs at LEP. The negative results
of LEP 1 searches for $Z^0 \rightarrow \chi^+ \chi^-$ and $Z^0 \rightarrow \chi
\chi'$ (where $\chi'$ denotes a generic heavier neutralino) already
established important limits on supersymmetric model parameters,
but left open the possibility that the lightest neutralino might
be massless~\cite{lep2}.
Subsequently, the data from higher-energy LEP runs, based on chargino and neutralino
pair production, complemented LEP~1 data and
excluded the possibility of a massless neutralino, at least if the input
gaugino masses $M_{\alpha}$ were assumed to be universal~\cite{achi,EFOS}.

 In Figure \ref{cplot020_long} \cite{efos2}, the constraints imposed by the
LEP chargino and neutralino and slepton
searches~\cite{achi,LEP2} (hatched region) at LEP 2
are shown. The hatched regions correspond to the limits in the MSSM. The thick
curve to the right of this region, corresponds to the slightly more
restrictive bound due to the assumption of universal Higgs masses at the GUT
scale (CMSSM). The distinction between these two cases is more apparent at
other values of $\tan
\beta$ \cite{EFOS,efos2}. The bounds shown here correspond to the run at LEP at 172
GeV center of mass energy. The D0 limit on the  gluino mass~\cite{D0} is also shown
(dotted line). Of particular importance are the bounds 
on supersymmetric Higgs production, for which we consider
both the $e^+e^- \rightarrow h Z$ and $e^+e^- \rightarrow h A$
reactions. The regions bounded by the lack of Higgs events are labeled nUHM
corresponding to the MSSM (non-Universal Higgs Mass) and UHM corresponding to the
UHM. Once again, the limits plotted in this figure (\ref{cplot020_long})
correspond to the 172 GeV run and have been greatly improved since. In
this case the improvement in the bound when restricting the value of $\mu$
to take its CMSSM value is clear.  At lower $\tan \beta$, the constraint
curve moves quickly to the left, i.e., to higher values of $m_{1/2}$
\cite{efos2}. 

\begin{figure}[hbtp]
	\centering
\vskip -.5in
	\includegraphics[width=10.truecm]{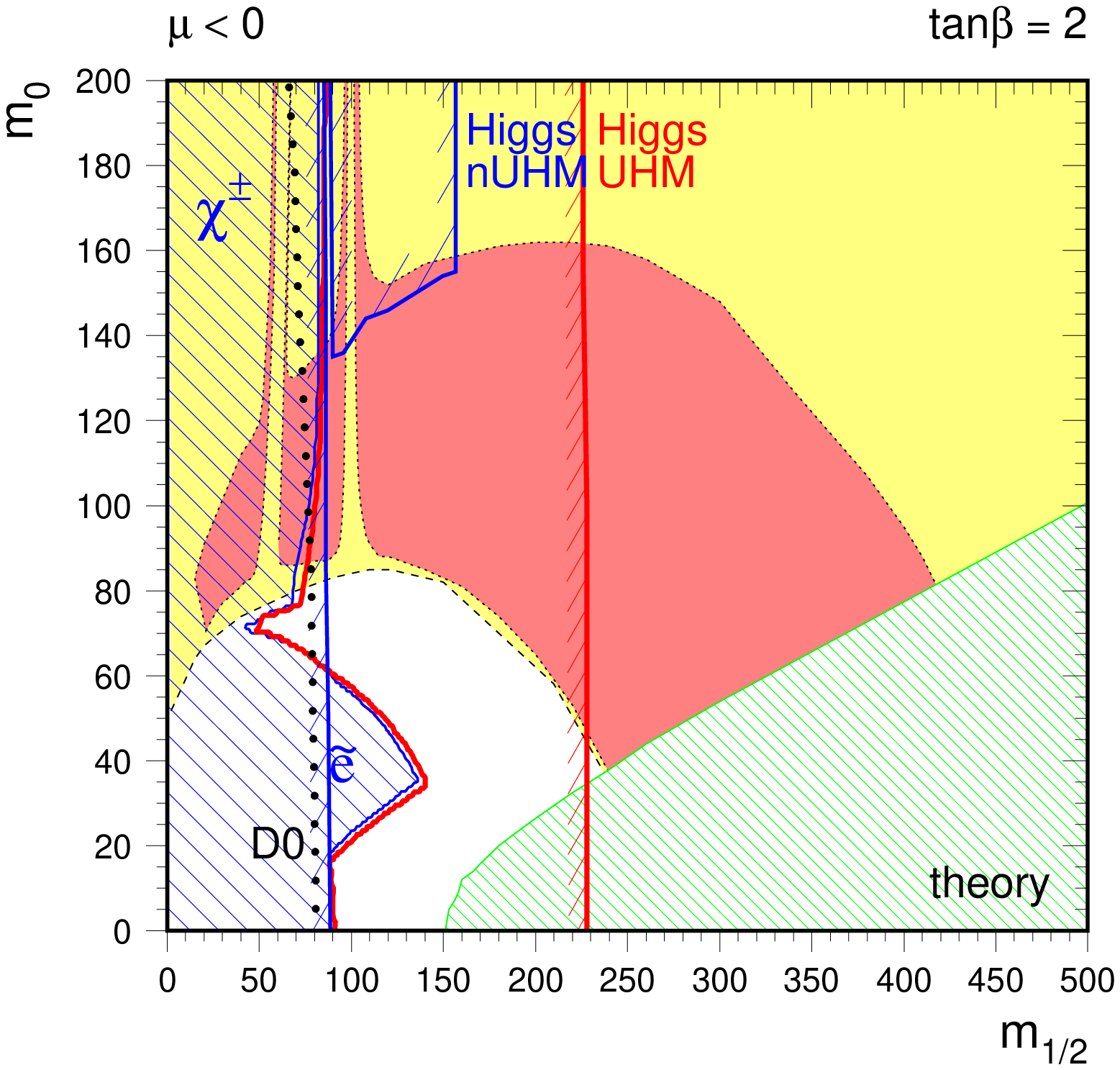}
\vskip -1in
	\caption{The domains of the $m_{1/2}$--$m_0$
  plane (masses in GeV)
  that are excluded by the LEP~2 chargino and selectron searches, 
  both without (hatched) and with the assumption of Higgs
  scalar-mass universality for $\mu < 0$ and $\tan\beta = 2$ \cite{efos2}.
Also displayed are 
  the domains that are excluded by Higgs
  searches (solid lines) with and without
  the assumption of universal scalar masses for Higgs bosons (UHM).
 The region marked theory  is
  excluded cosmologically because $m_{{\tilde \tau}_R} < m_{\chi}$.
  The domains that have relic neutralino densities in the favored
  range with (dark) and without (light shaded) the
  scalar-mass universality assumption.  
}
	\label{cplot020_long}
\end{figure}

 The region of the $m_{1/2} - m_0$ plane in which $0.1 < \Omega_{\chi} h^2<
0.3$ for some experimentally allowed value of $\mu<0$ is light-shaded in
Figure
\ref{cplot020_long}, and the region of
  the plane in which $0.1 < \Omega_{\chi} h^2< 0.3$ for $\mu$ determined by
  the CMSSM constraint on the scalar masses is shown dark-shaded.
The MSSM region extends to large values of $m_0$ since $\mu$ can be adjusted
to take low values so the the LSP is Higgsino like and the relic density
becomes  insensitive to the sfermion masses. 

As one can see from Figure \ref{cplot020_long}, the combined bounds from the
chargino and slepton searches, provide a lower bound to $m_{1/2}$ which can be
translated into a bound on the neutralino mass.  In the region of interest, we have
the approximate relation that $m_\chi \approx 0.4 m_{1/2}$.

\begin{figure}[ht]
 \hspace*{.7in}
\begin{minipage}{8in}
	\includegraphics[width=5.truecm]{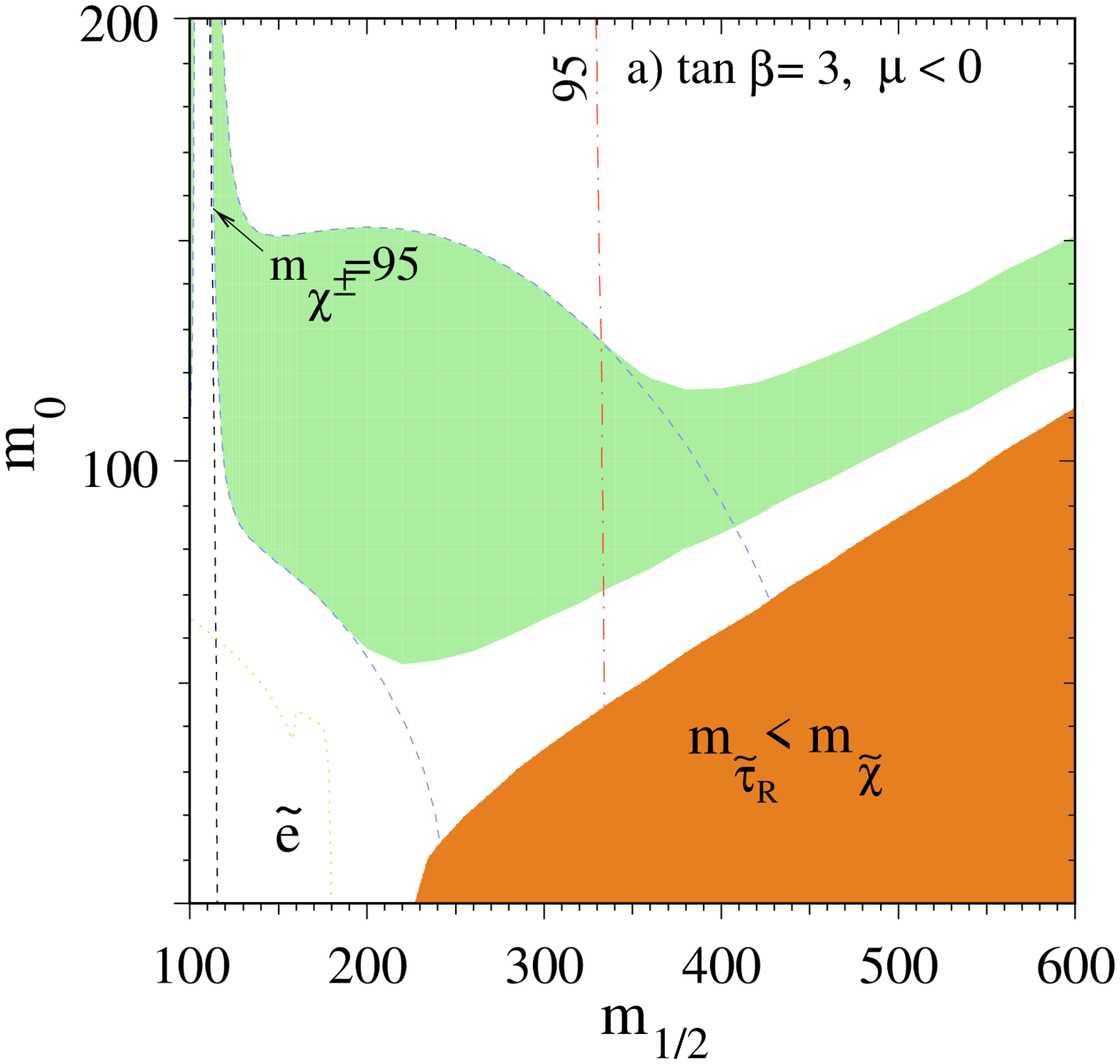} 
\hspace*{.2in}
\includegraphics[width=5.truecm]{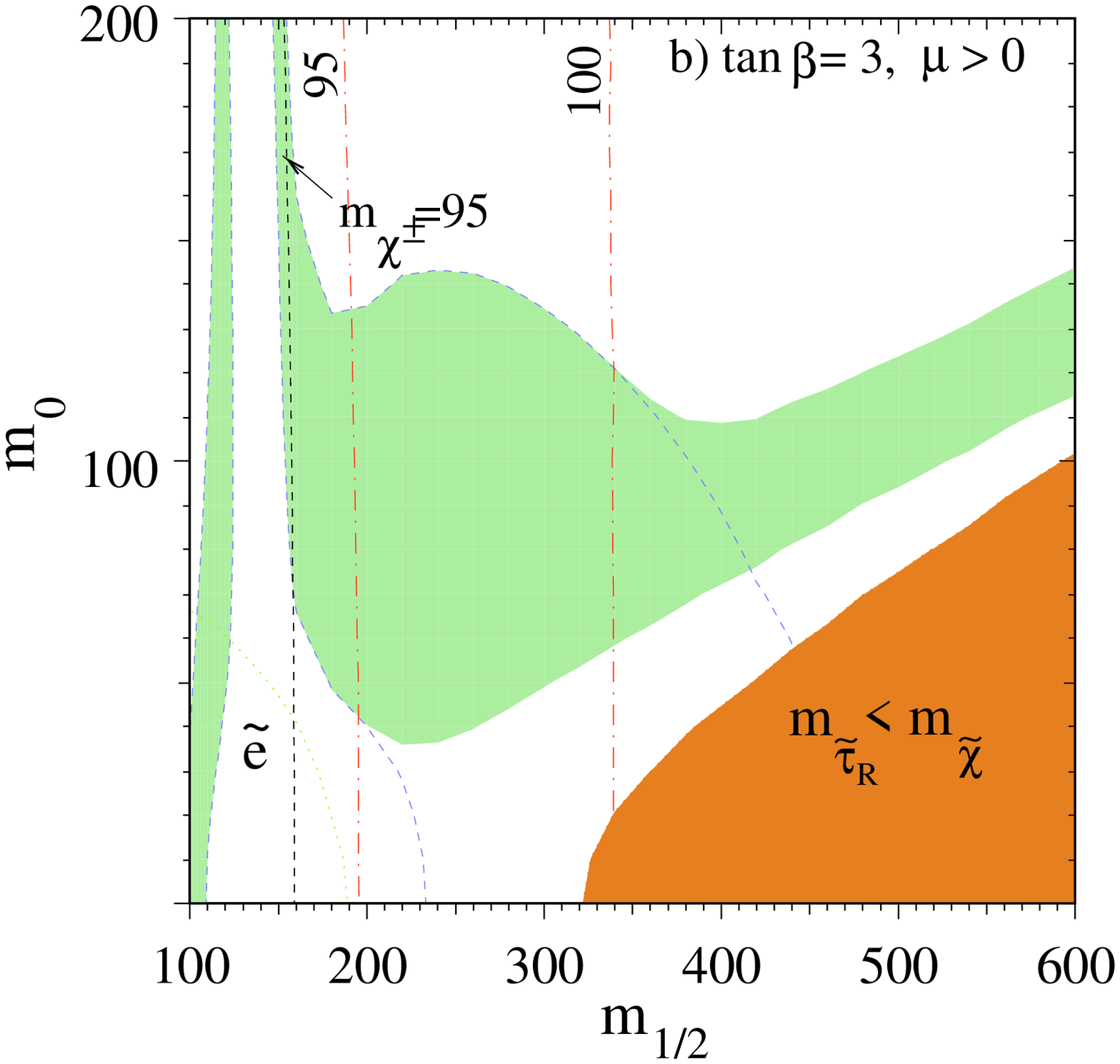} 
\end{minipage}
\begin{minipage}{8in}
\vspace*{0.4in}
 \hspace*{.7in}
	\includegraphics[width=5.truecm]{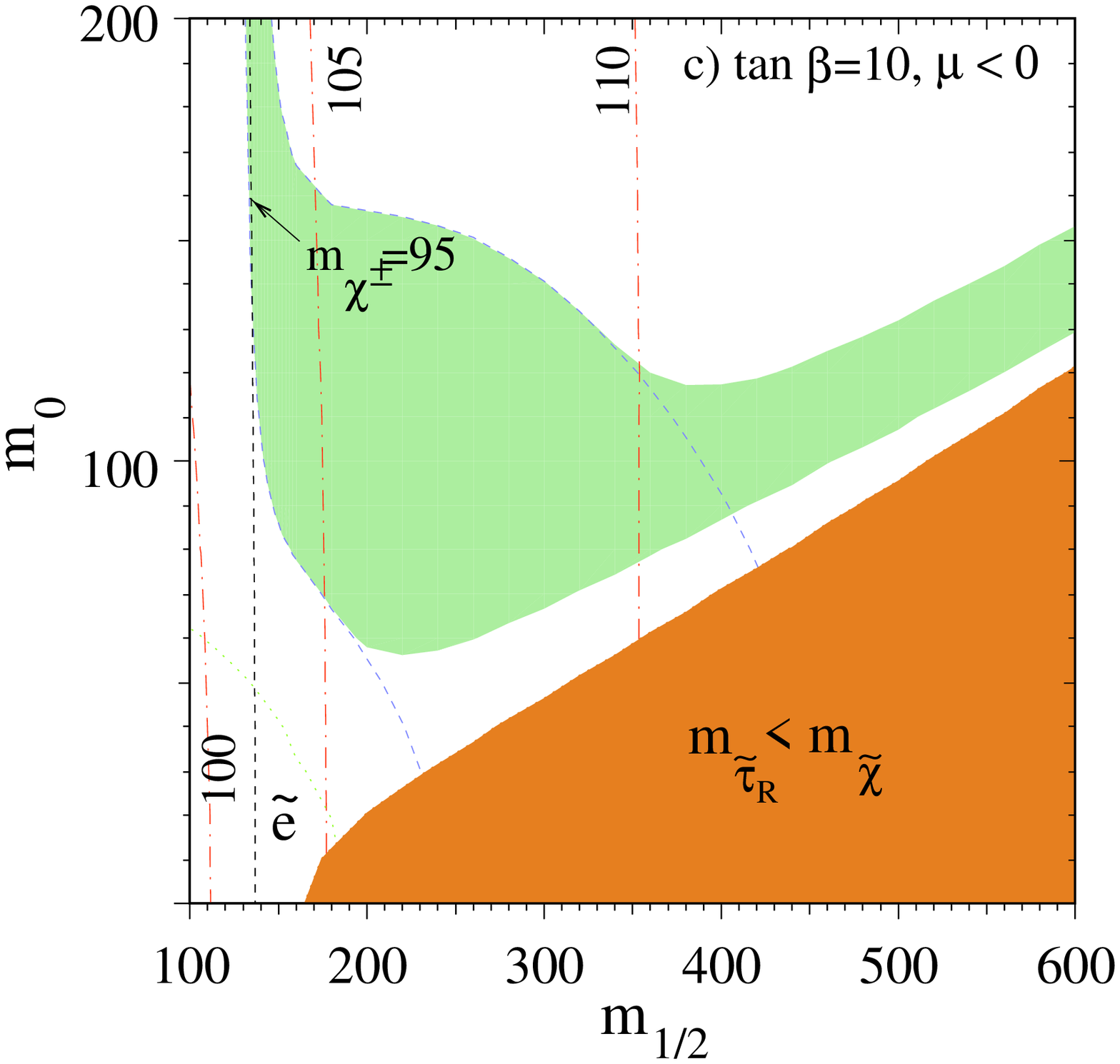} 
\hspace*{.2in}
\includegraphics[width=5.truecm]{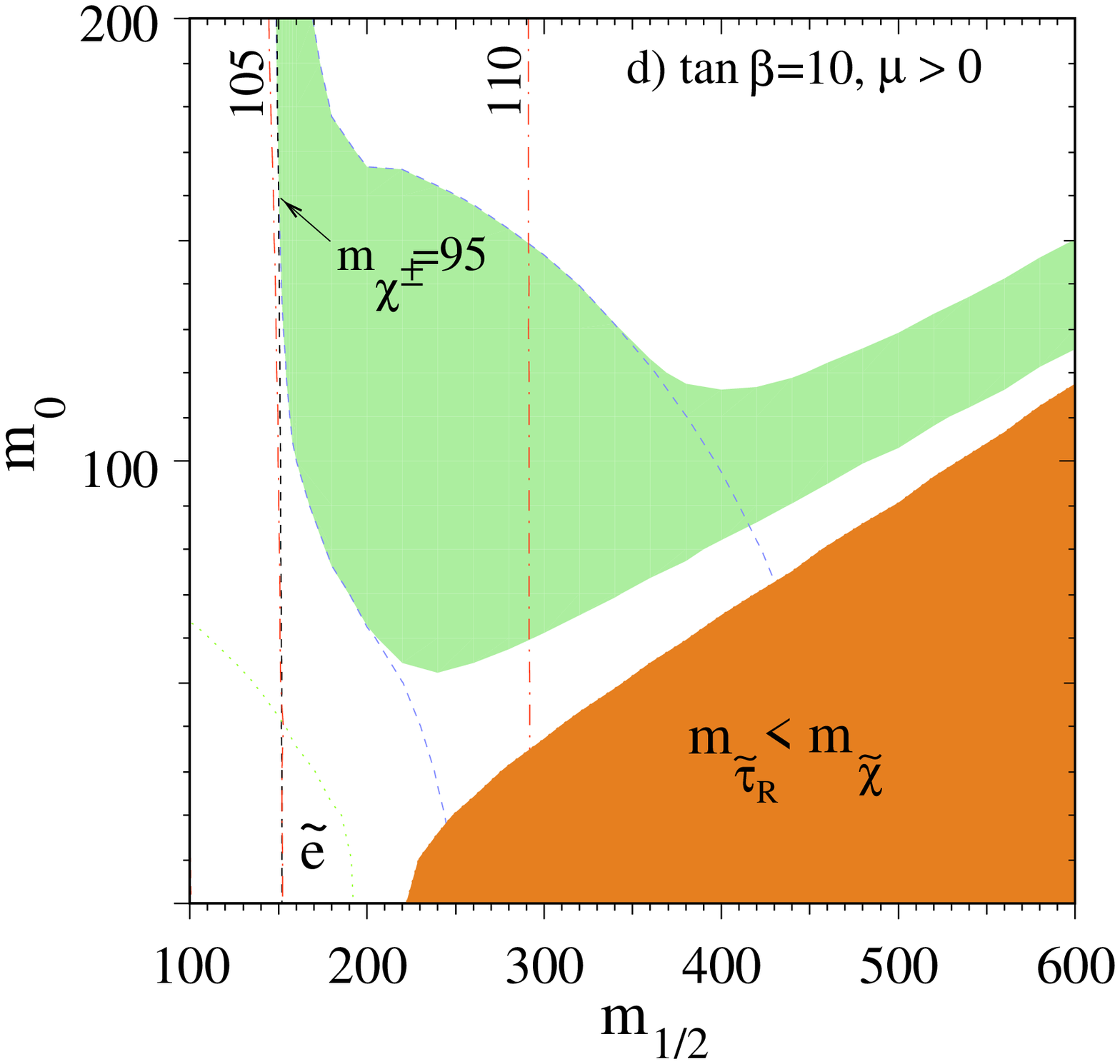} 
\end{minipage}
\caption{The light-shaded area is the cosmologically preferred 
region with \protect\mbox{$0.1\leq \Omega h^2 \leq 0.3$}.   The light dashed lines
show the location  of the cosmologically preferred region  if one
ignores  coannihilations with the light sleptons.  
In the dark shaded regions in the bottom right of each panel, the LSP is
the ${\tilde
\tau}_R$, leading to an unacceptable abundance
of charged dark matter.  Also shown are the isomass
contours $m_{\chi^\pm} = 95$~GeV and $m_h = 95,100,105,110$~GeV,
as well as an indication of the slepton bound from
LEP~\protect\cite{lep189}. These figures are adapted from those in
\protect\cite{efo}.}
\label{efoblock}
\end{figure}

Ultimately, the Higgs mass bound alone does not provide an independent bound on
$m_{1/2}$.  The reason is that the Higgs constraint
curves bend to the left at large $m_0$, where large sfermion masses lead to
greater positive radiative corrections to the Higgs mass, and the 
Higgs curve strikes the chargino bound
at sufficiently large $m_0$.  However, when combined with cosmological limit on the
relic density, a stronger constraint can be found. Recall that at large
$m_0$, the cosmological bound is satisfied by lowering $\mu$. At certain
values of $\tan
\beta$ and $m_{1/2}$, one can not lower $\mu$ and remain consistent with the relic
density limit and the Higgs mass limit.  This is seen in Figure
\ref{cplot020_long} by the short horizontal extension of the nUHM Higgs
curve.  At lower values of
$\tan
\beta$ this extension is lengthened.  In the UHM case, once again the UHM
Higgs curve bends to the right at large $m_0$. However, cosmology excludes
the large $m_0$ region in the CMSSM (CMSSM and UHM are treated synonymously
here).  In this case the mass bound on
$m_{1/2}$ or $m_\chi$ is much stronger.

The results from the 172 GeV run at LEP provided a bound on the chargino mass
of $m_{\chi^\pm} > 86$ GeV and corresponding bound on the neutralino mass of
$m_\chi \ga 40$ GeV \cite{efos2}. As noted above, the UHM Higgs mass bound becomes
very strong at low $\tan \beta$. In fact, for $\tan \beta < 1.7$, the UHM Higgs
curve moves so far to the left so as to exclude the entire dark shaded region
required by cosmology. Subsequent to the 183 GeV run at LEP \cite{latest}, the
chargino mass bound was pushed to $m_{\chi^\pm} > 91$ GeV, and a Higgs mass bound
was established to be $m_h \ga 86 $ GeV for $\tan \beta \la 3$, and $m_h \ga 76 $
GeV for $\tan \beta \ga 3$. These limits were further improved by the 189 GeV run,
so that at the kinematic limit the chargino mass bound is $m_{\chi^\pm} \ga 95$ GeV,
implying that $m_{1/2} \ga 110$ GeV and the neutralino mass limit becomes $m_\chi
\ga 50$ Gev. The Higgs mass bound is now (as of the 189 run) $m_h \ga 95 $ GeV for
low $\tan \beta$ \cite{lep189}.

\begin{figure}[h]
 \hspace*{.7in}
\begin{minipage}{8in}
	\includegraphics[width=5.truecm]{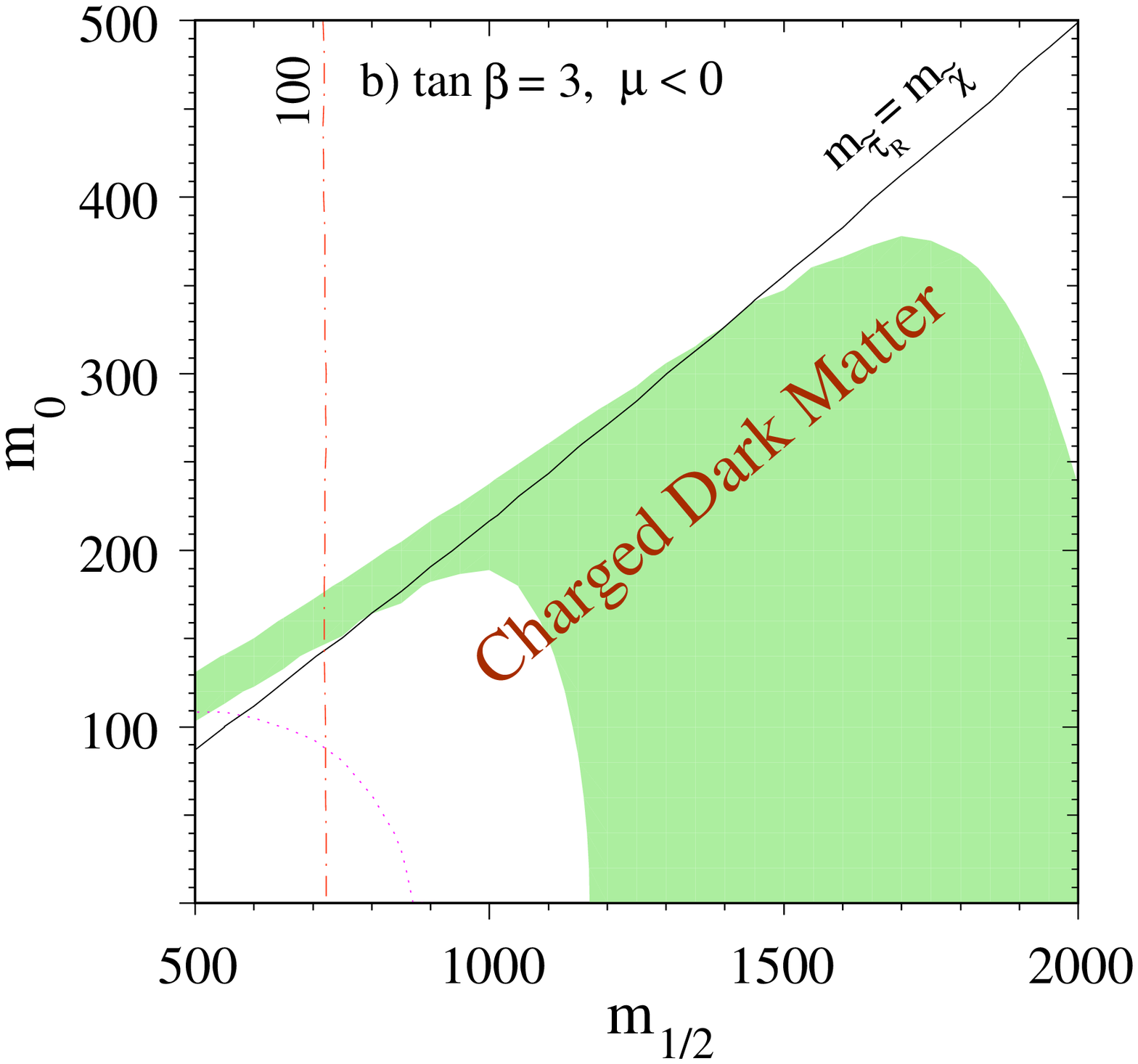} 
\hspace*{.2in}
\includegraphics[width=5.truecm]{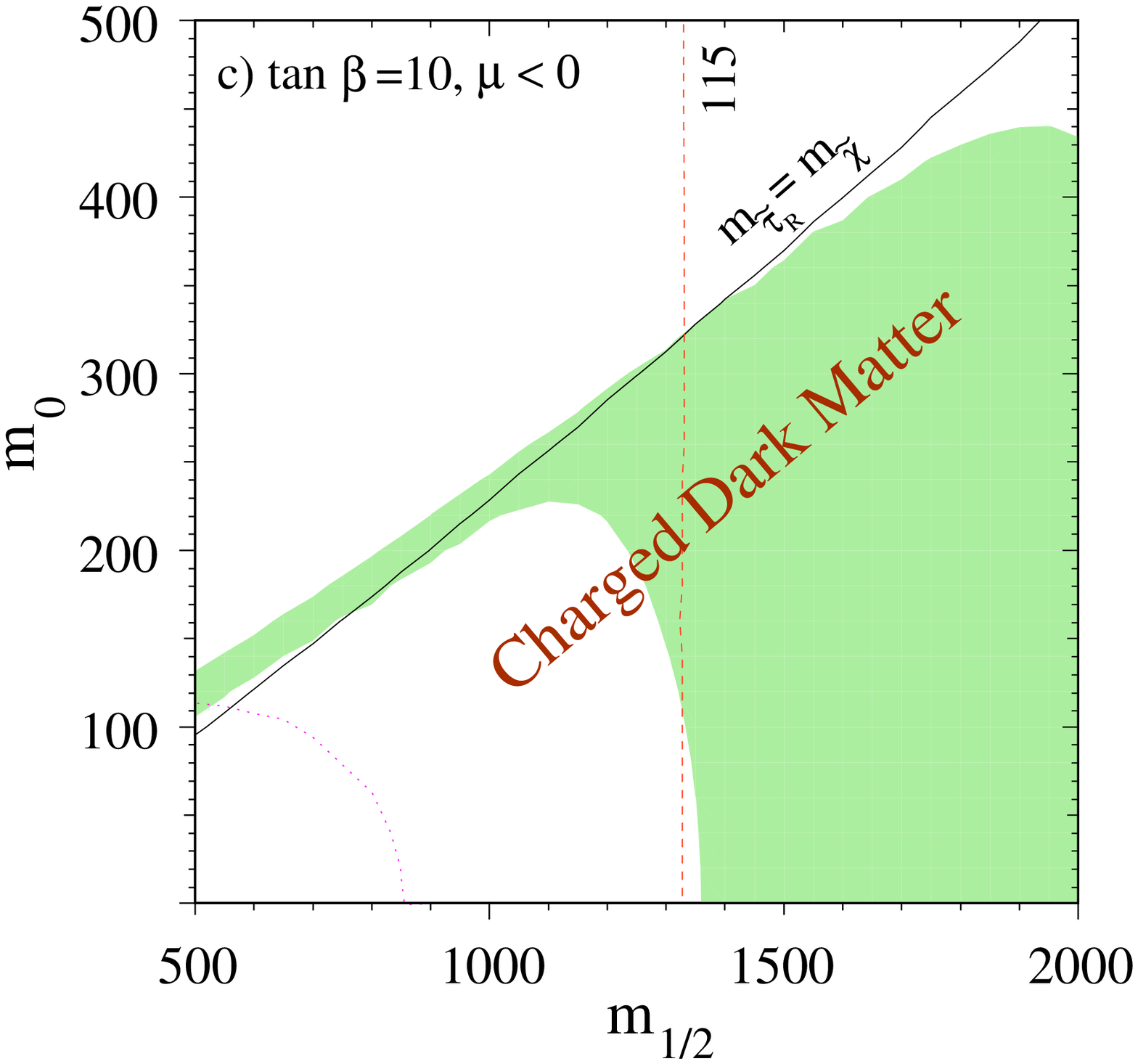} 
\end{minipage}
\caption{The
same as Fig.~\protect\ref{efoblock}, for $\mu<0$, 
but extended to larger $m_{1/2}$. These figures are adapted from those in
\protect\cite{efo}.}
\label{efoblock2}
\end{figure}

\begin{figure}
 \hspace*{.7in}
\begin{minipage}{8in}
	\includegraphics[width=5.truecm]{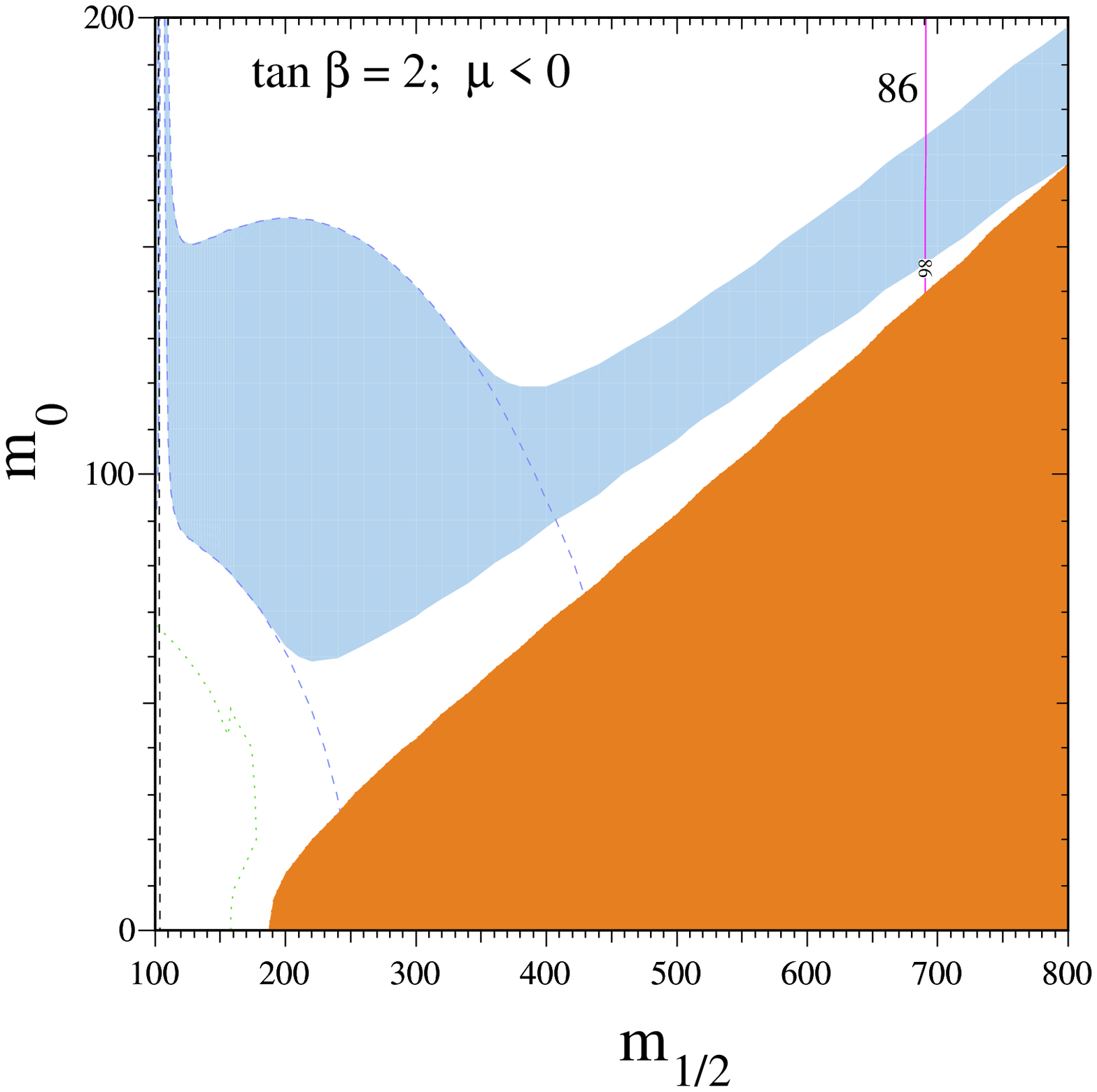} 
\hspace*{.2in}
\includegraphics[width=5.truecm]{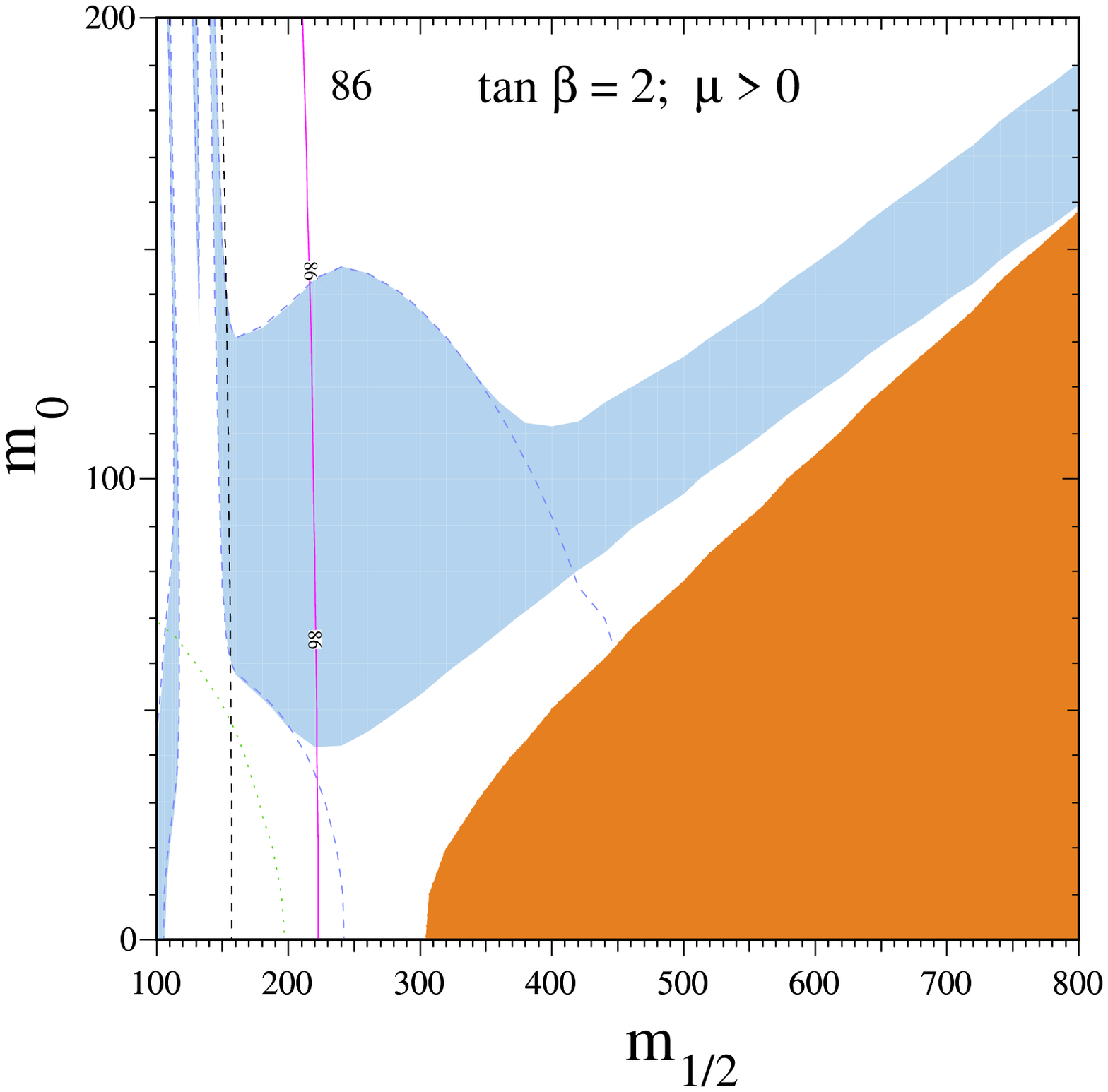} 
\end{minipage}
\begin{minipage}{8in}
\vspace*{0.4in}
\hspace{0.7in}
\includegraphics[width=5.truecm]{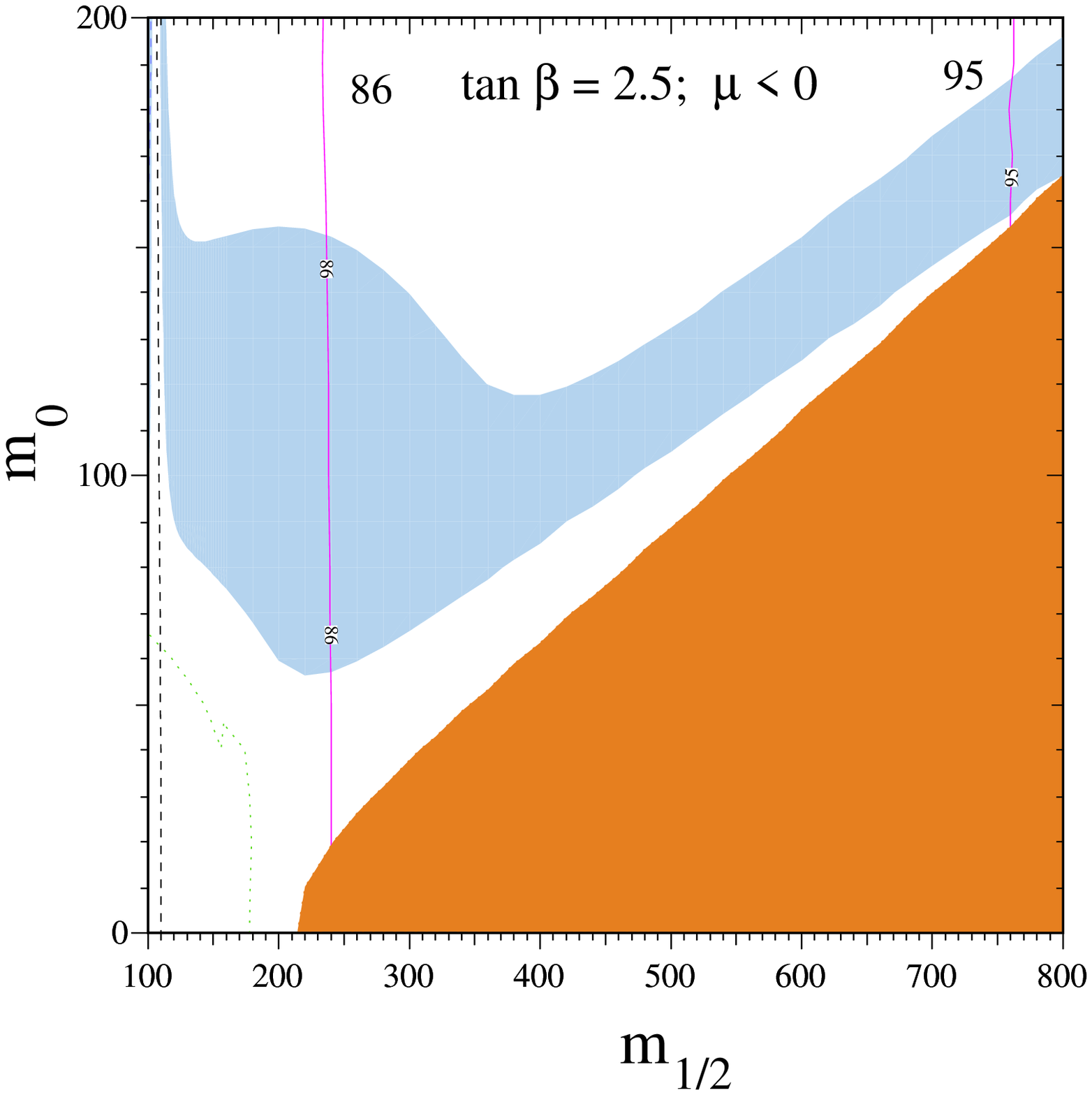} 
\hspace*{.2in}
\includegraphics[width=5.truecm]{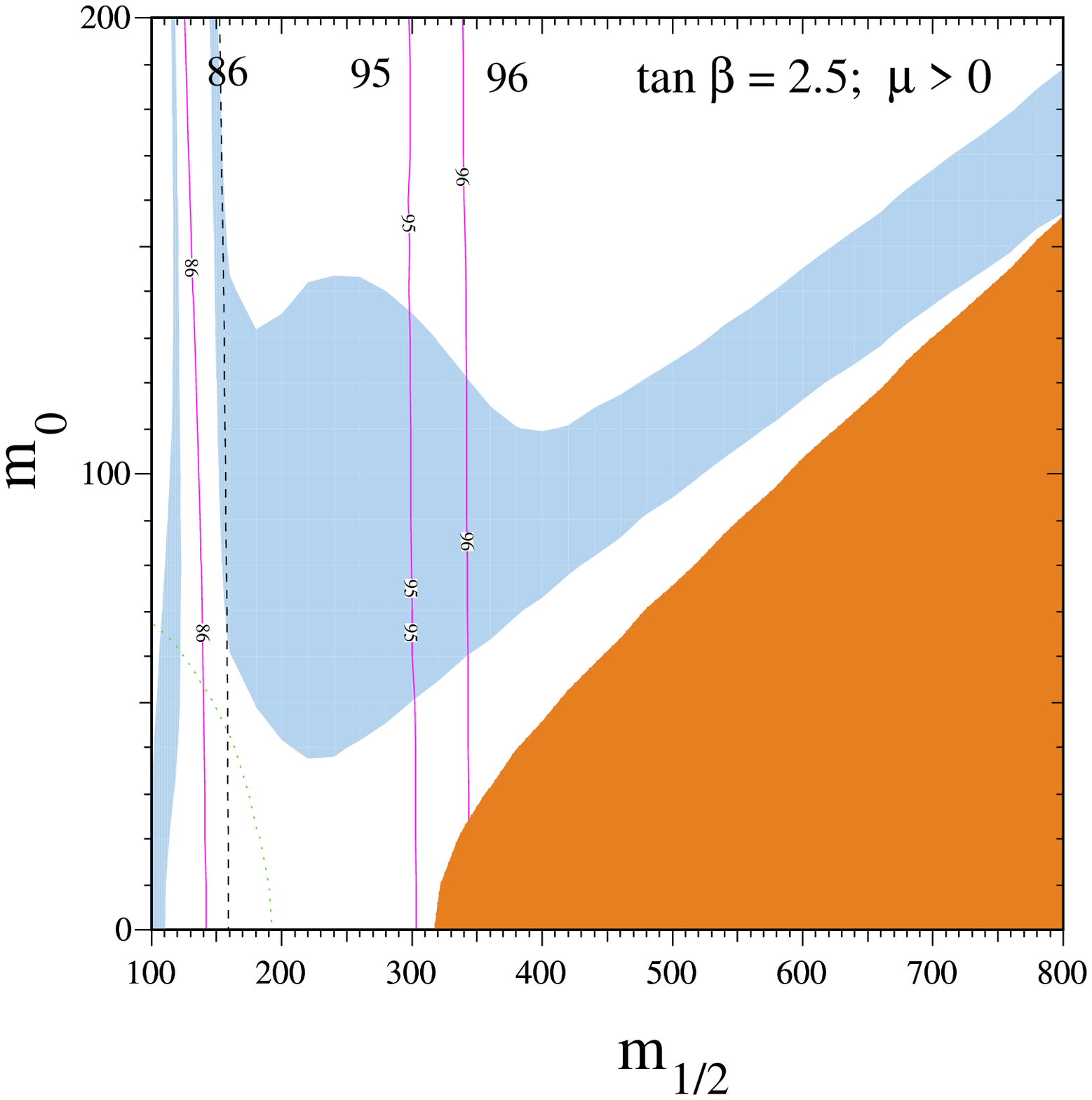} 
\end{minipage}
\begin{minipage}{8in}
\vspace*{0.4in}
\hspace{0.7in}
\includegraphics[width=5.truecm]{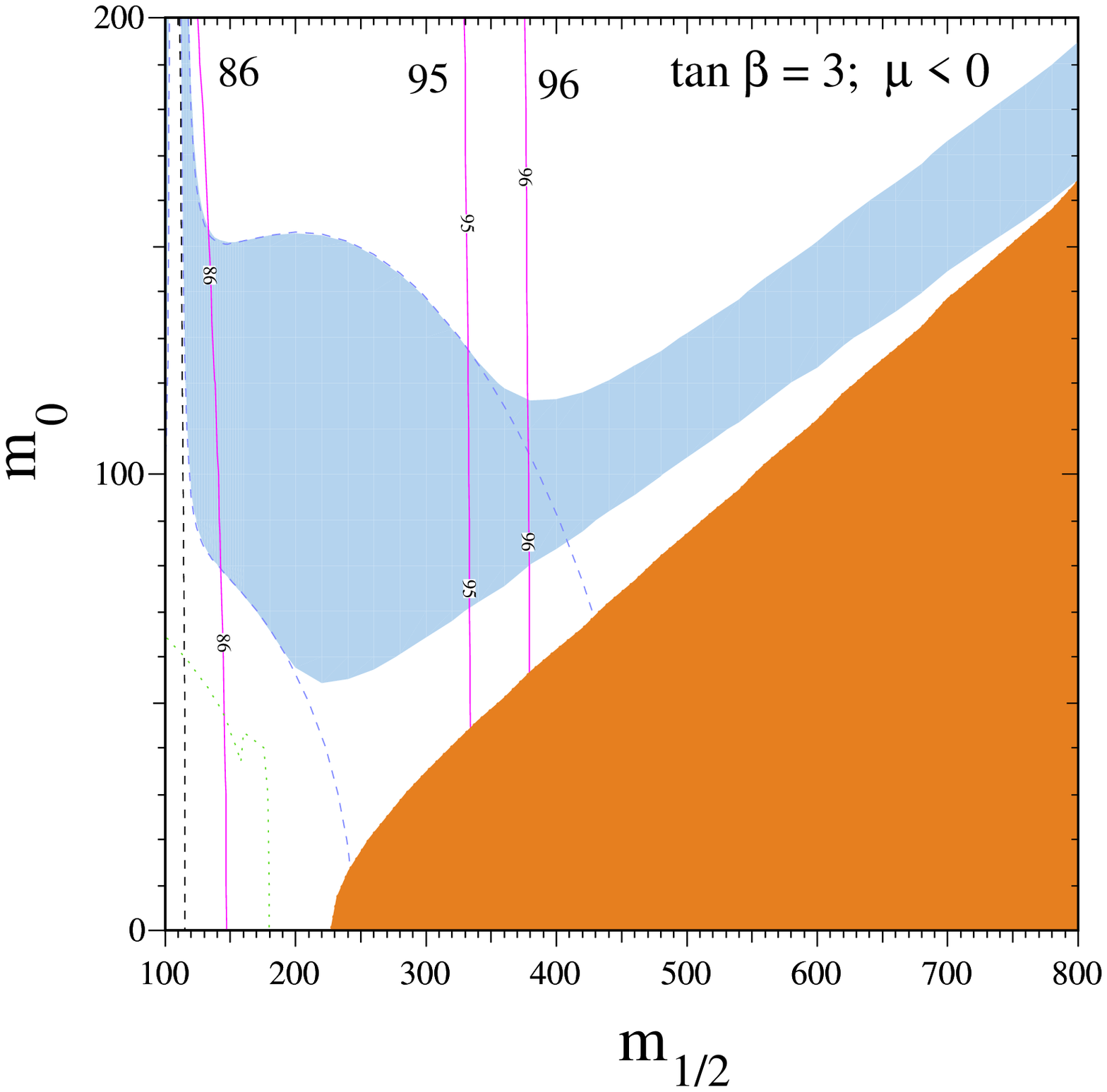} 
\hspace*{.2in}
\includegraphics[width=5.truecm]{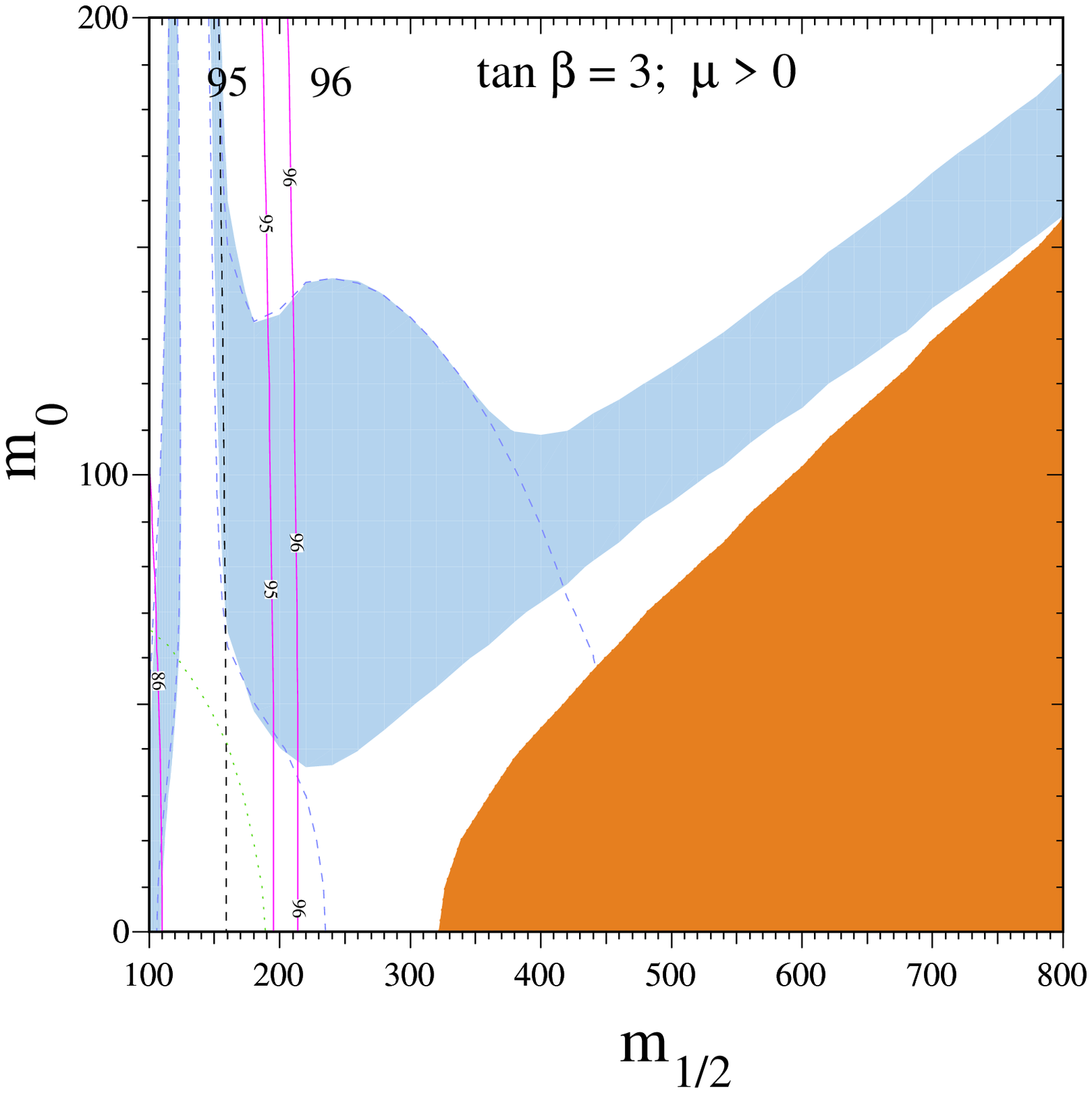} 
\end{minipage}
\caption{As in \protect\ref{efoblock} for $\tan \beta = 2, 2.5$, and 3, for
both $\mu > 0$ and $\mu <0$. Higgs mass contours of 86, 95, and 96 GeV, are
displayed to show the dependence on $\tan \beta$. I thank Toby Falk for
providing these figures.}
\label{series}
\end{figure}

In the MSSM, in the region of the $M_2$-$\mu$ plane, where $H_{(12)}$ is the LSP
(that is, at large $\mu$ and very large $M_2$), the next lightest neutralino or
NLSP, is the $H_{[12]}$ state and the two are nearly degenerate with masses close
to $\mu$. In this case, additional annihilation channels (or co-annihilation)
which involve both a $H_{(12)}$ and a $H_{[12]}$ in the initial state become
important \cite{gs,co2}. The enhanced annihilation of Higgsinos lowers the
relic density substantially and virtually eliminates the Higgsino as a viable
dark matter candidate. In the CMSSM, co-annihilation is also important, now
at large values of
$m_{1/2}$ \cite{efo}.  Recall, that previously we discussed a possible upper limit
to the mass of the neutralino in the CMSSM, where the cosmologically allowed region
of Figure \ref{rd2c} runs into the region where the $\t \tau_R$ is the LSP. 
Along the boundary of this region, the neutralino $\wt B$, and the $\t \tau_R$ are
degenerate. Therefore close to this boundary, new co-annihilation channels involving
both $\wt B$ and  $\t \tau_R$ (as well as the other right-handed sleptons which are
also close in mass) in the initial state become important.  As in the MSSM, the
co-annihilations reduce the relic density and as can be seen in Figures
\ref{efoblock} the upper limit to the neutralino is greatly relaxed \cite{efo}. 
As one can plainly see, the cosmologically allowed region is bent way from the 
$\t \tau_R$-LSP region extending to higher values of $m_{1/2}$.
Also shown in this set of figures are the iso-mass contours for charginos, sleptons
and Higgses, so that the LEP limits discussed above can be applied. In these
figures, one can see the sensitivity of the Higgs mass with $\tan \beta$.

Despite the importance of the coannihilation, there is still an upper limit to the
neutralino mass.  If we look at an extended version of Figures \ref{efoblock}, as
shown in Figures \ref{efoblock2}, we see that eventually, the cosmologically allowed 
region intersects the $\t \tau_R$-LSP region.  
The new upper limit occurs at $m_{1/2} \approx 1450$ GeV, implying that $m_\chi \la
600$ GeV.

As noted above, the Higgs mass bounds can be used to exclude low values of 
$\tan \beta$. However, when co-annihilations are included, the limits on $\tan
\beta$ is weakened. In the
$m_{1/2} - m_0$ plane, the Higgs iso-mass contour appears  nearly vertical
for low values of
$m_0$.  A given contour is highly dependent on
$\tan \beta$, and moves to the right quickly as $\tan \beta $ decreases.  This
behavior is demonstrated in the series of Figures \ref{series}, which show the
positions of the Higgs mass contours for $\tan \beta = 2,2.5,$ and 3, for both
positive and negative $\mu$. If we concentrate, on the 95 GeV contour, we see
that at $\tan \beta = 3$, the bulk of the cosmological region is allowed. At
$\tan
\beta = 2.5$, much of the bulk is excluded, though the trunck region is
allowed. At the lower value of $\tan \beta = 2$, the 95 GeV contour is off
the scale of this figure.  A thorough examination \cite{efo} yields a limit
$\tan \beta > 2.2$ for
$\mu < 0$. For positive $\mu$, the results are qualitatively similar. The Higgs
mass contours are farther to the left (relative to the negative $\mu$ case), and
the limit on $\tan \beta$ is weaker. Nevertheless the limit is $\tan \beta > 1.8$
for $\mu > 0$.

As the runs at LEP are winding down, the prospects for improving the
current limits or more importantly discovering supersymmetry are diminishing. 
Further progress will occur in the future with Run II at the Tevetron at ultimately
and the LHC. Currently, while we have strong and interesting limits on the
MSSM and CMSSM parameter space for LEP, much of the phenomenological and
cosmological parameter space remains viable.

\vskip .3in

{\large\bf Acknowledgments} I would like to thank B. Campbell, S. Davidson, J.
Ellis, K. Enqvist, T. Falk, M.K. Gaillard, G. Ganis, J. Hagelin, K. Kainulainen, R.
Madden, J. McDonald,  H. Murayama, D.V. Nanopoulos, S.J. Rey, M. Schmitt, M.
Srednicki, K. Tamvakis, R. Watkins for many enjoyable collaborations which
have been summarized in these lectures. I would also like to thank J. Ellis, 
T. Falk, S. Lee, and M. Voloshin for assistance in the preparation of these
lecture notes.
 This work was supported in part by 
DOE grant DE-FG02-94ER40823 at Minnesota.


\begin{thebibliography}{99}

\bibitem{GL} Y.A. Gol'fand and E.P. Likhtman, {\it Pis'ma Zh.Eksp.Teor.Fiz.}
{\bf 13} (1971) 323; \\
 P. Ramond, \PR {\bf D3} (1971) 2415;\\
A. Neveu and J.H. Schwarz, \PR {\bf D4} (1971) 1109; \\
D.V. Volkov and V.P. Akulov, \PL {\bf 46B} (1973) 109.

\bibitem{WZ} J. Wess and B. Zumino, \NP {\bf B70} (1974) 39.

\bibitem{hier} L. Maiani, {\it Proc. Summer School on Particle
Physics},
Gif-sur-Yvette, 1979 (IN2P3, Paris, 1980) p. 3;\\
G't Hooft, in:  G't Hooft et al., eds., {\it Recent Developments in Field
Theories}
(Plenum Press, New York, 1980);\\
E. Witten, \NP {\bf B188} (1981) 513;\\
R.K. Kaul, \PL {\bf 109B} (1982) 19.

\bibitem{EKN} J. Ellis, S. Kelley and D.V. Nanopoulos, \PL {\bf B249} (1990)
441; \\
J. Ellis, S. Kelley and D.V. Nanopoulos, \PL {\bf B260}
(1991) 131;\\
U. Amaldi, W. de Boer and H. Furstenau, \PL {\bf B260} (1991) 447;\\
P. Langacker and M. Luo, \PR {\bf D44} (1991) 817.

\bibitem{bd} J.D. Bjorken and S.D. Drell. {\it Relativistic Quantum
Mechanics} (McGraw Hill, New York, 1964).

\bibitem{FF} P. Fayet and S. Ferrara, {\it Phys.Rep.} {\bf 32} (1977) 251.

\bibitem{BW} J.~Wess and J.~Bagger,
{\em Supersymmetry and Supergravity},
(Princeton University Press, Princeton NJ, 1992)

\bibitem{ross} G.G.~Ross, {\em Grand Unified Theories},
(Addison-Wesley, Redwood City CA, 1985).

\bibitem{martin} S. Martin, hep-ph/9709356.

\bibitem{ellis} J. Ellis, hep-ph/9812235.

\bibitem{CM} S. Coleman and J. Mandula, \PR {\bf 159} (1967) 1251.

\bibitem{HLS} R. Haag, J. Lopusz\'anski and M. Sohnius, \NP {\bf B88}
(1975) 257.

\bibitem{gg} L.~Girardello and M.T.~Grisaru
{\NP}{\bf B194} (1982) {65}.


\bibitem{fay} P.~Fayet,
{\PL}{\bf B64} (1976) {159};
{\PL}{\bf B69} (1977) {489};
{\PL}{\bf B84} (1979) {416}.


\bibitem{MSSM} H.E. Haber and G.L. Kane, {\it Phys.Rep.} {\bf 117} (1985) 75.

\bibitem{susyHiggs} Y. Okada, M. Yamaguchi and T. Yanagida, 
{\it Progr.Theor.Phys.} {\bf 85} (1991) 1;\\
J. Ellis, G. Ridolfi and F. Zwirner, \PL {\bf B257} (1991) 83, \PL {\bf B262}
(1991) 477;\\
H.E. Haber and R. Hempfling, \PRL {\bf 66} (1991) 1815;\\
R. Barbieri, M. Frigeni and F. Caravaglios, \PL {\bf B258} (1991) 167;\\
Y. Okada, M. Yamaguchi and T. Yanagida, \PL {\bf B262} (1991) 54 \\
H.E. Haber, hep-ph/9601330.

\bibitem{moreradcorr} M. Carena, M. Quiros and C.E.M. Wagner, \NP {\bf
B461} (1996) 407; \\
H.E. Haber, R. Hempfling and A.H. Hoang, {\it Zeit.Phys.} {\bf C75} (1997) 539.

\bibitem{ER} J. Ellis and S. Rudaz, \PL {\bf 128B} (1983) 248.



\bibitem{ehnos} J. Ellis, J.S. Hagelin, D.V. Nanopoulos, K.A. Olive and M.
Srednicki, \NP {\bf B238} (1984) 453.

\bibitem{fi} P. Fayet and J. Iliopoulos, \PL {\bf 51B} (1974) 461.

\bibitem{oraif} L. O'Raifeartaigh, \NP {\bf B96} (1975) 331.

\bibitem{Rparity} G.R.~Farrar and P.~Fayet,
{\PL}{\bf B76} (1978) {575}.

\bibitem{gold} H. Goldberg, \PRL {\bf 50} (1983) 1419.

\bibitem{EN} J. Ellis and D.V. Nanopoulos, \PL {\bf 110B} (1982) 44.

\bibitem{DG} S. Dimopoulos and H. Georgi, \NP {\bf B193} (1981) 150.

\bibitem{Inoue} K. Inoue, A. Kakuto, H. Komatsu and S. Takeshita,
{\it Prog.Theor.Phys.} {\bf 68} (1982) 927; {\bf 71} (1984) 413.

\bibitem{DRW} S. Dimopoulos and H. Georgi, \NP {\bf B193} (1981)
50;\\
S. Dimopoulos, S. Raby and F. Wilczek, \PR {\bf D24} (1981)
1681;\\
L. Ib\`a\~nez and G.G. Ross, \PL {\bf 105B} (1981) 439.

\bibitem{rge} M. Drees and M.M. Nojiri, {\it Nucl. Phys.} {\bf B369} (1992)
54; \\
W. de Boer, hep-ph/9402266.

\bibitem{spm}
S.P.~Martin and M.T.~Vaughn, {\PR}{\bf D50} (1994) {2282}; \\
Y.~Yamada,  {\PR}{\bf D50} (1994) {3537}; \\
I.~Jack and D.R.T.~Jones, {\PL}{\bf B333} (1994) {372}; \\
I.~Jack, D.R.T.~Jones, S.P.~Martin, M.T.~Vaughn
and Y.~Yamada, {\PR}{\bf D50} (1994) {5481}; \\
P.M.~Ferreira, I.~Jack, D.R.T.~Jones,
{\PL}{\bf B387} (1996) {80}; \\
I. Jack, D.R.T. Jones and A. Pickering, \PL {\bf B432}
(1998) 114.

\bibitem{nrt}
A.~Salam and J.~Strathdee, {\PR}{\bf D11} (1975) {1521}; \\
M.T.~Grisaru, W.~Siegel and M.~Rocek, {\NP}{\bf B159} (1979) {429}.

\bibitem{rewsb}
L.E.~Ib\'a\~nez and G.G.~Ross, {\PL}{\bf B110} (1982) {215}; \\
L.E.~Ib\'a\~nez, {\PL}{\bf B118} (1982) {73}; \\
J.~Ellis, D.V.~Nanopoulos and K.~Tamvakis,
{\PL}{\bf B121} (1983) {123}; \\
J.~Ellis, J. Hagelin, D.V.~Nanopoulos and K.~Tamvakis,
{\PL}{\bf B125} (1983) {275}; \\
L.~Alvarez-Gaum\'e, J.~Polchinski, and M.~Wise,
{\NP}{\bf B221} (1983) {495}.

\bibitem{kane} G. Kane, C. Kolda, L. Roszkowski, J. Wells, {\it Phys. Rev.}
{\bf D49} (1994) 6173.


\bibitem{phases} M. Dugan, B. Grinstein and L. Hall, {\it Nucl. Phys.} {\bf
B255}, (1985) 413; \\ R. Arnowitt, J.L. Lopez and D.V. Nanopoulos, 
{\it Phys. Rev.} {\bf D42} (1990) 2423; \\ R. Arnowitt, M.J. Duff and K.S.
Stelle, {\it Phys. Rev.} {\bf D43} (1991) 3085; \\
Y. Kizukuri \& N. Oshimo, {\it Phys. Rev.} {\bf D45} (1992) 1806; {\bf D46}
(1992) 3025; \\
 T. Ibrahim and P. Nath, {\it Phys. Lett.} {\bf B418} (1998) 98;  {\it Phys.
Rev.} {\bf D57} (1998) 478;   {\bf D58} (1998) 111301; \\ M. Brhlik, G. J.
Good and G.L. Kane, {\it Phys. Rev.} {\bf D59} (1999) 115004; \\ A. Bartl, T.
Gajdosik, W. Porod, P. Stockinger, and H. Stremnitzer, {\it Phys. Rev.} {\bf
D60} (1999) 073003; \\
T. Falk, K.A. Olive, M. Pospelov, and R. Roiban, hep-ph/9904393.

\bibitem{fkos}T. Falk, K.A. Olive, and M. Srednicki, {\it Phys. Lett.}
{\bf B354} (1995) 99
\bibitem{fko} T. Falk and K.A. Olive, {\it Phys. Lett.} {\bf B375}
 (1996) 196; {\bf B439} (1998) 71.


\bibitem{sugr} D.Z. Freedman, P. Van Nieuwenhuizen and S. Ferrara, \PR {\bf
D13} (1976) 3214; \\
S. Deser and B. Zumino, \PL {\bf 62B} (1976) 335; \\
D.Z.~Freedman and  P.~van Nieuwenhuizen, {\PR}{\bf D14} (1976) {912}; \\
see also P. Van Nieuwenhuizen, {\it Phys. Rep.} {\bf 68C} (1981) 189.

\bibitem{sugr2}
E. Cremmer, B. Julia, J. Scherk, S. Ferrara, L. Girardello and P. Van
     Nieuwenhuizen, \PL {\bf 79B} (1978) 231; and \NP {\bf B147}
     (1979) 105; \\ E. Cremmer, S. Ferrara, L. Girardello and A. Van Proeyen,
     \PL {\bf 116B} (1982) 231; and \NP {\bf B212} (1983) 413; \\ R.
     Arnowitt, A.H. Chamseddine and P. Nath, \PRL {\bf 49} (1982) 970;
     {\bf 50} (1983) 232; and \PL {\bf 121B} (1983) 33; \\ J. Bagger and E.
     Witten, \PL {\bf 115B} (1982) 202; {\bf 118B} (1982) 103; \\ J.
     Bagger,  \NP {\bf B211} (1983) 302.


\bibitem{polonyi} J. Polonyi, Budapest preprint KFKI-1977-93 (1977).

\bibitem{superh} D.V. Volkov and V.A. Soroka, {\it JETP Lett.} {\bf 18}
(1973) 312; \\ S. Deser and B.
     Zumino, \PRL {\bf 38} (1977) 1433.

\bibitem{carlos} R. Barbieri, S. Ferrara, and C.A. Savoy, \PL {\bf 119B}
(1982) 343.

\bibitem{mark} H.-P. Nilles, M. Srednicki, and D. Wyler, \PL {\bf 120B}
(1983) 345; \\
     L.J. Hall, J. Lykken and S. Weinberg, \PR {\bf D27} (1983) 2359.

\bibitem{noscale} 
	E. Cremmer, S. Ferrara, C. Kounnas, and D.V. Nanopoulos, \PL
     {\bf 133B} (1983) 61; \\ J. Ellis, C. Kounnas and D.V. Nanopoulos, \NP
     {\bf B241} (1984) 429; \\
J. Ellis, A.B. Lahanas, D.V. Nanopoulos, and K.
     Tamvakis, {\it Phys. Lett.} {\bf 134B} (1984) 429.

\bibitem{sns} E. Witten, {\it Phys.\ Lett.} {\bf 155B} (1985) 151; \\
S. Ferrara, C. Kounnas and M. Porrati, {\it Phys. Lett.} 
{\bf 181B} (1986) 263; \\
L.J. Dixon, V.S. Kaplunovsky and J. Louis, {\it Nucl. Phys.} 
{\bf B329} (1990) 27; \\ S. Ferrara, D. L\"ust, and S. Theisen, 
{\it Phys. Lett.} {\bf 233B} (1989) 147.


\bibitem{nsguts} 
J. Ellis, C. Kounnas, and D.V. Nanopoulos,  {\it Nucl. Phys.}
     {\bf B247} (1984) 373; \\ for a review see: A.B. Lahanas 
and D.V. Nanopoulos, {\it Phys. Rep.} {\bf 145} (1987) 1.

\bibitem{nshid} J.D. Breit, B. Ovrut, and G. Segr\'e, \PL {\bf
162B}  (1985) 303; \\
P.\ Bin\'etruy and M.K. Gaillard, \PL {\bf 168B} (1986)  347; \\
P.\ Bin\'etruy, S.\ Dawson M.K. Gaillard and I.\ Hinchliffe, 
\PR {\bf D37} (1988) 2633 and references therein.



\bibitem{pprob} G.D. Coughlan, W. Fischler, E.W. Kolb, S. Raby and G.G. Ross,
\PL {\bf 131B} (1983) 59.

\bibitem{ks} E.W. Kolb and R. Scherrer, \PR {\bf D25} (1982) 1481.

\bibitem{grprob} S. Weinberg, \PRL {\bf 48} (1982) 1303.

\bibitem{oss} G. Steigman, K.A. Olive and D.N. Schramm, \PRL {\bf 43}
     (1979) 239; \\ K.A. Olive, D.N. Schramm and G. Steigman, \NP {\bf B180}
     (1981) 497.

\bibitem{grinf} J. Ellis, A.D. Linde and D.V. Nanopoulos, \PL {\bf 118B}
(1982) 59.

\bibitem{grnuc} J. Ellis, J.E. Kim and D.V. Nanopoulos, \PL {\bf 145B}
(1984) 181; \\
	J. Ellis, D.V. Nanopoulos and S. Sarkar, \NP {\bf B259} (1985) 175; \\
R. Juszkiewicz, J. Silk and A. Stebbins, \PL {\bf 158B} (1985) 463; \\
     D. Lindley, \PL {\bf B171} (1986) 235; \\ M. Kawasaki and K. Sato, \PL
{\bf B189} (1987) 23.

\bibitem{enor} J. Ellis, D.V. Nanopoulos, K.A. Olive, and S.J. Rey, 
{\it Astropart. Phys.} {\bf 4} (1996) 371.

\bibitem{km} M. Kawasaki and T. Moroi,
{\it Prog. Theor. Phys.} {\bf 93} (1995) 879.

\bibitem{infl} A.D. Linde, {\it Particle
Physics And Inflationary Cosmology} (Harwood, 1990); \\ K.A. Olive,
Phys. Rep. {\bf 190} (1990) 181; \\ D. Lyth and A. Riotto,
{\it Phys. Rept.} {\bf 314} (1999) 1.

\bibitem{pri}J. Ellis, D. V. Nanopoulos, K. A. Olive and K. Tamvakis,
{\it Nucl. Phys.} {\bf B221} (1983) 224; {\it Phys. Lett.} {\bf 118B} (1982)
335.



\bibitem{new}A.D. Linde, {\it Phys. Lett.} {\bf 108B} (1982) 389; \\
A. Albrecht and P.J. Steinhardt, {\it Phys. Rev. Lett.} {\bf 48} (1982) 1220.


\bibitem{pert}W.H. Press, {\it Phys. Scr.} {\bf 21} (1980) 702; \\
	V.F. Mukhanov and G.V. Chibisov, {\it JETP Lett.} {\bf  33} (1981) 532;\\
	S.W. Hawking, {\it Phys. Lett.} {\bf 115B} (1982) 295;\\
	A.A. Starobinsky,  {\it Phys. Lett.} {\bf 117B} (1982) 175;\\
	A.H. Guth and S.Y. Pi, {\it Phys. Rev. Lett.} {\bf 49} (1982) 1110;\\
	J.M. Bardeen, P.J. Steinhardt and M.S. Turner, {\it Phys. Rev.} 
{\bf D28} (1983) 679.

\bibitem{lin2}A.D. Linde, {\it Phys. Lett.} {\bf 116B} (1982) 335.

\bibitem{dens}J. Ellis, D.V. Nanopoulos, K.A. Olive and K. Tamvakis,
{\it Phys. Lett.} {\bf 120B} (1983) 334.

\bibitem{nost1}D. V. Nanopoulos, K. A. Olive, M. Srednicki and K. Tamvakis,
{\it Phys. Lett.} {\bf 123B} (1983) 41.
\bibitem{hrg}R. Holman, P. Ramond and G. G. Ross,
{\it Phys. Lett.} {\bf 137B} (1984) 343.

\bibitem{eenos}J. Ellis, K Enquist, D. V. Nanopoulos,
 K. A. Olive and M. Srednicki, {\it Phys. Lett.} {\bf 152B} (1985) 175.

\bibitem{bg}P. Binetruy and M.K. Gaillard, \PR {\bf  
D34}(1986)3069.


\bibitem{cobe}G.F. Smoot et al. {\it Ap.J} {\bf 396} (1992) L1; \\
E.L. Wright et al. {\it Ap.J.} {\bf 396} (1992) L13.

\bibitem{cdo}B. Campbell, S. Davidson, and K.A. Olive, 
{\it Nucl.Phys.} {\bf B399}
(1993) 111.

\bibitem{eeno}J. Ellis, K. Enqvist, D.V. Nanopoulos, and K.A. Olive,  
{\it Phys. Lett.} {\bf B191}(1987) 343.



\bibitem{sak} A.D. Sakharov, {\it JETP Lett.} {\bf 5} (1967) 24.

\bibitem{ww} S. Weinberg, {\it Phys. Rev. Lett.} {\bf 42}, (1979) 850; \\
	D. Toussaint, S. B. Treiman, F. Wilczek, and A. Zee, {\it Phys. Rev.}
 {\bf D19} (1979) 1036.

\bibitem{dlnos} A.D. Dolgov, and A.D. Linde, {\it Phys. Lett.} {\bf B116}
(1982) 329;\\ D.V. Nanopoulos, K.A. Olive, and M. Srednicki, 
{\it Phys. Lett.} {\bf B127} (1983) 30.

\bibitem{sy} N. Sakai and T. Yanagida, \NP {\bf B197} (1982) 533; \\
S. Weinberg, \PR {\bf D26} (1982) 287.

\bibitem{enr} J. Ellis, D.V. Nanopoulos, and S. Rudaz, {\it Nucl. Phys.}
{\bf B202} (1982) 43; \\
S. Dimopoulos, S. Raby and F. Wilczek, \PL {\bf 112B} (1982) 133.

\bibitem{nt} D.V. Nanopoulos and K. Tamvakis, {\it Phys. Lett.}
{\bf B114} (1982) 235.
\bibitem{ad}I. Affleck and M. Dine, {\it Nucl. Phys.} {\bf B249} (1985) 361.
\bibitem{lin} A.D. Linde, {\it Phys. Lett.} {\bf B160} (1985) 243.

\bibitem{drt} M.Dine, L. Randall, and S. Thomas,
{\it Phys. Rev. Lett.} {\bf 75} (1995) 398; {\it Nucl.Phys.}
{\bf B458} (1996) 291.

\bibitem{heis} P.~Binetruy and M.K.~Gaillard, \PL {\bf B195}
(1987) 382.

\bibitem{gmo} M.K. Gaillard, H. Murayama, and K.A. Olive, \PL
{\bf B355} (1995) 71.

\bibitem{tr} M.K.~Gaillard and V.~Jain, \PR {\bf D49}
(1994) 1951; \\
M.K.~Gaillard, V.~Jain and K. Saririan, \PL {\bf B387}
 (1996) 520 and \PR {\bf D55}  (1997) 833.

\bibitem{eno} J. Ellis, D.V. Nanopoulos, and K.A. Olive,  
\PL {\bf  B184} (1987) 37.

\bibitem{cgmo} B.A. Campbell, M.K. Gaillard, H. Murayama, and K.A.
Olive, {\it Nucl. Phys.} {\bf B538} (1999) 351. 

\bibitem{KRS} V.A. Kuzmin, V.A. Rubakov, and M.E. Shaposhnikov,  Phys. 
Lett. {\bf 191B} 171 (1987); \\ see also, H. Dreiner and G. Ross, \NP
{\bf B410} (1993) 188; \\ S. Davidson, K. Kainulainen, and K.A. Olive, \PL 
{\bf B335} (1994) 339. 

\bibitem{dm}see:  J.R. Primack  in {\it Enrico Fermi.  Course 92}, 
ed. N. Cabibbo (North Holland, Amsterdam, 1987), p. 137; \\
	V. Trimble, {\it Ann. Rev. Astron. Astrophys.} {\bf 25} (1987) 425; \\
	J. Primack, D. Seckel, and B. Sadulet, {\it Ann. Rev. Nucl. Part. Sci.} 
{\bf 38} (1988) 751; \\
	{\it Dark Matter}, ed. M. Srednicki (North-Holland, Amsterdam,1989).

\bibitem{osw2} K.A. Olive, G. Steigman, and T.P. Walker, {\it Phys. Rep. }
(in press), astro-ph/9905320 

\bibitem{hio} D.J. Hegyi, and K.A. Olive, {\it Phys. Lett.}
{\bf 126B} (1983) 28;
 {\it Ap. J.} {\bf 303} (1986) 56.

\bibitem{snov} S. Perlmutter et al., {\it Nature} {\bf 391} (1998) 51; \\
A.G. Riess et al., {\it Astron. J.} {\bf 116} (1998) 1009.

\bibitem{isotopes} J. Rich, M. Spiro and J. Lloyd-Owen, {\it Phys.Rep.} {\bf
151} (1987) 239; \\
P.F. Smith, {\it Contemp.Phys.} {\bf 29} (1998) 159; \\
T.K. Hemmick et al., \PR {\bf D41} (1990) 2074.

\bibitem{snu}L.E. Ibanez, {\it Phys. Lett.} {\bf 137B} (1984) 160; \\
	J. Hagelin, G.L. Kane, and S. Raby, 
{\it Nucl., Phys.} {\bf B241} (1984) 638; \\
T. Falk, K.A. Olive, and M. Srednicki, 
{\it Phys. Lett.} {\bf B339} (1994) 248.

\bibitem{dir}S. Ahlen, et. al., {\it Phys. Lett.} {\bf B195} (1987) 603; \\
	D.D. Caldwell, et. al., {\it Phys. Rev. Lett.} {\bf 61} (1988) 510; \\
M. Beck et al., {\it Phys. Lett.} {\bf B336} (1994) 141.

\bibitem{indir} see e.g.	K.A. Olive and M. Srednicki,
 {\it Phys. Lett.} {\bf 205B} (1988) 553.

\bibitem{accel} The LEP Collaborations ALEPH, DELPHI, L3, OPAL
and the LEP Electroweak Working Group, CERN preprint PPE/95-172 (1995).

\bibitem{EFOS} J. Ellis, T. Falk, K. Olive and M. Schmitt, \PL {\bf B388}
(1996) 97.

\bibitem{uground} H.V. Klapdor-Kleingrothaus and Y. Ramachers,
{\it Eur.Phys.J.} {\bf A3} (1998) 85.

\bibitem{osi3}K.A. Olive and M. Srednicki, {\it Phys. Lett.} {\bf B230}
 (1989) 78;
{\it Nucl. Phys.} {\bf  B355} (1991) 208.




\bibitem{lep2} ALEPH collaboration, D. Decamp et al., 
Phys. Rep. {\bf 216} (1992) 253; \\
L3 collaboration, M. Acciarri et al., 
Phys. Lett. {\bf B350} (1995) 109; \\
OPAL collaboration, G. Alexander et al., 
Phys. Lett. {\bf B377} (1996) 273.

\bibitem{lep15}ALEPH collaboration, ALEPH Collaboration, D.Buskulic et al.,
Phys. Lett. {\bf B373} (1996) 246; \\
OPAL Collaboration, G. Alexander et al.,
\PL {\bf B377} (1996) 181; \\
L3 Collaboration, M. Acciarri et al.,
\PL {\bf B377} (1996) 289; \\
DELPHI Collaboration, P. Abreu et al.,
\PL {\bf B382} (1996) 323.

\bibitem{LEP2} ALEPH collaboration, R. Barate et al.,
\PL {\bf B412} (1997) 173; {\it Eur. Phys. J.} {\bf C2} (1998) 417; \\ DELPHI
collaboration, P. Abreu, {\it Eur. Phys. J.} {\bf C1} (1998) 1; {\it Eur.
Phys. J.} {\bf C2} (1998) 1;\\ L3 collaboration, M. Acciarri et al.,
\PL {\bf B411} (1997) 373;  {\it Eur. Phys. J.} {\bf C4} (1998) 207;\\ OPAL
collaboration, K. Ackerstaff et al., {\it Eur. Phys. J.} {\bf C1} (1998) 425;
{\it Eur. Phys. J.} {\bf C2} (1998) 213.


\bibitem{phot}
	L.M. Krauss, {\it Nucl. Phys.} {\bf B227} (1983) 556.

\bibitem{wso}R. Watkins, M. Srednicki and K.A. Olive, 
{\it Nucl. Phys.} {\bf  B310}
 (1988) 693.

\bibitem{lw}P. Hut, {\it Phys. Lett.} {\bf  69B} (1977) 85; \\
	B.W. Lee and S. Weinberg, {\it Phys. Rev. Lett.} {\bf  39} (1977) 165.

\bibitem{mcdos}J. McDonald, K. A. Olive and M. Srednicki, {\it Phys. Lett.}
 {\bf B283} (1992) 80.

\bibitem{dn}M. Drees and M.M. Nojiri, {\it Phys. Rev.} {\bf D47} (1993) 376.

\bibitem{jkg}
 G.\ Jungman, M.\ Kamionkowski,
and K.\ Griest, \PR {\bf 267}, 195 (1996).

\bibitem{bb} H. Baer and M. Brhlik, \PR {\bf D53} (1996)
597.

\bibitem{fkmos}T. Falk, R. Madden, K.A. Olive, and M. Srednicki,
 {\it Phys. Lett.}
{\bf B318} (1993) 354.


\bibitem{efgos} J. Ellis, T. Falk, G. Ganis, K.A. Olive and M. Schmitt,
\PR {\bf D58} (1998) 095002.

\bibitem{gs}K. Griest and D. Seckel, \PR {\bf D43} (1991) 3191.

\bibitem{co2} S. Mizuta and M. Yamaguchi, \PL {\bf B298} (1993)
120.

\bibitem{dnr} M. Drees, M.M. Nojiri, D.P. Roy, and Y. Yamada,
\PR {\bf D56} (1997) 276.

\bibitem{gkt}K. Greist, M. Kamionkowski, and M.S. Turner, {\it Phys. Rev.}
{\bf D41} (1990) 3565.

\bibitem{efo} J. Ellis, T. Falk, and K. Olive, \PL {\bf B444} (1998)
367; \\
J. Ellis, T. Falk, K. Olive, and M. Srednicki, {\it Astr. Part. Phys.} (in
press), hep-ph/9905481. 


\bibitem{achi}ALEPH Collaboration, D. Buskulic et al.,
{\it Z. Phys.} {\bf C72} (1996) 549.

\bibitem{efos2} J. Ellis, T. Falk, K.A. Olive and M. Schmitt,
\PL {\bf B413} (1997) 355. 



\bibitem{D0}
D0 collaboration, S. Abachi et al., \PRL {\bf 75} (1995) 618;
\\
CDF collaboration, F. Abe et al., \PRL {\bf 76} (1996) 2006;
\PR {\bf D56} (1997) 1357.

\bibitem{latest} DELPHI
collaboration, P. Abreu, \PL {\bf B446} (1999) 75; \\ OPAL
collaboration, G. Abbiendi et al., {\it Eur. Phys. J.} {\bf C8} (1999) 255; \\
see also:
The LEP Working Group for Higgs Boson Searches, ALEPH, DELPHI, L3 and
OPAL, CERN-EP/99-060.

\bibitem{lep189} Recent official compilations of LEP limits on supersymmetric
particles are available from:
{\tt http://www.cern.ch/LEPSUSY/}.


\end{thebibliography}
\end{document}